\documentclass[11pt,fleqn]{article} 
\usepackage{graphicx}
\usepackage{textcomp} 
\usepackage[pdftex]{pict2e}
\usepackage{epstopdf}
\usepackage{latexsym}
\usepackage{amsmath}
\usepackage{amssymb}
\usepackage{bm}
\usepackage[mathcal]{euscript}
\usepackage{slashed}
\usepackage[boxsize=0.3em]{ytableau}

\usepackage{tikz}
\usetikzlibrary{decorations.markings}
\usetikzlibrary{arrows}
\usetikzlibrary{backgrounds}
\usepackage[compat=1.1.0]{tikz-feynman}

\usepackage[top=30mm, bottom=30mm, left=25mm, right=25mm]{geometry}

\numberwithin{equation}{section}

\usepackage{marvosym}

\newcommand\blfootnote[1]{
  \begingroup
  \renewcommand\thefootnote{}\footnote{#1}
  \addtocounter{footnote}{-1}
  \endgroup
}

\usepackage{hyperref}
\hypersetup{
        unicode,	
        colorlinks,
        citecolor=blue,
        linkcolor=blue, 
        urlcolor=red,
        bookmarksopen=true,
        bookmarksopenlevel=\maxdimen,
      }

\usepackage{genyoungtabtikz}
\Yboxdim{5pt}
\Ylinethick{0.8pt}
\newcommand{\yngRed}[1]{%
    {\Ylinecolour{red}
    {\gyoung(#1)}}%
}
\newcommand{\yngBlack}[1]{%
    {\Ylinecolour{black}
    {\gyoung(#1)}}%
}
\newcommand{\yngBlue}[1]{%
    {\Ylinecolour{blue}\gyoung(#1)}%
}

\newcommand{\bluefill}{\Yfillcolour{blue}}

\def\gl#1#2{\ifmmode \mathrm{GL}(#1; {\bf #2}) \else $\mathrm{GL}(#1; {\bf #2})$\fi}
\def\sl#1#2{\ifmmode \mathrm{SL}(#1; {\bf #2}) \else $\mathrm{SL}(#1; {\bf #2})$\fi}
\def\so#1{\ifmmode \mathrm{SO}({#1}) \else $\mathrm{SO}(#1)$\fi}

\def\sp#1#2{\ifmmode \mathrm{Sp}(#1; {\bf #2}) \else $\mathrm{Sp}(#1; {\bf #2})$\fi}
\def\usp#1{\ifmmode \mathrm{USp}(#1) \else $\mathrm{USp}(#1)$\fi}
\def\spin#1{\ifmmode \mathrm{Spin}(#1) \else $\mathrm{Spin}(#1)$\fi}
\def\su#1{\ifmmode \mathrm{SU}({#1}) \else $\mathrm{SU}(#1)$\fi}

\def\double #1{#1{\hbox{\kern-2pt $#1$}}}

\mathcode`\*="702A                  
\def\half{{\textstyle{1\over{\raise.1ex\hbox{$\scriptstyle{2}$}}}}}

\def \a{\alpha}

\def \r{\rho}

\def\btau{\boldsymbol\tau}

\def\blueBullet{\textcolor{blue}{\bullet}}
\def\redBullet{\textcolor{red}{\bullet}}

\newcommand\x[1]{\textcolor{red}{x_{#1}}}
\newcommand\C[1]{\left(C^{#1}\right)}
\tikzfeynmanset{warn luatex=false}
\begin{document}

\vspace{10pt}
\begin{center}
{\LARGE Unfolding the Exotic Supergravity Multiplet}

\vspace{20pt}
Carlo Iazeolla${}^{\heartsuit,\ast}$, Per Sundell${}^{\clubsuit}$, Brenno Carlini Vallilo${}^{\spadesuit}$
\vspace{10pt}

{\em  
\vspace{5pt}
${}^\heartsuit$ Dipartimento di Scienze Ingegneristiche,  Universit\`a degli Studi Guglielmo Marconi, \\  Via Plinio 44, 00193, Roma, Italy \& \\Sezione INFN Roma Tor Vergata \\ Via della Ricerca Scientifica 1, 00133, Roma, Italy\\
\vspace{5pt}
${}^{\clubsuit}$  Instituto de Ciencias Exactas y Naturales, Universidad Arturo Prat,\\ Playa Brava 3265, 1111346 Iquique, Chile \& \\
Facultad de Ciencias, Universidad Arturo Prat,\\ Avenida Arturo Prat Chacón 2120, 1110939 Iquique, Chile\\
\vspace{5pt}
${}^{\spadesuit}$ Departamento de Física y Astronimía, Facultad de Ciencias Exactas,\\ Universidad Andres Bello, 
Sazié 2212, Santiago, Chile\\
}

\vspace{25pt}
{\bf Abstract}
\end{center}

\noindent We provide an unfolded formulation of linearised 6D exotic supergravity with manifest superconformal symmetry including conformal duality transformations. The construction is based on a superoscillator realisation of the metaplectic supergroup $MOSp(16|16)$ containing the Howe dual pair $MOSp(8^\ast|8)\times SU(2)$. Exotic graviton potentials arise in conformally dual pairs of two-forms in direct products of $SU(2)$-triplets and  supertraceless graded-symmetric rank-two supertensors of $OSp(8^\ast|8)$ generated from a hypercharge by the supercharges, providing an analogue of the adjoint representation of $\mathfrak{psu}(2,2|4)$. The potentials are sourced via relative Chevalley-Eilenberg cocycles by a conformally dual pair of chiral and anti-chiral Weyl zero-forms valued in $SO(1,1)\times Spin(1,5)\times USp(8)$-covariant monomorphic restrictions of an extended supersingleton, i.e., an infinite-dimensional $SU(2)$-invariant Hermitian left-module of $MOSp(8^\ast|8)$ containing various irreps associated with different boundary conditions, including the unitarizable exotic supergravity multiplet. Two natural quadratic observables are i) an on-shell closed zero-form built from the chiral Weyl zero-form and its dual, akin to the on-shell value of the second Chern class on the internal twistor $Z$-space of 4D Vasiliev systems; and ii) an off-shell topological six-form built from the conformally dual pair of two-form potentials, reducing on-shell to a generating function for chiral two-point functions in super-Poincar\'e backgrounds.

\blfootnote{
${}^\heartsuit$ \href{mailto:}{c.iazeolla@gmail.com}, 
${}^{\clubsuit}$ \href{mailto:per.anders.sundell@gmail.com}{per.anders.sundell@gmail.com}, 
${}^{\spadesuit}$ \href{mailto:vallilo@unab.cl}{vallilo@unab.cl} }
\blfootnote{${}^\ast$ Member of INDAM-GNFM.}

\setcounter{page}0
\thispagestyle{empty}

\newpage

\tableofcontents

\parskip = 0.1in

\section{Introduction}

A lasting lesson of the second superstring revolution is that the space of
consistent quantum field theories is larger than the corners visible in
perturbative Lagrangian constructions.  String dualities, M-theory, and brane
dynamics provide strong evidence for interacting fixed points in five and six
dimensions for which no conventional local Lagrangian description is known
\cite{Witten:1995ex,Witten:1995zh,Seiberg:1996qx,Aharony:1997th,Seiberg:1997ax}.
Six-dimensional superconformal field theories (SCFT) based on the $(2,0)$ algebra are
the archetypal examples: their compactifications produce lower-dimensional
Lagrangian gauge theories together with their duality structures, while the
higher-dimensional origins are characterised only through symmetry, protected
observables, extended objects, and duality orbits
\cite{Witten:2007ct}.\footnote{For the holomorphic factorization of a
non-chiral $2k$-form into chiral sectors, and for the Hilbert-space
factorization of a six-dimensional two-form, see
\cite{Henningson:1999dm,Henningson:2001wh}.}  A still sharper version of this
relationship is posed by the conjectural $(4,0)$ theory: the multiplet contains an
exotic graviton rather than a metric, so even the field that encodes lower-dimensional
gravitational interactions lies outside the usual geometric description in the higher dimension.

Hull proposed the $(4,0)$ theory as the strong-coupling limit of
five-dimensional maximal supergravity
\cite{Hull:2000zn,Hull:2000rr}.  If this limit exists, its circle compactification  should
reproduce five-dimensional $\mathcal N=8$ supergravity, including its $U$-duality
structure, with  Planck length equal to the circle radius.
The six-dimensional multiplet contains $42$ scalars, $27$ chiral
two-forms and, in place of the ordinary graviton, a $(2,2)$ Young-projected tensor gauge field with self-dual traceless $(3,3)$ Young-projected curvature. 
The proposal therefore replaces the metric by a
chiral tensor gauge system while preserving maximal supersymmetry.  This is why
the putative theory cannot be treated as an ordinary covariant uplift of
five-dimensional ungauged supergravity
\cite{Hull:2000zn,Hull:2000rr,Hull:2022vlv}.

The conjecture concerns an interacting theory, and its complete
six-dimensional definition remains unknown.  A supersymmetric action 
has nevertheless been constructed for the free $(4,0)$ multiplet, showing
that the absence of an ordinary metric does not by itself obstruct a quadratic
action. Superspace formulations and recent
loop-space constructions provide complementary descriptions of the free
kinematics \cite{Cederwall:2020dui,Howe:2024ojq}.  
The outstanding challenge is therefore to formulate the nonlinear theory in a framework that accommodates the exotic gauge symmetry, self-duality and maximal supersymmetry while retaining the expected
relation to five-dimensional supergravity.

Higher-spin gravity provides a useful structural analogy. Nonabelian higher-spin theories are difficult to formulate through conventional
local Lagrangian densities on spacetime. 
Rather, they can be formulated directly at the level of field equations using the  \emph{unfolded formalism} \cite{Vasiliev:1988sa}, which provides a tool for formulating and solving partial differential equations originating ideawise from Cartan's structure equations for on-shell gravity and free-differential-algebra formulation of on-shell supergravities \cite{DAuria:1982uck}; see  \cite{Vasiliev:1999ba,Bekaert:2004qos,Didenko:2014dwa} for reviews,
\cite{Shaynkman:2004vu,Vasiliev:2007yc,Joung:2021bhf} for unfolded conformal
modules and geometries, and \cite{Misuna:2024ccj,Misuna:2024dlx,TwoCommaZero,Misuna:2026bhy} for a few recent applications. 

Unfolding a field theory amounts to formulating its field equations as a Cartan-integrable system 
\begin{align}
R^\mathcal{I}:=dX^\mathcal{I}+Q^\mathcal{I}(X)\approx 0\ ,
\end{align}
for a set of locally defined differential forms $X^\mathcal{I}$ corresponding\footnote{The forms $X^\mathcal{I}$ of an unfolded system on a manifold $\boldsymbol{M}$ are isomorphic to zero-forms on $T[1]\boldsymbol{M}\equiv \boldsymbol{X}$ given by pull-backs $x^\mathcal{I}\circ \varphi$ of coordinates $x^\mathcal{I}$ on a graded manifold $\boldsymbol{Y}$ induced by a function $\varphi:\boldsymbol{X}\to \boldsymbol{Y}$, i.e., $X^{\mathcal{I}}=V^{-1}\circ x^{\mathcal{I}}\circ \varphi$, using the canonical isomorphism $V:{\Omega}(\boldsymbol{M})\to {C}(\boldsymbol{X})$.
Thus, the unfolded system is a classical truncation of an AKSZ sigma model with dynamical variable $\varphi$.} to coordinates on a target $Q$-manifold, i.e., a graded manifold equipped with a homological vector field 
\begin{align}
\vec Q:=Q^\mathcal{I}(X)\frac{\vec\partial}{\partial X^\mathcal{I}}\ ,\qquad \vec Q^2\equiv 0\ .
\end{align}
The nilpotency of $\vec Q$ is
equivalent to that the Cartan curvatures $R^\mathcal{I}$ obey generalised Bianchi identities, viz.,
\begin{align}
dR^\mathcal{I}-R^\mathcal{J} \partial_\mathcal{J} Q^\mathcal{I}\equiv 0\ ,
\end{align}
i.e., the differential constraints are compatible with $d^2=0$ 
\begin{itemize}
\item[--] without any further algebraic constraints on the forms; and
\item[--] independently of the dimension of the source manifold on which the forms live.
\end{itemize}
The nilpotency of $\vec Q$ furthermore implies that the constraint surface is preserved under nonabelian gauge transformations, viz.
\begin{align}
\delta_\epsilon X^\mathcal{I}=d\epsilon^\mathcal{I}-\epsilon^\mathcal{J} \partial_\mathcal{J} Q^\mathcal{I}\quad \Rightarrow\quad \delta_\epsilon R^\mathcal{I}\approx0\ ,
\end{align}
i.e., each differential form in strictly positive degree brings an algebraically independent gauge parameter. 
As a consequence of these lemmas, an unfolded system exhibits two basic properties:
\begin{itemize}
\item[--] Letting exponentiated finite Cartan gauge transformations act on zero-form integration constants yields locally defined solution spaces (without the need for any integration).
\item[--] Since the pull-back operation commutes with exterior differentiation and multiplication using wedge products, unfolded equations are invariant under diffeomorphisms acting on the manifold where the forms live.
\end{itemize}
In fact, their diffeomorphism invariance extends to smooth maps\footnote{In particular, since supermanifolds are homotopic to their bodies, the Cartan integrable system arising upon unfolding the component formulation of a supersymmetric field theory contains a superspace formulation of the theory.} between sources,
paving the way towards a proposal for an alternative approach to quantum field theory based on combining unfolding with the Alexandrov-Kontsevich-Schwarz-Zaboronsky (AKSZ) formalism \cite{Alexandrov:1995kv}, to be described below. 

Classical solution spaces\footnote{Fixing the source to be a Cartan integration manifold with specific topology (including dimension), classical moduli spaces can be explored along two routes: either by i) embedding a soldering one-form into the unfolded system, which triggers a homotopy contraction to locally defined fields obeying higher-order partial differential equations subject to (asymptotic) boundary conditions; or by ii) Cartan integration using gauge functions and integration constants in representations dual to the boundary conditions used in (i). } of unfolded systems consist of 
locally defined configurations glued together into globally defined configurations using transition functions from a structure group.
These functions represent additional topological degrees of freedom of the fully nonlinear theory, and the structure-group connection is assumed to be embedded into the locally defined form content of the unfolded system.
In classical perturbation theory, we thus choose a background solution that contains the structure-group connection.
Letting $\mathfrak{g}$ denote the symmetry Lie algebra of the background, i.e., the space of Cartan gauge transformations preserving it, the resulting linearised system is organised into differential forms valued in $\mathfrak{g}$-irreps, with background covariant derivatives equated on-shell to cocycles, i.e., linear maps built from the background connection that are 
\begin{itemize}
\item[--] covariant under structure-group transformations; and
\item[--] compatible with Cartan integrability.
\end{itemize}
As a consequence, the linearised system exhibits three types of symmetries
\begin{itemize}
\item[--] abelian Cartan gauge symmetries with parameters associated to the linearised forms in strictly positive degrees; 
\item[--] nonabelian Cartan gauge symmetries with parameters from $\mathfrak{g}$ associated to the background connection; and 
\item[--] nonabelian transition functions from the structure group defined on overlaps.
\end{itemize}

Beyond linearised dynamics, the unfolded formalism converts the problem of finding consistent interactions in a perturbative expansion around said background to the cohomological problem of deforming $Q$ while maintaining manifest structure-group covariance \cite{Vasiliev:1988sa,Vasiliev:1989yr,more,Sharapov:2017yde,Sharapov:2020quq}. Equivalently, this amounts to consistently deforming the homotopy algebra encoded into the $Q$-structure. Back to the higher-spin context, this strategy is realized through a cohomological resolution, obtained by enlarging the base manifold with a non-commutative twistor space, and endowing the extended system with associative star products representing strict operator algebras, which is the noncommutative analogue to imposing boundary conditions in commutative geometry; for details, see, e.g., \cite{more,Vasiliev:1999ba,Bekaert:2004qos,Didenko:2014dwa,Sharapov:2017yde} and also \cite{Sharapov:2022awp,Didenko:2022qga,Korybut:2025vdn} for the special case of chiral higher-spin gravity.

As already remarked, unfolded dynamics admits a natural extension beyond classical field theory by viewing the fundamental differential forms as living not only on a set of sources of a single dimension but also including a larger class of sources of different dimensions connected by smooth maps, that can be treated as boundaries of sources of AKSZ sigma models with targets equipped with graded Poisson structures compatible with the original $Q$-structure; for initial progress in this direction for the Vasiliev system, see \cite{Boulanger:2011dd,Boulanger:2015kfa,Bonezzi:2016ttk}, and for the emergence of spacetime witin the target of the sigma model, see \cite{FSG1,FSG2,TwoCommaZero} and the outline in Section \ref{aksz}. 
Thus, according to the ensuing proposal, solution spaces of unfolded systems subject to boundary conditions on higher-dimensional spaces (including spacetime manifolds and non-commutative twistor spaces) can be projected to boundaries of two-dimensional Poisson sigma models with targets containing groups such that the (real) Cartan integration constants are deformed into generators of boundary operator algebras; in quadratic approximations, the resulting BRST cohomologies contain quantised fields on ordinary spacetime manifolds arising as cosets in said groups \cite{FSG1,FSG2}. 

The proposed combination of unfolded dynamics and AKSZ quantisation provides a framework for quantum field theory that is a version of second-quantisation as originally envisaged, enhanced by differential algebra structures.
It remains to be seen how it relates to the conventional metric-like framework when both exist.
However, the unfolded approach appears less constrained by obstacles that cause the metric-like approach to stall:  spacetime nonlocalities can be controlled by associative structures underlying noncommutative AKSZ sigma models; and chiral bosons, which cause the metric-like formalism to stall already at the level of quadratic actions, can be quantised provided the target of the AKSZ sigma model can be equipped with non-trivial Poisson structures.

In this paper we do not assume that Hull's theory is a higher-spin gravity. The point is methodological: when a conventional Lagrangian description is unavailable\footnote{For an off-shell formulation of linearised exotic supergravity based on a Kaluza-Klein-type 5+1 split of coordinates
and fields, see \cite{Bertrand:2022pyi}.}, or when it obscures the underlying gauge structure, unfolding the system offers a structurally transparent, richer geometry containing auxiliary variables that may turn out to be crucial for deformations to exist.  Casting the linearised $(4,0)$ system in unfolded form  separates the curvature modules, containing the local degrees of freedom, from the gauge potential modules and reduces the
linear problem to finding a cocycle that glues them consistently in a given
supergeometry. The resulting data --- the zero-form module, the two-form
gauge module and the glueing cocycle --- then provide a starting point for a cohomological deformation problem, severely constrained by the Cartan-integrability condition. 

However, as envisaged already in \cite{TwoCommaZero} for the $(2,0)$ superconformal theory, it is nonetheless natural to further enrich the unfolded system by viewing the background one-form connection as the expectation value of topological conformal higher-spin connection receiving nonlinear corrections from the exotic supermultiplet (viewed as superconformal matter) initiated by higher-spin currents at quadratic order, along the lines of the three-dimensional system obtained in \cite{FSG1,FSG2} (see also \cite{Iazeolla:2025lwj}).

\subsection{Protected multiplets and non-local deformations}
\label{sec:protected-multiplets-nonlocal-deformations}

The rigidity of six-dimensional superconformal representations puts this
interaction problem into perspective.  A conformal field theory (CFT) can be
modified in three physically distinct ways: by perturbing the action with an
integrated local operator,
\begin{align}
    \delta S=g\int d^d x\,\mathcal O(x),
    \label{eq:local-operator-deformation}
\end{align}
by gauging a global symmetry, or by moving along a moduli space of vacua.  
Only the first operation is an infinitesimal deformation of the local operator
algebra.  
It assumes a unitary local CFT but not a Lagrangian; the relevant
superconformal multiplets and their deformations were classified by C\'ordova, Dumitrescu and Intriligator
\cite{Cordova:2016xhm,Cordova:2016emh}.

In six dimensions, long type-A multiplets that reach a unitarity bound split
into threshold type-A multiplets and shorter multiplets of types B, C and D.
Some of these short multiplets, including the stress-tensor and
fundamental-field multiplets, are absolutely protected against recombination.
In particular, unitary $(1,0)$ and $(2,0)$ theories have no supersymmetric
relevant or marginal deformations generated by Lorentz-invariant local
operators.  They can have supersymmetric irrelevant deformations, and in
$(1,0)$ theories familiar supersymmetric flows instead arise by moving onto a
moduli space where superconformal symmetry is spontaneously broken to
Poincar\'e supersymmetry
\cite{Cordova:2015fha,Cordova:2016xhm}.  These statements constrain local
conformal perturbation theory.  They neither identify gauging with a local
operator deformation nor, by themselves, prove or disprove the existence of
an interacting $(4,0)$ theory.

Three-dimensional theories with $\mathfrak{osp}(8|4)$ symmetry illustrate the
distinction.  Their relevant and marginal deformations and their multiplet
recombination rules are highly constrained.
Yet interacting BLG 
models exist, and at appropriate Chern--Simons levels, the superconformal symmetry of the
ABJ(M) models is enhanced from $\mathfrak{osp}(6|4)$ to $\mathfrak{osp}(8|4)$
\cite{Cordova:2016xhm,Chester:2014fya}.  Isolation under infinitesimal local
deformations is therefore not a theorem against an interacting fixed point.
Ranks, levels, gaugings and choices of vacuum can label theories that are not
connected by a continuous local perturbation.

Higher-spin symmetry makes the local obstruction more explicit.  In a unitary
three-dimensional CFT, one conserved current of spin greater than two implies
an infinite tower of higher-spin currents with free-theory correlators
\cite{Maldacena:2011jn}.  Interacting large-$N$ vector models instead realize a
slightly broken higher-spin symmetry, in which current divergences are
multi-trace operators \cite{Maldacena:2012sf}.  If $J_s$ is conserved in the
undeformed theory, a local perturbative deformation must modify its shortening
condition schematically as
\begin{align}
    \partial\mathbin{\cdot}J_s
    =g\,\mathcal K_{s-1}+\mathcal O(g^2),
    \label{eq:local-current-recombination}
\end{align}
where $\mathcal K_{s-1}$ is a local composite operator of the undeformed CFT.
Unitarity makes the dilation operator Hermitian in radial quantization, so a
short representation can disappear only by joining the appropriate partner at
the boundary of a continuum of long multiplets.  If that partner is absent,
Eq.~\eqref{eq:local-current-recombination} cannot describe a local unitary
deformation with the assumed supersymmetry
\cite{Cordova:2016xhm,Cordova:2016emh}.

The unfolded system suggests where an enlarged deformation problem may differ.
We work with horizontal master fields on a correspondence space with a
graded-commutative base and a graded noncommutative fibre.  Gauging global
symmetries is then part of the Cartan-integrable system, and eliminating
auxiliary or topological fields may produce vertices that are non-local from
the spacetime viewpoint \cite{Vasiliev:1999ba}.  A concrete model is the
three-dimensional coloured conformal higher-spin system
\cite{FSG1,FSG2}, which couples unfolded conformal matter to topological
conformal higher-spin and colour gauge fields.  At leading order the
higher-spin fields are sourced by colour-singlet bilinear currents, whereas
the colour connection is sourced by spacetime-non-local composites.  Beyond
the leading current coupling one expects a conservation law of the form
\begin{align}
    D\mathbin{\cdot}J_s[C,\overline C]
    =g\,\mathcal K_{s-1}[C,\overline C;W,V],
    \label{eq:unfolded-current-recombination}
\end{align}
where $C$ and $\overline C$ denote a complex conjugated pair of zero-forms containing the coloured conformal scalar field, while $W$ and $V$ denote the conformal higher-spin and colour connections.
After solving for the auxiliary and topological sectors, the right-hand side
is expected to become a spacetime-non-local matter composite, schematically
quartic at the first induced self-interaction order.  Equation
\eqref{eq:unfolded-current-recombination} is then outside the local operator
problem assumed in Eq.~\eqref{eq:local-current-recombination}; applying the
usual recombination argument would first require a definition of the enlarged
operator space and of the action of dilations on it.

For the conformally dual zero-form sectors used below, the expected bilinear
currents have the schematic structure 
\begin{align}
    J_s\sim E\wedge E\,C^{(+)}C^{(+)}
    +\widetilde E\wedge\widetilde E\,C^{(-)}C^{(-)}\ ,
    \label{eq:intro-dual-current}
\end{align}
where $E$ and $\widetilde E$ are the frame field and dual frame field, respectively, while mixed composites $C^{(+)}C^{(-)}$ are natural candidates for the
threshold type-A partners required by recombination.  This is a representation-theoretic possibility rather than a result of the present linear analysis: it
requires an explicit decomposition of the composite zero-form module.

The same distinction may persist after quantization.  In an AKSZ formulation
the Cartan-integrable system descends from a Batalin--Vilkovisky master action,
so its integrability can be promoted to Ward identities when the quantum
master equation is anomaly free \cite{Alexandrov:1995kv,FSG1}.  This offers a
possible route beyond the assumptions of local conformal perturbation theory,
but it is not a proof of quantum consistency.  One must still identify the
physical observables, construct the relevant inner product and show that the
non-local operator algebra has a consistent quantum completion.  If the colour
sector admits an interpretation in terms of brane-like or tensionless extended
sources, the departure from the local operator framework would be sharper
still \cite{Sundborg:2000wp,FSG2}.

For the putative $(4,0)$ theory, this is the role of the unfolded construction.
The six-dimensional protection results explain why an ordinary local
supersymmetric perturbation of a free exotic multiplet is too restrictive.
They do not exclude an interacting theory defined through a larger
correspondence-space algebra, non-local composite sources and gauged
topological sectors.  The linear system constructed here identifies the
modules and cocycles that any such deformation must preserve.  A nonlinear
completion would test whether these data extend consistently beyond the local
superconformal deformation problem.

\subsection{Methodology and main results}

The framework described above motivates the unfolding of linearised Hull's six-dimensional $(4,0)$ gravity, for which we provide an output in this paper essentially by extending the corresponding treatment of the $(2,0)$ tensor multiplet in \cite{TwoCommaZero}. In more detail, we provide a superoscillator realisation of a linearised unfolded system with
\begin{itemize}
\item[--] structure group 
\begin{align}
H=SO(1,1)\times Spin(1,5)\times USp(8) ;   
\end{align}
\item[--] background connection valued in the superconformal $(4,4)$ algebra (and not just its chiral $(4,0)$ super-Poincar\'e subalgebra); and
\item[--] invariance under discrete conformal duality transformations akin to spacetime parity transformations\footnote{The discrete symmetries act on second-quantised component fields by means of a first-quantised generator given by the inner realisation in the metaplectic group of one of the Weyl reflections of the conformal group.}, forming a $\mathbb{Z}_4$ symmetry reducing to $\mathbb{Z}_2$ on bosons.
\end{itemize}
To this end, following the principles of unfolding, we 

i) take the background to be a one-form valued in the superconformal algebra realisable as graded anti-symmetric bilinears in superoscillators, and an additional Howe-dual Lie algebra,  

ii) use a six-dimensional Newman--Penrose transformation to map the linearised system's local degrees of freedom, i.e., the exotic supergravity multiplet, to zero-forms valued in $H$-covariant irreps of the superconformal algebra, for which there are two choices, $\mathsf{S}^{(\pm)}$, referred to as the chiral and anti-chiral Weyl zero-form modules, which are each other's algebraic duals\footnote{The anti-singleton sector that appears in this conformally dual description is
not peculiar to the present construction.  Non-unitary modules of the
background isometry algebra arise naturally inside composite adjoint modules
and in fibre realizations of higher-spin systems \cite{fibre}; related sectors
also naturally occur in the analysis of conformally covariant unfolded field equations \cite{Vasiliev:2007yc} and in Coxeter higher-spin theories 
\cite{Vasiliev:2018zer,Tarusov:2025sre}.}, i.e., $(\mathsf{S}^{(\pm)})^\ast \cong \mathsf{S}^{(\mp)}$.

iii) embed the set of Skvortsov two-form potentials required for unfolding the exotic graviton and gravitini, into a two-form master field valued in an irreducible finite-dimensional representation of the superconformal algebra constructed as graded symmetric superoscillator bilinears;

iv) construct $H$-covariant chiral and anti-chiral cocycles compatible with universal Cartan integrability on general superconformal backgrounds;

v) impose covariance under conformal duality transformations that exchange $\mathsf{S}^{(\pm)}$ and $\mathsf{S}^{(\mp)}$ and the chiral and anti-chiral cocycles, which requires a duality-doublet of two-form master fields;

vi) though not a main theme of this work, we embed $\mathsf{S}^{(\pm)}$ as mutually compatible coordinate into a globally defined abelian Weyl zero-form module $\mathsf{S}$ of the metaplectic extension (see, e.g., \cite{CSM,Folland,Guillemin:1990ew,Woit:2017vqo} and references therein) of the superconformal \emph{group}, referred to as the extended supersingleton.

\noindent Step (iv), which represents the paper's main result, is taken by first expanding the chiral and anti-chiral Weyl zero-forms into $H$-tensors which have opposite conformal weights and eigenvalues under the six-dimensional Hodge-duality operation.
On a super-Poincar\'e background, the conformal chiral primaries are
\begin{align}
R_{\alpha\beta\gamma\delta}\ ,\quad
\Psi_{\alpha\beta\gamma I}\ ,\quad
H_{\alpha\beta IJ}\ ,\quad
\Theta_{\alpha IJK}\ ,\quad
\Phi_{IJKL}\ .
\label{eq:intro-curvature-multiplet}
\end{align}
Introducing a zero-form valued in a Howe-dual Lie algebra, we construct a chiral three-form cocycle that glues \eqref{eq:intro-curvature-multiplet} to a two-form master field (which thus contains a pure gauge sector on the super-Poincar\'e background).
The Cartan integrability condition then decomposes into algebraic embedding conditions, referred to as $t$-equations, and differential constraints, referred to as $s$-equations, which imply the fermionic
descendant chain among the primaries in \eqref{eq:intro-curvature-multiplet}. Finally, we give the superconformal completion using the superoscillator realisation, which packages the conformally dual modules.
The resulting fermionic ascent chain among the primaries in \eqref{eq:intro-curvature-multiplet} ensures that the cocycle remains universally Cartan-integrable in the superconformal background.  

Taking the source manifold of the unfolded system to be a supermanifold of superdimension $(6|32)$ and going to the Wess--Zumino gauge, the unfolded system implies that the primaries in  \eqref{eq:intro-curvature-multiplet} obey on-shell superspace
constraints implying Bargmann--Wigner equations on the spacetime submanifold.
The corresponding exotic graviton and gravitini, and chiral two-form potentials, occupy slots in the chiral two-form master field; we are omitting their superspace constraints in this work.

We leave the proposed strongly coupled higher-spin uplifts and the
construction of a nonlinear completion for future work.

\subsection{Notation and conventions}

\paragraph{Spaces and functions.}
The ring structures arising on manifolds, groups, and Lie algebras are bi-graded objects equipped with 

-- $\mathbb{Z}$-gradings arising as ghost numbers of first- and second-quantised AKSZ sigma models;

\noindent and independent 

-- $\mathbb{Z}_2$-gradings in concordance with spacetime supersymmetry;

\noindent for further conventions, see \cite{TwoCommaZero}.
We denote
\begin{itemize}
\item[i)] manifolds using boldfaced capital Roman letters, e.g.,\\
-- real or $\mathbb{Z}_2$-graded manifolds $\boldsymbol{M}$;\\
-- the source $\boldsymbol{X}$ and target $\boldsymbol{Y}=\boldsymbol{\mathcal{A}}[1]$ of the second-quantised sigma model; \\
-- the $\mathbb{Z}$-graded correspondence space $\boldsymbol{F}\hookrightarrow T[1]\boldsymbol{C}\to \boldsymbol{X}$ of the target of the first-quantised sigma model/domain of the second-quantised superconnection;
 
\item[ii)] differential graded associative algebras using boldfaced capital calligraphic letters, e.g.,\\
-- the finite-dimensional graded Frobenius algebra $\boldsymbol{\mathcal{F}}$ generated by ghosts and twisted boundary conditions of the first-quantised sigma model;\\
-- the gauge algebra $\boldsymbol{\mathcal{A}}=C(\boldsymbol{F})\otimes \boldsymbol{\mathcal{F}}$ of the second-quantised sigma model;\\
-- for the algebras $\Omega(\boldsymbol{M})$ of differential forms on $\boldsymbol{M}$, idem, $\Omega(\boldsymbol{X})$ and $\Omega(\boldsymbol{Y})$, and the algebra $C(\boldsymbol{F})$ of functions on the fibre $\boldsymbol{F}$ (with trivial differential), we make notational exceptions;\\

\item[iii)] Lie groups and algebras using capital Roman and minuscule gothic letters, respectively, e.g.,\\
-- the superconformal matrix group $\pi(G)\cong OSp(8^\ast|8)$ with its algebra $\mathfrak{g}\equiv \mathfrak{osp}(8^\ast|8)$ and metaplectic extension $G\equiv MOSp(8^\ast|8)$;\\
-- the conformal matrix group $\pi(G_{C})=Spin(2,6)$ with its algebra $\mathfrak{g}_C\equiv \mathfrak{so}(2,6)$ and metaplectic extension $G_{C}=MSpin(2,6)$;\\
-- the R-symmetry group $G_R\equiv USp(8)$ with its algebra $\mathfrak{g}_R\equiv \mathfrak{usp}(8)$;\\
-- the structure group $H=SO(1,1)\times Spin(1,5)\times G_R$ with its algebra $\mathfrak{h}\equiv \mathfrak{so}(1,1) \oplus\mathfrak{so}(1,5)\oplus \mathfrak{g}_R$;\\
-- the Howe dual group $\widetilde K=SU(2)$ with its algebra $\tilde{\mathfrak{k}}$;

\item[iv)] modules of Lie groups and algebras using capital Roman letters on sans serif font, e.g.,\\
-- the extended supersingleton module $\mathsf{S}$ of $G$ containing the unitarizable exotic supergravity multiplet;\\
-- the $\pi(G)\times \widetilde{K}$-tensor $\mathsf{T}$ containing the two-form potentials;\\
-- the $H$-tensorial $\mathfrak{g}$-modules $\mathsf{S}^{(\pm)}$  containing the superconformal primaries;\\
-- $H$-tensors $\boldsymbol{\tau}$, for which we make a notational exception;

\item[v)] elements in $\Omega(\boldsymbol{X})\otimes \boldsymbol{\mathcal{A}}[1]$ using boldfaced minuscule letters, e.g.,\\
-- the superconnection $\boldsymbol{x}$;\\
-- the Weyl zero-form $\boldsymbol{c}$ valued in $\mathsf{S}$,\\
-- the conformal-duality doublet $\boldsymbol{b}^{(\pm)}$ of two-forms valued in $\mathsf{T}$;\\
-- the chiral and anti-chiral Weyl zero-forms\footnote{In \cite{TwoCommaZero}, the chiral and anti-chiral Weyl zero-forms of the unfolded $(2,0)$ tensor multiplet were denoted by $\boldsymbol{h}^{(\pm)}$.} $\boldsymbol{c}^{(\pm)}$ valued in $\mathsf{S}^{(\pm)}$;\\
-- the background connection $\boldsymbol{\Omega}$ valued in $\mathfrak{g}$, for which we make a notational exception;

\item[vi)] superconnection components using  capital Roman letters, e.g., the conformally dual pairs\\
--  $\left(R_{\alpha\beta\gamma\delta},R^{\alpha\beta\gamma\delta}\right)$ of exotic Weyl tensors;\\
--  $(B^{(\pm)})^{ij}_{AB}$ of two-form potentials;\\
--  $(E^{\alpha\beta},F^{\alpha I})$ and $(\widetilde{E}_{\alpha\beta},\widetilde{F}_{\alpha I})$ of superframe fields;

\item[vii)] configuration spaces using calliographic capital letters, e.g., \\
-- the space $\mathcal{M}$ of AKSZ field configurations with its classical moduli subspaces $\mathcal{C}$.

\end{itemize}

\paragraph{$Spin(2,6)$ matrices.}

The $SO(2,6)$-invariant symmetric tensor 
\begin{align}
\hat\eta_{\hat{a}\hat{b}}={\rm diag}(-1,-1,+1,\dots,+1)\ ,\qquad \hat a=0',0,1,\dots ,6\ ;
\end{align}
the charge conjugation matrix $\widehat{C}^{\hat{\alpha}\hat{\beta}}$ and the gamma matrices $(\widehat{C}\widehat{\Gamma}^{\hat{a}})^{\hat{\alpha}\hat{\beta}}$ are assumed to be symmetric.
In the Weyl basis,
\begin{align}
\widehat C^{\hat{\alpha}\hat{\beta}}=\left[\begin{array}{cc} C^{\underline{\alpha\beta}}&0\\ 0& C^{\underline{\dot\alpha\dot\beta}}\end{array}\right]\ ,   \qquad \widehat \Gamma_{\hat{\alpha}}{}^{\hat{\beta}}=\left[\begin{array}{cc} \delta_{\underline{\alpha}}^{\underline{\beta}}&0\\ 0& -\delta_{\underline{\dot\alpha}}^{\underline{\dot\beta}}\end{array}\right]\ ,\qquad \underline{\alpha},\dot{\underline{\alpha}}=1,\dots,8\ .
\end{align}

\paragraph{$Spin(1,5)$ matrices.}

The $SO(1,5)$-invariant symmetric tensor
\begin{align}
{\eta}_{{ab}}={\rm diag}(-1,+1,\dots,+1)\ ,\qquad {a}=0,1,\dots ,5\ ;
\end{align}
the charge conjugation matrix and gamma matrices
\begin{align}
C^{\underline{\alpha\beta}}=\left[\begin{array}{cc}0&\delta^\alpha_\beta\\\delta^\beta_\alpha&0\end{array}\right]\ ,\qquad (\Gamma^a)_{\underline{\alpha}}{}^{\underline{\beta}}=\left[\begin{array}{cc}0&(\sigma^a)_{\alpha\beta}\\(\tilde\sigma^a)^{\alpha\beta}&0\end{array}\right]\ ,\qquad \alpha=1,\dots,4\ ,
\end{align}
where the sigma matrices
obey
\begin{align}
(\sigma^{a})_{\alpha\beta}(\tilde\sigma^{b})^{\beta\gamma}&=\eta^{ab}\delta_\alpha^\gamma+(\sigma^{ab})_\alpha{}^\gamma\ ,\qquad (\tilde\sigma^{a})^{\alpha\beta}(\sigma^{b})_{\beta\gamma}=\eta^{ab}\delta_\alpha^\gamma+(\tilde\sigma^{ab})^\alpha{}_\gamma\ ,\\
(\sigma_a)_{\alpha\beta}&=\frac12 \epsilon_{\alpha\beta\gamma\delta}(\tilde\sigma_a)^{\gamma\delta}\ ,\qquad(\sigma^{ab})_\alpha{}^\beta=-(\tilde\sigma^{ab})^\beta{}_\alpha\ ,\\
(\sigma_{abc})_{\alpha\beta}&=-\frac16 \epsilon_{abcdef}(\sigma^{def})_{\alpha\beta}\ ,\qquad (\tilde\sigma_{abc})^{\alpha\beta}=\frac16 \epsilon_{abcdef}(\tilde\sigma^{def})^{\alpha\beta}\ .
\end{align} 
Chiral (+) and anti-chiral (-) spinors and selfdual (+) and anti-selfdual (-) rank-three tensors, respectively, obey
\begin{align}
\Psi^{(\pm)}_{\underline{\alpha}}&=(\Pi^{(\pm)})_{\underline{\alpha}}{}^{\underline{\beta}}\Psi^{(\pm)}_{\underline{\beta}}\ ,\qquad \Pi^{(\pm)}=\frac12(1\pm \Gamma)\ ,\qquad \Gamma=\Gamma^{01\dots 5}\ ,\\
T^{(\pm)}_{abc}&=(\Pi^{(\pm)})_{abc}{}^{def} T^{(\pm)}_{def}\ ,\qquad (\Pi^{(\pm)})_{abc}{}^{def}=\frac12\left(\delta_{[abc]}^{def}\pm \frac16 \epsilon_{abc}{}^{def}\right)\ ,
\end{align}
such that symmetric chiral and anti-chiral bi-spinors correspond to selfdual and anti-selfdual tensors, respectively, viz.,
\begin{align}
T^{(+)}_{abc}=(\tilde{\sigma}_{abc})^{\alpha\beta} T^{(+)}_{\alpha\beta}\ ,\qquad T^{(-)}_{abc}=({\sigma}_{abc})_{\alpha\beta} T^{(-)\alpha\beta}\ .
\end{align}
In the Weyl basis,
\begin{align}
\Gamma_{\underline{\alpha}}{}^{\underline{\beta}}=\left[\begin{array}{cc}\delta_\alpha^\beta&0\\ 0&-\delta_\beta^\alpha\end{array}\right]\ ,\qquad \Psi^{(+)}_{\underline{\alpha}}=\left[\begin{array}{c}\Psi_\alpha\\ 0\end{array}\right]\ ,\qquad \Psi^{(-)}_{\underline{\alpha}}=\left[\begin{array}{c}0\\ \Psi^{\alpha}\end{array}\right]\ ,
\end{align}

\paragraph{Conformal weights.}
Letting $\rho_{\mathsf{M}}:\mathfrak{g}\otimes \mathsf{M}\to \mathsf{M}$ denote the representation of $\mathfrak{g}$ in a module $\mathsf{M}$, the linearised superconnection $\boldsymbol{x}=\btau_{\mathcal{I}} X^{\mathcal{I}}$, where $\btau^{\mathcal{I}}$ span one- and two-sided $\mathfrak{g}$-modules arising within $\boldsymbol{\mathcal{A}}$, and $X^{\mathcal{I}}\equiv\boldsymbol{\tau}^{\ast}{} ^{\mathcal{I}}(\boldsymbol{x})$ are locally defined coordinates on the configuration space $\mathcal{M}$, with $\mathfrak{g}$-module structure defined by 
\begin{align} \left(\rho_{\boldsymbol{\mathcal{A}}}(x)+\rho_{\mathcal{M}}(x)\right)\boldsymbol{x}:=0\ ,\qquad \rho_{\mathcal{M}}:\boldsymbol{\mathcal{A}}\to {\rm Vec}(\mathcal{M})\ .
\end{align}
We assume that the dilation operator $D\in\mathfrak{g}$ acts diagonally on $\btau_{\mathcal{I}}$, viz, 
\begin{align}
\rho_{\boldsymbol{\mathcal{A}}}(D)\btau_{\mathcal{I}}=i \Delta_{\boldsymbol{\mathcal{A}}}(\btau_{\mathcal{I}})\btau_{\mathcal{I}}\ ,
\end{align}
referring to $\Delta_{\boldsymbol{\mathcal{A}}}(\btau_{\mathcal{I}})$ as the (first-quantised) conformal weight of $\btau_{\mathcal{I}}$, and 
\begin{align}
\Delta_{\mathcal{M}}(X^{\mathcal{I}}):=-\Delta_{\boldsymbol{\mathcal{A}}}(\btau_{\mathcal{I}})\ ,
\end{align}
as the (second-quantised) conformal weight of $X_{\mathcal{I}}$, indicated by subscripts, viz., 
\begin{align}
\btau_{\mathcal{I}}\equiv 
\left(\btau_{\Delta_{\boldsymbol{\mathcal{A}}}(\btau_{\mathcal{I}})}\right)_{\mathcal{I}}\ ,\qquad 
X^{\mathcal{I}}\equiv \left(X_{\Delta_{\mathcal{M}}}\right)^{\mathcal{I}}\ .
\end{align}
If $\boldsymbol{x}'$ is a projection of $\boldsymbol{x}$ spanned by first-quantised elements of fixed conformal weight $\Delta$, viz.,
\begin{align}
\rho_{\boldsymbol{\mathcal{A}}}(D) \boldsymbol{x}'=i\Delta \boldsymbol{x}'\ ,\qquad \rho_{\mathcal{M}}(D)\boldsymbol{x}'= -i\Delta \boldsymbol{x}'\ ,
\end{align}
then we write
\begin{align}
\boldsymbol{x}'\equiv \boldsymbol{x}'_{\Delta}\ .
\end{align}

\paragraph{R-symmetry indices.}

We raise and lower $USp(8)$ indices $I=1,\dots,8$ using the invariant tensor $\eta^{IJ}$ following the convention 
\begin{align}
x^I:= \eta^{IJ}x_J\ ,\qquad x_I:=x^J\eta_{JI}\ ,\qquad \eta^{IJ}\eta_{KJ}=\delta^I_K\ .   
\end{align}

\paragraph{Young diagrams.}
\label{TableConventions}

We use Young diagrams with \\[5pt]
-- black cells $\yngBlack{;}$ for vectors of $SO(1,5)$;\\[5pt]
-- blue cells $\yngBlue{;}$ for chiral spinors of $Spin(1,5)\cong SU^\ast(4)$; \\[5pt]
-- solid blue cells $\yngBlue{!\bluefill;}$ for anti-chiral spinors of $Spin(1,5)\cong SU^\ast(4)$; and\\[5pt]
-- \textcolor{red}{red} cells $\yngRed{;}$ for the  fundamental representation of $USp(8)$.\\[5pt]
\noindent We give tensors of $SU^\ast(4)$ and $USp(8)$ on symmetric and anti-symmetric bases, respectively, with indices from adjacent rows and columns separated by commas, with the exception of the Lie superalgebra generators, and of the totally antisymmetric tensors $\epsilon_{\alpha\beta\gamma\delta}$ and $\epsilon_{IJKLMNOP}$. We use $\eta$-subscripts to indicate tracelessness of R-symmetry tensors, and vertical bars to separate groups of indices not related by Young projectors; examples from the body of the paper are
\begin{align}
\btau_{\alpha_1\alpha_2\alpha_3\alpha_4,\beta_1\beta_2}&\in\yngBlue{;;;;,;;}\ ,\qquad \btau_{\alpha_1\alpha_2,\alpha_3\alpha_4|\beta_1\beta_2}\in\yngBlue{;;,;;}\otimes \yngBlue{;;}\ ,\\
\btau_{I_1I_2I_3,J}&\in\yngRed{;;,;,;}_\eta\ , \qquad \eta^{I_2 I_3}\btau_{I_1I_2I_3,J}=0\ ,
\end{align}
and 
\begin{align}
\C{\Psi}^\alpha_{\beta\rho,\lambda|\gamma I}\in \yngBlue{;;,;}\otimes\yngBlue{;}^\yngBlue{!\bluefill;}\otimes\yngRed{;} \ .
\end{align}

\paragraph{On-shell equations.} We let $\approx$ denote classical equations of motion; following the variational principle, these should describe stationary points of action functionals on spaces of configurations obeying off-shell boundary conditions. 

\subsection{Organization}
The paper is organized as follows:

Section \ref{Sec:2} introduces the field-theoretic and algebraic underpinning of the paper.

Section \ref{algebraicstructures} describes the superconformal modules constituting the linearized Cartan integrable system.

Section \ref{unfoldedeom} defines the linearised unfolded system and the superconformally and conformal-duality invariant zero-form charge and exotic second Chern class.

In Section \ref{Sec:5} we show that the three-form cocycle is Cartan integrable super-Poincar`e and superconformal backgrounds.

Section \ref{Sec:6} describes the superoscillator realisation of the conformally dual chiral and anti-chiral Weyl zero-forms. 

In Section \ref{Sec:7}, we conclude and comment on deformations of the
linear system towards an interacting theory.

Spinor and superalgebra conventions are spelled out in Appendices \ref{App:spinors} and \ref{Sec:conventions}.

\section{Field theoretic and algebraic ingredients}
\label{Sec:2}

This section outlines three basic ingredients of the unfolded exotic supergravity multiplet:
\begin{itemize}
\item[--] the Newman--Penrose transformation mapping the supermultiplet to Lorentz-covariant chiral primaries;
\item[--] the corresponding conformally extended set of Skvortsov potentials; and
\item[--] the AKSZ approach to quantising unfolded systems, i.e., second-quantising first-quantised geometries.
\end{itemize}

\noindent all of which admit natural superoscillator realisations facilitating our construction.

\subsection{Newman--Penrose transform of exotic supergravity multiplet}
\label{newmanpenrose}

In the absence of central charges, the supercharges of the chiral $(4,0)$ algebra is a set 
\begin{align}
Q_\alpha{}^I\ ,\qquad  \alpha=1,\dots,4\ ,\qquad I=1,\dots,8\ ,
\end{align}
of symplectic Majorana--Weyl (SMW) spinors of $SO(1,5)$ transforming as $8$-plets of an internal $USp(8)$ serving as R-symmetry group, and obeying 
\begin{align}
[Q_\alpha{}^I,Q_\beta{}^J] = \eta^{IJ} T_{\alpha\beta}\ ,  
\end{align}
where $T_{\alpha\beta}=-T_{\beta\alpha}$ are the translation generators of $\mathfrak{iso}(1,5)$.
The superconformal $(4,4)$-extension $\mathfrak{osp}(8^\star|8)$ of the chiral algebra contains an additional set 
\begin{align}
S^\alpha{}_I\ , \qquad \alpha=1,\dots,4\ ,\qquad I=1,\dots,8\ ,
\end{align}
of anti-chiral conformal supercharges obeying 
\begin{align}
[S^\alpha{}_I,S^\beta{}_J] ={}& \eta_{IJ} K^{\alpha\beta}\ ,  \\
[Q_\alpha{}^I,S^\beta{}_J] ={}& \delta_J^I (M_{\alpha}{}^\beta+\tfrac12\delta_\alpha^\beta D)+\delta_\alpha^\beta N_J^I \ ,
\end{align}
where $M_{\alpha}^{\beta}$ are the Lorentz generators of $\mathfrak{iso}(1,5)$; $K^{\alpha\beta}$ and $D$ are the special conformal and dilation generators of $\mathfrak{iso}(2,6)$; and $N_I^J$ are the generators of the R-symmetry algebra $\mathfrak{usp}(8)$. 
The exotic supergravity multiplet arises on-shell as the unique CPT self-conjugate massless multiplet of the $(4,4)$-algebra.
It consists of states 
\begin{align}
|p;(J,J);[4-2J]\rangle\ ,\qquad J=0,\tfrac12,1,\tfrac32,2\ , 
\end{align}
carrying
\begin{itemize}
\item[i)] massless spacetime momenta\footnote{The inner product $p^a_1 p^b_2\eta_{ab}$ of a pair of Lorentz vectors is proportional to $p_1^{\alpha\beta}p_2^{\gamma\delta}\epsilon_{\alpha\beta\gamma\delta}$.}
\begin{align}
p^{\alpha\beta}
=\omega\,
\tilde\lambda^\alpha{}_{\tilde\alpha'}\tilde\lambda^\beta{}_{\tilde\beta'}\epsilon^{\tilde\alpha'\tilde\beta'}\ ,
\end{align}
stabilised by the little group\footnote{The unirreps of $Spin(4)$ are labelled by highest weights $(J_1,\pm J_2)$ with $J_1,J_2\in \{0,\tfrac12,1,\dots\}$ obeying $J_1\geqslant J_2$, corresponding to  unirreps $(J)_{\rm l}\otimes (\tilde J)_{\rm r}$ of $SU(2)_{\rm l}\times SU(2)_{\rm r}$ with $J=\tfrac12(J_1\pm J_2)$ and $\tilde J=J_1\mp J_2$.} 
\begin{align}
Spin(4)\cong SU(2)_{\rm l}\times SU(2)_{\rm r}\ ,
\end{align} 
and 
factorised using twistorial momenta
\begin{align}
\tilde\lambda^\alpha{}_{\tilde\alpha'}|_{Spin(1,5)\times Spin(4)}\in (\tfrac12,\tfrac12,-\tfrac12)\otimes (\tfrac12,-\tfrac12)\ ,
\end{align} 
where thus $\tilde\alpha'=1,2$ labels components of anti-chiral SMW spinors of $Spin(4)$;
\item[ii)] polarisation tensors
valued in direct products of chiral  $(J,J)\cong (J)_{\rm l}\otimes (0)_{\rm r}$ of $Spin(4)$; and
\item[iii)] traceless totally antisymmetric R-symmetry tensors $[r]$ of rank $r=4-2J$.
\end{itemize}
Thus, 
under $SU(2)_{\rm l}\times SU(2)_{\rm r}\times USp(8)$, $|p;(J,J);[4-2J]\rangle$ decomposes into 
\begin{equation}
(\underline{\mathbf{5}},\underline{\mathbf{1}};\underline{\mathbf{1}})
\oplus
(\underline{\mathbf{4}},\underline{\mathbf{1}};\underline{\mathbf{8}})
\oplus
(\underline{\mathbf{3}},\underline{\mathbf{1}};\underline{\mathbf{27}})
\oplus
(\underline{\mathbf{2}},\underline{\mathbf{1}};\underline{\mathbf{48}})
\oplus
(\underline{\mathbf{1}},\underline{\mathbf{1}};\underline{\mathbf{42}})\ ,
\end{equation}
using $\underline{\mathbf{n}}$ to label unirreps of dimension $n$, i.e., an exotic graviton, 8 exotic gravitini, 27 chiral bosons, 48 chiral fermions, and 42
scalars, splitting into 
\begin{equation}
5\times 1+3\times 27+1\times 42
=
4\times 8+2\times 48
=
128
\end{equation} 
bosonic and fermionic degrees of freedom, respectively.
Expanding $|p;(J,J);[4-2J]\rangle$ into polarization tensors
\begin{equation}
R_{{\alpha}'_1 {\alpha}'_2{\alpha}'_3{\alpha}'_4}(p),
\qquad
\Psi_{{\alpha}'_1 {\alpha}'_2{\alpha}'_3I}(p),
\qquad
H_{{\alpha}'_1 {\alpha}'_2IJ}(p),
\qquad
\Theta_{{\alpha}'IJK}(p),
\qquad
\Phi_{IJKL}(p)\ ,
\end{equation}
carrying symmmtrised indices $\alpha'=1,2$ labelling components of chiral SMW spinors of $Spin(4)$, and introducing dual Newman--Penrose spin-frames
\begin{align}
\lambda^\alpha{}_{\hat \alpha}:=(\lambda^{\alpha}{}_{\alpha'},\lambda^\alpha{}_{\tilde \alpha'})\ ,\qquad \tilde\lambda_\alpha{}^{\hat \alpha}:=(\tilde\lambda_{\alpha}{}^{\alpha'},\lambda_\alpha{}^{\tilde \alpha'})\ ,\qquad \lambda^\alpha{}_{\hat \alpha} \tilde\lambda_\alpha{}^{\hat \beta}:=\delta_{\hat\alpha}^{\hat{\beta}}\ ,
\end{align}
where thus 
\begin{align}
\tilde\lambda^\alpha{}_{\hat\alpha}|_{Spin(1,5)\times Spin(4)}&\in (\tfrac12,\tfrac12,\tfrac12)\otimes ((\tfrac12,-\tfrac12)\oplus  (\tfrac12,-\tfrac12))\ ,\\
\lambda_\alpha{}^{\hat\alpha}|_{Spin(1,5)\times Spin(4)}&\in (\tfrac12,\tfrac12,\tfrac12)\otimes ((\tfrac12,\tfrac12)\oplus  (\tfrac12,-\tfrac12))\ ,
\end{align} 
the polarisation tensors provide $SO(1,5)\times USp(8)$-covariant Fourier modes 
\begin{align}
R_{\alpha_1\alpha_2\alpha_3\alpha_4}(p)&:= \lambda_{\alpha_1}{}^{ \alpha'_1}\lambda_{\alpha_2}{}^{ \alpha'_2}\lambda_{\alpha_3}{}^{ \alpha'_3}\lambda_{\alpha_4}{}^{ \alpha'_4}R_{{\alpha}'_1 {\alpha}'_2{\alpha}'_3{\alpha}'_4}(p)\ ,\\
\Psi_{\alpha_1\alpha_2\alpha_3 I}(p)&:= \lambda_{\alpha_1}{}^{ \a'_1}\lambda_{\alpha_2}{}^{ \a'_2}\lambda_{\alpha_3}{}^{ \a'_3}\Psi_{{\alpha}'_1 {\alpha}'_2{\alpha}'_3I}(p)\ ,\\
H_{\alpha_1\alpha_2IJ}(p)&:= \lambda_{\alpha_1}{}^{ \alpha'_1}\lambda_{\alpha_2}{}^{ \alpha'_2}
H_{{\alpha}'_1 {\alpha}'_2IJ}(p) \ ,\\
\Theta_{\alpha IJK}(p)&:= \lambda_{\alpha}{}^{ \alpha'}
\Theta_{{\alpha}'IJK}(p)\ ,
\end{align}
combining with $\Phi_{IJKL}(p)$ into a chirally-projected supertraceless graded-symmetric $OSp(8*|8)$-tensor
\begin{align}
C_{A_1 A_2 A_3 A_4}(p)=\Pi^{(+)}_{(A_1|}{}^B C_{B|A_2 A_3 A_4)}(p)\ ,\qquad \zeta^{AB}C_{ABCD}(p)=0\ ,
\end{align}
using the chiral projector $\Pi^{(+)}_{A}{}^B$ and $OSp(8*|8)$-invariant rank-two tensor $\zeta^{AB}$ defined in Section \ref{Sec:3.10}.
These modes span solution spaces to the mass-shell condition\footnote{In momentum space, the equation of motion of a massless chiral tensor-spinor $\phi_{\alpha_1\dots \alpha_{2J}}$ of Lorentz spin $(J,J,J)$ reads $p^{\alpha \beta} \phi_{\beta\gamma_1 \dots\gamma_{2J-1}}(p)\approx 0$ for $J\geqslant \tfrac12$.}
\begin{align}
p^{AB} C_{B C_1 C_2 C_3}(p)\approx 0\ ,\qquad p^{AB}=p^{[A|C}\Pi^{(+)}_{C}{}^{|B]}\ ;
\end{align}
for $J\geqslant 1$,  the corresponding wave-equations are the Bargmann--Wigner equations for the abelian curvature tensors of the multiplet, i.e., the
chiral three-form curvature
\begin{align}
H^{[3]^{(+)}}_{abc}:=(\tilde\sigma_{abc})^{\beta\gamma}H_{\alpha\beta IJ}\ ;
\end{align}
the exotic gravitini curvatures
\begin{align}
\Psi^{[3]^{(+)}_{\tiny{\frac12}}}_{abc,\alpha,I}:=(\tilde\sigma_{abc})^{\beta\gamma}\Psi_{\alpha\beta\gamma I}\ ; 
\end{align}
and the exotic Weyl tensor
\begin{align}
R^{[3,3]^{(+)}}_{abc,def}:=(\tilde \sigma_{abc})^{\alpha\beta}(\tilde \sigma_{def})^{\gamma\delta} R_{\alpha\beta\gamma\delta}\ .
\end{align}
The Newman--Penrose transform arises naturally starting from the superoscillator realisation of the exotic supergravity multiplet.

\subsection{Conformally extended  Skvortsov system}
\label{skvortsovpotentials}

In the metric-like approach, the exotic Weyl tensor arises as the self-dual abelian curvature of a mixed-symmetry tensor-gauge field 
$h_{\mu\nu,\rho\sigma}$, alias, Hull's dual graviton potential, which upon reduction to five dimensions yields the graviton of 
five-dimensional maximal supergravity.
In a six-dimensional Poincar\'e background with frame-field $E^a$ and Lorentz covariant exterior derivative $\nabla$, the exotic graviton potential arises upon elimination of auxiliary fields and gauge parameters starting from the unfolded
Skvortsov system  \cite{Skvortsov:2008vs} describing the glueing of
the exotic Weyl tensor $R^{[3,3]^{(+)}}_{abc,def}$ via a cocycle to a two-form $\omega^{[2]^{(+)}}_{def}$, viz.,
\begin{align}
\nabla \omega^{[2]^{(+)}}_{def}+E^a\wedge E^b\wedge E^c R^{[3,3]^{(+)}}_{abc,def}\approx 0\ ,
\end{align}
forming an $\mathfrak{iso}(1,5)$-irrep together with a two-form $h^{[2]}_{ab}$ providing an exotic frame field, viz.,
\begin{align}
\nabla h^{[2]}_{ab} +E^c\wedge \omega^{[2]^{(+)}}_{abc}\approx 0\ ;
\end{align}
equivalently, in the Newman--Penrose basis, the universal Cartan integrability of the system 
\begin{align}
\nabla h_\alpha{}^\beta -4E^{\beta\gamma}\wedge\omega_{\alpha \gamma}&\approx0\ ,\\
\nabla\omega_{\alpha\beta}+ \tfrac14 E^a\wedge E^b\wedge E^c (\sigma_{abc})^{\gamma\delta} R_{\gamma\delta|\alpha\beta}&\approx0\ ,
\end{align}
implies $R_{\gamma\delta|\alpha\beta}\in \yngBlue{;;;;}$; for details, see Section \ref{poincareBackground}.
The Skvortsov system admits a conformally covariant extension built on conformal backgrounds with 
\begin{itemize}
\item[--] frames $E^a$ and conformally dual frames $\widetilde E^a$;
\item[--] $SO(1,1)\times SO(1,5)$-covariant exterior derivative $\nabla$;
\item[--] chiral exotic Weyl tensor $R^{[3,3]^{(+)}}_{abc}$ and anti-chiral exotic Weyl tensors $\widetilde{R}^{[3,3]^{(-)}}_{abc}$; and
\item[--] a two-form valued in the finite-dimensional irrep of the conformal group $SO_0(2,6)$ given by a chiral totally anti-symmetric rank-four tensor.
\end{itemize}
The resulting system, viz.,
\begin{align}
\nabla \omega^{[2]^{(+)}}_{def}+E^a\wedge E^b\wedge E^c R^{[3,3]^{(+)}}_{abc,def}&\approx 0\ ,\\
\nabla h^{[2]}_{ab} +E^c\wedge \omega^{[2]^{(+)}}_{abc}+\widetilde E^c\wedge \omega^{[2]^{(-)}}_{abc}\approx 0\ ,\\
\nabla \omega^{[2]^{(-)}}_{def}+\widetilde E^a\wedge \widetilde E^b\wedge \widetilde E^c \widetilde{R}^{[3,3]^{(-)}}_{abc,def}&\approx 0\ ,
\end{align}
take the following form in Newman--Penrose basis:
\begin{align}
\nabla\omega^{(+)}_{\alpha\beta}+ \tfrac14 E^a\wedge E^b\wedge E^c (\sigma_{abc})^{\gamma\delta} R_{\gamma\delta\alpha\beta}&\approx0\ ,\\
\nabla h_\alpha{}^\beta -4E^{\beta\gamma}\wedge\omega^{(+)}_{\alpha \gamma}-4\widetilde{E}_{\alpha\gamma}\wedge\omega^{(-)\beta \gamma}&\approx0\ ,\\
\nabla\omega^{(-)\alpha\beta}+ \tfrac14 \widetilde{E}^a\wedge \widetilde{E}^b\wedge \widetilde{E}^c (\tilde\sigma_{abc})_{\gamma\delta}\widetilde{R}^{\gamma\delta\alpha\beta}&\approx0\ .
\end{align}
The exotic Weyl tensor and its dual, which are algebraically independent in locally defined solution spaces, arise as primaries and anti-primaries, respectively, of chiral and anti-chiral Weyl zero-forms belonging to $SO(1,1)\times SO(1,5)$-covariant $\mathfrak{so}(2,6)$-irreps given by projections of a globally defined Weyl zero-form belonging to an $SO_0(2,6)$-module comprising various $SO_0(2,6)$-irreps encoding different boundary conditions.
Also these conformally dual structures fit naturally into the superoscillator realisation of the exotic supergravity multiplet.

\subsection{First-quantised geometry and second-quantised fields}
\label{aksz}

Starting from classical field theory on a metric space $(\boldsymbol{M},ds^2)$ whose equations of motion follow from applying the variational principle to a spacetime local action functional of a set of fields, its solution spaces can be quantised essentially by summing over these fields off-shell using functional integral methods. 
Whether this metric-like approach to quantum field theory exists or not, we propose an alternative approach by viewing unfolded systems as saddle points of a specific class of topological AKSZ sigma models.
The key difference between the two approaches is the role of the spacetime manifold itself.
In the metric-like approach, this manifold is the domain of the fundamental fields.
In the AKSZ approach, however, this manifold appears in sources of classical solution spaces, alias, Cartan integration modules, as well as in cosets of groups in the model's target.
The proposal, which is currently being investigated, is to quantise the system by \cite{FSG1,TwoCommaZero}:
\begin{itemize}
\item[i)] unfolding the classical equations of motion on $(\boldsymbol{M},ds^2)$ into a universally Cartan-integrable system (possibly with zero-form constraints), i.e., a space of maps 
\begin{align}
\varphi:\{\boldsymbol{X}\}\to \boldsymbol{Y}\ ,
\end{align}
from a set of $Q$-manifolds, referred to as sources, into a fixed $Q$-manifold, referred to as target, obeying the embedding, or $Q$-morphism, condition \cite{Sharapov:2017yde,Sharapov:2020quq}
\begin{align}
\varphi_\ast \vec Q_{\boldsymbol{X}}\approx \vec Q_{\boldsymbol{Y}}\ ,
\end{align}
described locally by flat superconnections\footnote{Letting $\mathcal{M}=\{\varphi:\boldsymbol{X}\to\boldsymbol{Y}\}$;  ${\rm ev}:\mathcal{M}\times \boldsymbol{X}\to \boldsymbol{Y}$ be the evaluation map; $x:\boldsymbol{Y}'\to \boldsymbol{\mathcal{A}}[1]$ be a local target coordinate; $\boldsymbol{\tau}_\alpha$ be a basis for $\boldsymbol{\mathcal{A}}$; and $V^I$ be a basis for $\Omega(\boldsymbol{X}')$; the superconnection $\boldsymbol{x}:={\rm ev}^\ast x\equiv  X^\alpha \boldsymbol{\tau}_\alpha$, where ${\rm deg}(X^\alpha)=1-{\rm deg}(\boldsymbol{\tau}_\alpha)$, and $X^\alpha \equiv \nu^\alpha_I V^I$, where $\nu^\alpha_I$ coordinatise a chart of $\mathcal{M}$ and ${\rm gh}(\nu^\alpha_I)={\rm deg}(X^\alpha)-{\rm deg}(V^I)$. Letting $\{m_n\}_{n=1}^\infty$ denote the $n$-ary products of $\boldsymbol{\mathcal{A}}$, and assuming $C(\boldsymbol{X})$ to be graded associative, $\vec Q_{\mathcal{M}} \boldsymbol{x}:=\vec Q_{\boldsymbol{X}} \boldsymbol{x}- \sum_{n=1}^\infty m_n(\boldsymbol{x},\cdots,\boldsymbol{x})\equiv R^\alpha \boldsymbol{\tau}_\alpha$ defines a $Q$-structure on $\mathcal{M}$, and the embedding condition implies the flatness condition $\vec Q_{\mathcal{M}} \boldsymbol{x}\approx 0$, defining a shell $\mathcal{C}\subset \mathcal{M}$ dual to a free-differential subalgebra of $\Omega(\boldsymbol{X})$ generated by $X^\alpha$ subject to $R^\alpha\approx 0$.} $\boldsymbol{x}$ valued in parity-shifts $\boldsymbol{\mathcal{A}}[1]$ of homotopy Lie algebras arising as charts of $\boldsymbol{Y}$;
\item[ii)] extending (i) into a universal Cartan integration module consisting of $Q$-morphisms
\begin{align}
\varphi:\{\boldsymbol{X}\}\to \boldsymbol{Y}\times \boldsymbol{G}\times \boldsymbol{Y}\ ,
\end{align}
where $\boldsymbol{G}$ is the Cartan gauge group, with local descriptions in terms of pairs $(\boldsymbol{x},\boldsymbol{x}')$ of flat superconnections acting from the left and right on a covariantly constant function $\boldsymbol{g}$ valued in $\boldsymbol{G}$;
\item[iii)] constructing classical solution spaces on $\boldsymbol{M}$ inside the Cartan integration module (ii), i.e., spaces of $Q$-morphisms
\begin{align}
\bar\varphi:T[1]\boldsymbol{M}\to \overline{\boldsymbol{Y}}\ ,\qquad \bar\varphi_\ast \vec Q_{T[1]\boldsymbol{M}}=\vec Q_{ \overline{\boldsymbol{Y}}}\ ,
\end{align}
where $(\overline{\boldsymbol{Y}},\vec Q_{\overline{\boldsymbol{Y}}})$ are consistent truncations of  $({\boldsymbol{Y}},\vec Q_{{\boldsymbol{Y}}})$ described by embeddings 
\begin{align}
\imath: \overline{\boldsymbol{Y}}\hookrightarrow {\boldsymbol{Y}}\ ,\qquad \imath_\ast\vec Q_{\overline{\boldsymbol{Y}}}=\vec Q_{{\boldsymbol{Y}}}\ ,
\end{align}
which encodes boundary conditions into  integration constants for $\overline{\boldsymbol{g}}$ and $\overline{\boldsymbol{x}}'$, providing spacetime coordinates and Fourier modes, respectively, for harmonic expansions of linearised configurations in $\overline{\boldsymbol{x}}$;
\item[iv)] equipping $(\boldsymbol{Y},\vec Q_{\boldsymbol{Y}})$ with a  compatible symplectic structure of degree\footnote{$\hat p=3-\hat s$ where $\hat s$ is the intrinsic degree of the supertrace ${\rm STr}_{\boldsymbol{\mathcal A}}$ used to construct the symplectic structure on $\boldsymbol{Y}$.} $\hat p-1$, $\hat p\in \{0,1,\dots\}$ lifting to an anomalous Hamiltonian $Q$-structure on $\boldsymbol{Y}\times T^\ast[\hat p-1]\boldsymbol{G}\times \boldsymbol{Y}$ whose anomaly vanishes on the zero-section of $T^\ast[\hat p-1]\boldsymbol{G}$;
\item[v)] quantising (iv) using AKSZ sigma models with off-shell configuration spaces
\begin{align}
\hat\varphi:\{T[1]{\boldsymbol{\Sigma}}\}\to T^\ast[\hat p]\left(\boldsymbol{Y}\times T^\ast[\hat p-1]\boldsymbol{G}\times \boldsymbol{Y}\right)\ ,\qquad {\rm dim}({\boldsymbol{\Sigma}})=\hat p+1\ ,
\end{align}
on open sources subject to anomaly-cancelling boundary conditions including (iii), viz.,
\begin{align}
\hat\varphi|_{T[1]\partial{\boldsymbol{\Sigma}}} = \imath \circ \bar\varphi\circ \hat{\imath}\ ,
\end{align}
using a consistent truncation 
\begin{align}
\hat{\imath}: T[1]\partial{\boldsymbol{\Sigma}}\to T[1]\boldsymbol{M}\ ,\qquad \hat{\imath}_\ast \vec Q_{ T[1]\partial{\boldsymbol{\Sigma}}}\hookrightarrow \vec Q_{T[1]\boldsymbol{M}}\ ;
\end{align}
\item[vi)] For $\hat p=1$, the resulting two-dimensional sigma model actions have quadratic approximations producing boundary operator algebras whose BRST cohomologies contain quantum fields built from

-- creation and annihilation operators for particle states arising as quantised integration constants for Weyl zero-forms contained in $\overline{\boldsymbol{x}}'$; and

-- commutative coset elements obtained from integration constants contained in $\overline{\boldsymbol{g}}$ playing the role of spacetime coordinates.
\end{itemize}

\noindent The AKSZ approach to a relativistic quantum field theory is akin to second-quantisation of a first-quantised system, whereby
\begin{itemize}
\item[--] the spacetime manifold enters actively at the level of constructing classical solution spaces within first-quantised operator algebras coordinatised by integration constants;
\item[--] the integration constants  provide expectation values for fundamental fields of a  topological sigma model second-quantising the original first-quantised noncommutative geometry; 
\item[--] the Hamiltonian structure of the fundamental relativistic quantum fields emerges within the graded target of the second-quantised topological sigma model rather than time-slicing of any Lagrangian density written directly on the original spacetime manifold; and 
\item[--] the BRST cohomology of the second-quantised model contains elements given by traces over the underlying first-quantised operator algebras valued in the second-quantised algebra, including functionals given by integrals over the spacetime manifold and its submanifolds, interpretable as effective actions and charges\footnote{ For example, the Einstein--Cartan action arises as such a cohomological element in the AKSZ approach to gravity.}. 
\end{itemize}
This approach is particularly useful in quantising classical relativistic field theories that are  
\begin{itemize}
\item[--] spacetime nonlocal viewed as Lagrangian field theories on $(\boldsymbol{M},g)$;
\item[--] non-Lagrangian viewed as classical field theories on $(\boldsymbol{M},g)$; or
\item[--] non-Lagrangian and spacetime nonlocal viewed as classical field theories on $(\boldsymbol{M},g)$.
\end{itemize}
Theories of these types arise naturally as universally Cartan integrable systems of horizontal superconnections on noncommutative fibre bundles, referred to as correspondence spaces, viewable as targets of first-quantised AKSZ sigma models.
Among these, Vasiliev's four-dimensional higher-spin gravities \cite{more,Vasiliev:1999ba,Bekaert:2004qos,Didenko:2014dwa} arise on correspondence spaces associated to first-quantised conformal particles with 
\begin{itemize}
\item[--] twistorial fibres inducing four-dimensional Newman--Penrose transformations of  sending states in induced representations to Lorentz covariant fields on-shell \cite{fibre,2011,2017,corfu19}; and 
\item[--]  Cartan integration submodules with sources $\boldsymbol{M}$ comprising spacetimes and additional noncommutative twistor spaces supporting closed and central elements in positive form degrees inducing nontrivial deformations \cite{more,FSG1,FSG2,Sharapov:2022awp,Didenko:2022qga}.
\end{itemize}
The noncommutative higher-spin geometries have (non-compact) Chern classes with perturbative expansions in terms of Weyl zero-forms, which are operators  represented in Hermitian spaces encoding local degrees of freedom of linearised configurations obeying boundary conditions in asymptotically locally constantly curved commutative spacetime geometries; for example, expanding the zero-form charges using unfolded boundary-to-bulk propagators yields correlation functions of holographically dual CFTs.

The aforementioned formalism applies naturally to linearised exotic supergravity, to which we turn next.

\section{Superconformal structure of linearised system}
\label{algebraicstructures}

In this section, we review the unfolded system's algebraic structure centered around its field content  fitted into $\mathbb{Z}\times \mathbb{Z}_2$-graded representations of the superconformal group.

\subsection{Induced representations and cocycles}

In an unfolded system, a spacetime background arises as a flat one-form connection valued in a Lie algebra $\mathfrak{g}$ and with holonomies in a corresponding group $G$, referred to as the vacuum gauge group. 
Linearisation yields locally defined $G$-modules consisting of $G$-irreps in distinct form degrees glued together via cocycles\footnote{Cocycles of vanishing form-degree encode linearised Stuckelberg shift symmetries; contracting these symmetries, the remaining linear $G$-module in form-degree zero decomposes under $G$ into a direct sum of $G$-irreps encoding local degrees of freedom.}.
Globally defined configurations arise by glueing together $G$-modules using transition functions from a structure group $H\subset G$ with Lie algebra $\mathfrak{h}$.
To this end, it is assumed that the $G$-modules admit monomorphic projections to $\mathfrak{g}$-modules consisting of $H$-irreps that are thus glued together
\begin{itemize}
\item[a)] on overlaps using transition functions valued in $H$; and 
\item[b)] on charts using $H$-covariant cocycles compatible with Cartan integrability, i.e., cohomological elements of $\mathfrak{h}$-relative Chevalley-Eilenberg cochain complexes built from $\mathfrak{g}/{\mathfrak{h}}$-valued background one-forms, alias, frame fields.
\end{itemize}
The locally defined $G$-modules can be built using Cartan integration, i.e., by subjecting integration constants\footnote{The integration constants may arise in zero-forms or via delta-forms in strictly positive degrees.} valued in $G$-irreps induced from subgroups $H'\subset G$ encoding various boundary conditions to finite Cartan gauge transformations
The resulting locally defined forms are glued together chartwise using $H'$-covariant cocycles built from $\mathfrak{g}/{\mathfrak{h}}'$-valued background one-forms compatible with Cartan integrability, i.e., cocycle elements of ${\mathfrak{h}}'$-relative Chevalley-Eilenberg cochain complexes.

\subsection{Vacuum gauge group and Howe-dual group}

We assume that the background holonomies are valued in the vacuum gauge group 
\begin{align}
G=MOSp(8^\ast| 8)\ ,
\end{align}
defined as the metaplectic double cover 
\begin{align}
\mathbb{Z}_2(-1)\hookrightarrow MOSp(8^\ast| 8)\stackrel{\pi}{\to} OSp(8^\ast| 8)
\end{align}
of the matrix supergroup $OSp(8^\ast| 8)$ via its superoscillator realisation.
The body, i.e., the underlying bosonic Lie group, coincides with the maximal bosonic subgroup, viz., 
\begin{align}
{\rm Body}(G)=G_C\times G_R\ ,\qquad G_C=MSpin(2,6)\ ,\qquad G_R=USp(8)\ , 
\end{align}
where 
\begin{itemize}
\item[--] the R-symmetry group  
\begin{align}
USp(8):=Sp(8;\mathbb{C})\cap U(8)\ ,   
\end{align}
which is compact and simply connected, implying $\pi(G_R)=G_R$, is realised using fermionic oscillators; and
\item[--] the conformal group\footnote{For $n\geqslant 3$, $O(2,n)=SO(2,n)\rtimes \mathbb{Z}_2(\sigma_r)$ and $SO(2,n)=SO_0(2,n)\rtimes \mathbb{Z}_2(\sigma_{+}\sigma_{-})$, where $\sigma_r$ is any reflection and $\sigma_{\epsilon}$, $\epsilon=\pm$, are reflections in directions of signature $\epsilon$. One has $\pi_0(SO_0(2,n))=1$ and   $\pi_1(SO_0(2,n))\cong \pi_1(SO(2))\times \pi_1(SO(n))$ with $\pi_1(SO(2))=\mathbb{Z}$ and $\pi_1(SO(n))=\mathbb{Z}_2$. The spin-cover $Spin(2,n)$ has $\pi_1(Spin(2,n))\cong \pi_1(SO(2))$. For $n=3,4,6$, $Spin(2,n)$ is the conformal group in $n$ spacetime dimensions, and it admits projective representations using bosonic spinorial oscillators inducing the metaplectic doublings $\mathbb{Z}_2(-1)\hookrightarrow MSp(2k_n;\mathbb{R})\stackrel{\pi}{\to} Sp(4k_n;\mathbb{R})$, $k_n=2,4,8$, containing the Howe-dual pairs $\widetilde K_n\times MSpin(2,n)$ with $\widetilde K_3=1$, $\widetilde K_4=U(1)$, and $\widetilde{K}_6=SU(2)$.}
\begin{align}
\mathbb{Z}_2(-1)\hookrightarrow MSpin(2,6)\stackrel{\pi}{\to} Spin(2,6)\ , 
\end{align}
which is non-compact and non-simply connected,
is the metaplectic doubling of the spin-cover $Spin(2,6)$ of $SO_0(2,6)$ realised via bosonic spinorial oscillators as a subgroup of the metaplectic double cover
\begin{align}
\mathbb{Z}_2(-1)\hookrightarrow Mp(16;\mathbb{R})\stackrel{\pi}{\to}  Sp(16;\mathbb{R})\ , 
\end{align}
of the symplectic matrix group $Sp(16;\mathbb{R})$. 
\end{itemize} 
In the superoscillator realisation, the superconformal group $G$ arises within the metaplectic supergroup 
\begin{align}
\mathbb{Z}_2(-1)\hookrightarrow MOSp(16|16)\stackrel{\pi}{\to}OSp(16|16)\ ,
\end{align}
with body $Spin(16;\mathbb{R})\times MSp(16;\mathbb{R})$, 
as a member of a Howe-dual pair together with an
\begin{align}
SU(2)_{\widetilde K}\equiv \widetilde{K}\subset MOSp(16|16)\ ,
\end{align}
viz.,
\begin{align}
G={\rm Stab}_{MOSp(16|16)}(\widetilde{K})\ ,\qquad 
\widetilde{K}={\rm Stab}_{MOSp(16|16)}(G)\ .
\end{align}
Letting $U$ denote the metaplectic superoscillator representation of $\pi(G)\equiv OSp(8^\ast|8)$, which yields a representation of its group algebra with associative product $\star$, inducing a superoscillator representation of $\mathfrak{g}$ that we also denote by $U$, the conformal duality transformation 
\begin{align}
\kappa_\sigma := U(\sigma) \in G_C\ ,
\end{align}
is the metaplectic uplift of 
\begin{align}
\sigma := e^{i\pi E}\in Spin(2,6)\subset Sp(16;\mathbb{R})\ ,\qquad E\equiv M_{00'}\ ,
\end{align}
 i.e., 
\begin{align}
\sigma=\sigma_{0}\sigma_{0'}\ ,
\end{align}
where $\sigma_{0}$ and $\sigma_{0'}$ are co-images under the vectorial representation of the reflections in the time-like $0$ and $0'$ directions.
In the fundamental matrix representation of $Sp(16;\mathbb{R})$, one has 
\begin{align}
\sigma=J_8\otimes i\sigma^2\ ,\qquad J_8=i\sigma^2\otimes {\rm Id}_{{\rm Mat}_4}\ ,\qquad U(E)=\frac 12 w\ ,\qquad w:=\sum_{\underline{\alpha}=1}^8 w_{(\underline{\alpha})}\ ,
\end{align}
where $w_{(\underline{\alpha})}$ are Weyl-ordered harmonic-oscillator Hamiltonians \cite{meta,Iazeolla:2022singularities}, i.e., $\kappa_\sigma$ acts by Fourier transformation in all $8$ conjugate pairs\footnote{The $\pi$-rotation in the $00'$-plane corresponds to a combined $\pi/2$-rotation in all $8$ canonical symplectic planes of the harmonic oscillators in twistor space.}. It follows that $\kappa_\sigma=\exp_\star(i\pi w/2)$, hence 
\begin{align}
\kappa_\sigma\star \kappa_\sigma=(-1)^w_\star\ ,\qquad (\kappa_\sigma)^{\star 4}=1\ ,
\end{align}
and that ${\rm Ad}^\star_{\kappa_\sigma}(\cdot):=(\kappa_\sigma)^{-1}\star (\cdot)\star \kappa_\sigma$ acts on polynomials in superoscillators by exchanging chiral and anti-chiral spinor indices on $\pi(G)$-tensors \cite{TwoCommaZero,wip:2026}\footnote{The conformal duality transformations preserve the overall chirality defined by the Pfaffian Casimir of $Spin(2,6)$.
Thus, the chirality matrices acting in the old and new bases differ by a sign; defining new gamma matrices $(\sigma^{\prime a})_{\alpha'\beta'}=\delta_{\alpha'\alpha}\delta_{\beta'\beta}(\tilde\sigma^{ a})^{\alpha\beta}$ and $(\tilde\sigma^{\prime a})^{\alpha'\beta'}=\delta^{\alpha'\alpha}\delta^{\beta'\beta}(\sigma^{ a})_{\alpha\beta}$, the new chirality matrix $(\Gamma')_{\underline\alpha'}{}^{\underline\beta'}=-\delta_{\underline\alpha'}^{\underline\alpha}\delta_{\underline\beta}^{\underline\beta'}(\Gamma)_{\underline\alpha}{}^{\underline\beta}$, such that ${\rm Ad}^\star_{\kappa_\sigma}(\pi_\sigma(\Gamma Q_I))=\Gamma' {\rm Ad}^\star_{\kappa_\sigma}(Q_I)$ and ${\rm Ad}^\star_{\kappa_\sigma}(\Gamma S_I)= \Gamma'{\rm Ad}^\star_{\kappa_\sigma}(S_I)$.}; in particular, 
\begin{align}
{\rm Ad}^\star_{\kappa_\sigma} U(x) =U(\pi_\sigma(x))\ ,\qquad x\in \mathfrak{g}\ ,
\end{align}
where the Weyl superreflection $\pi_\sigma$ is the cyclic map of order four
characterised by\footnote{Eq. \eqref{3.16} corrects an error in \cite{TwoCommaZero}.} 
\begin{align}\label{3.16}
\pi_\sigma(D)=-D\ ,\qquad \pi_\sigma(T_a)=K_a\ ,\qquad \pi_\sigma(Q_\alpha^I)=\delta_{\alpha\alpha'}S^{\alpha' I}\ ,\qquad \pi_\sigma(S^{\alpha I})=-\delta^{\alpha\alpha'}Q_{\alpha'}^I\ .
\end{align}
As we shall see, this $\mathbb{Z}_4$-action  

-- exchanges conformally dual chiral and anti-chiral Weyl zero-forms containing self-dual and anti-self-dual three-form curvatures; and 

-- extends to a discrete symmetry of the linearised unfolded system including two-form potentials and cocycle, and its quadratic invariants in form-degree zero and six.

\subsection{Structure group}

Six-dimensional conformal geometries are characterised by their structure subgroup $H\subset G$:
\begin{itemize}
\item[--] Conformal backgrounds are generalised BTZ-like  geometries built from gauge functions valued in $G/H$; 
\item[--] Classical moduli spaces consist of locally defined  induced representations encoding boundary conditions glued together by transition functions from $H$ into global configurations; and
\item[--] Local classical observables, alias, abelian $p$-forms, are given by integrals of $H$-invariant on-shell closed composite forms over topologically non-trivial cycles (which may have boundaries attached to corners of manifolds themselves being boundaries).
\end{itemize}
We assume that
\begin{align}
\pi_1(H)=1\ ,
\end{align}
implying that the metaplectic representation provides a single cover of the structure matrix group $\pi(H)$, i.e., $H\cong \pi(H)$; to this end, we take
\begin{align}
H=H_C\times G_R\ ,\qquad H_C=O(1,1)\times Spin(1,5)\ .
\end{align}
The vacuum connection thus contains a structure group connection valued in 
\begin{align}
\mathfrak{h}=\mathfrak{h}_C\oplus \mathfrak{g}_R\ ,\qquad \mathfrak{h}_C=\mathfrak{so}(1,1)\oplus\mathfrak{so}(1,5)\ ,\qquad \mathfrak{g}_R=\mathfrak{usp}(8)\ ,
\end{align}
and a superconformal frame valued in $\mathfrak{g}_C/\mathfrak{h}_C$ splitting into a dual pair of super-Poincar\'e frames exchanged by ${\rm Ad}_{\kappa_\sigma}$, which 
is thus broken in super-Poincaré vacua.

\subsection{Field content }

The (hypothetical) full system is assumed to admit a linearisation and a further consistent truncation to an unfolded system consisting of the following $G\times {\widetilde{K}}$-modules\footnote{The real forms of $\boldsymbol{\Omega}$ and $\boldsymbol{k}$ are chosen using the Hermitian conjugation operation of the superoscillator algebra.
Reality conditions on $\boldsymbol{c}$ can be imposed by 
realising $\mathsf{S}$ using Hermitian left modules within the holomorphic extension of the group algebra of $G$ equipped with a trace operation and separate Hermitian and linear conjugation operations; see \cite{FSG1,FSG2} for analogous constructions in the context of three-dimensional CFT.
The reality condition on $\boldsymbol{c}$ triggers a real form of $\boldsymbol{b}^{(\varepsilon)}$ to be studied elsewhere.}: 
\begin{itemize}
\item[i)] a vacuum connection 
\begin{align}
\boldsymbol{\Omega}\equiv \boldsymbol{\Omega}_{\mathfrak{g}}+\boldsymbol{\Omega}_{\tilde{\mathfrak{k}}} \in  \Omega_{[1]}(\boldsymbol{M}')\otimes (\mathfrak{g}\oplus \tilde{\mathfrak{k}})\ ,\qquad \mathfrak{g}\equiv \mathfrak{osp}(8^\ast| 8)\ ,\qquad \tilde{\mathfrak{k}}\equiv \mathfrak{su}(2)\ ,
\end{align}
defined on charts $\boldsymbol{M}'$ of $\boldsymbol{M}$;
\item[ii)] a globally defined abelian curvature zero-form 
\begin{align}
\boldsymbol{c}\in  \Omega_{[0]}(\boldsymbol{M})\otimes \left(\mathsf{S}|_G\otimes \boldsymbol{(0)}|_{\widetilde{K}}\right)_{[1]}\ ,
\end{align}
alias, the Weyl zero-form, where $\mathsf{S}$ is the extended supersingleton making up the spectrum of local degrees of freedom,  including the unitarizable exotic graviton multiplet; 
\item[iii)] a conformally dual pair of locally defined linearised two-form potentials
\begin{align}
\boldsymbol{b}^{(\varepsilon)}\in \Omega_{[2]}(\boldsymbol{M}')\otimes \left(\mathsf{T}|_G\otimes \boldsymbol{(1)}|_{{\widetilde{K}}}\right)_{[-1]}\ ,\qquad \varepsilon=\pm\ , 
\end{align}
where $\mathsf{T}\equiv\boldsymbol{(2)}$ is the supertraceless graded-symmetric rank-two supertensor of $G$.
\end{itemize}

\noindent Viewed as $\mathfrak{g}$-irreps, the zero- and two-form modules share quadratic super-Casimirs, viz.,
\begin{align}\label{3.7}
C_2\left(\mathfrak{g}\left|\mathsf{S}\right.\right)=C_2\left(\mathfrak{g}\left|\mathsf{T}\right.\right)=0\ ,
\end{align}
as required for the existence of cocycles glueing $\boldsymbol{c}$ to $\boldsymbol{b}^{(\varepsilon)}$ in general superconformal backgrounds.
The matching of Howe-dual quantum numbers requires 
\begin{itemize}
\item[iv)] a background zero-form 
\begin{align}
\boldsymbol{k}\in \Omega_{[0]}(\boldsymbol{M}')\otimes\left( \bullet|_G\otimes \boldsymbol{(1)}_{{\widetilde{K}}}\right)_{[1]}\ ,
\end{align}
referred to as the Howe-dual zero-form.
\end{itemize}

\subsection{Chiral and anti-chiral Weyl zero-form modules}

The extended supersingleton is a left $G$-module consisting of a spectrum of $G$-irreps, viz.,  
\begin{align} 
\mathsf{S}\downarrow_G \,=\bigoplus_{\xi\equiv  (\xi_{C},\xi_R)} \mathsf{S}(\xi)|_G\ ,
\end{align}
labeled by polarizations of  ${\rm Body}(G)$, viz., 
\begin{align}
\mathsf{S}(\xi)\downarrow_{{\rm Body}(G)}\,=\bigoplus_{(\xi_C,\xi_R)}\mathsf{H}(\xi_C)|_{G_C}\otimes \mathsf{R}(\xi_R)|_{G_R}\ ,
\end{align}
where 
\begin{itemize}
\item[--] $\xi_C$ labels infinite-dimensional Hermitian $G_C$-irreps $\mathsf{H}(\xi_C)$ arising as $\widetilde{K}$-invariant left modules in the algebra of the holomorphic complexified metaplectic group $Mp(16;\mathbb{C})$, which thus acts as a spectrum-generating group \cite{meta,FSG1}; and
\item[--] $\xi_R$ labels finite-dimensional representations $\mathsf{R}(\xi_R)$ of $USp(8)$ arising in its fermionic oscillator realization, which are hence unitarizable. 
\end{itemize}
Reference states $\psi_\xi \in \mathsf{S}(\xi)$ induce $G$-orbits $\mathsf{S}(\xi|\psi_\xi)$ inducing $\mathfrak{g}$-irreps
\begin{align}
\left.\mathsf{S}(\xi|\psi_\xi)\right|_{\mathfrak{g}}\subseteq \mathsf{S}(\xi)\ ,
\end{align}
generated from $\psi_\xi$, providing coordinatisations of $\mathsf{S}(\xi|\psi_\xi)$; these are global if $\mathsf{S}(\xi)$ is unitarizable.

The structure-group assumption requires that $\mathsf{S}$ consists of a set of $H$-covariant coordinate charts, i.e.,
\begin{align}
\mathsf{S}=\bigcup_{\Xi} \mathsf{S}^{(\Xi)}|_\mathfrak{g}\ ,
\end{align} 
equipped with $\mathfrak{g}$-epimorphisms 
\begin{align}
t^{(\Xi)}: \mathsf{S}^{(\Xi)}\to\mathsf{T}(\lambda_\Xi) \ ,\qquad \mathsf{T}(\lambda)\downarrow_H= \bigoplus_{\lambda'} \mathsf{T}(\lambda|\lambda')|_H\ ,
\end{align}
with monic inverses
\begin{align}
s^{(\Xi)}: \mathsf{T}(\lambda_\Xi)\to\mathsf{S}^{(\Xi)}\ ,\end{align}
that are
\begin{itemize}
\item[i)] universal, i.e., independent of $\xi$; and
\item[ii)] local, i.e., $G$ acts faithfully in $\mathsf{S}$ but not in $\mathsf{S}^{(\Xi)}$,  corresponding to singularities in $H$-tensorial fields.
\end{itemize}
\noindent In addition, we assume that they are 
\begin{itemize}
\item[iii)] analytical, i.e., a state in $\mathsf{S}$ can be recovered globally from its local $G$-orbit close to a regular configuration in $\mathsf{S}^{(\Xi)}$ by means of  analytical continuation in the metaplectic group; and
\item[iv)] conformal-duality covariant, i.e., 
\begin{align}
\kappa_\sigma:  \mathsf{S}^{(\Xi)}\to\mathsf{S}^{(\sigma(\Xi))}\ ,
\end{align}
intertwines the $\mathfrak{g}$-actions.
\end{itemize}
\noindent These conditions are met by the conformally dual pair 
\begin{align}
\mathsf{S}^{(\pm)}\stackrel{t^{(\pm)}}{\cong} \mathsf{T}((\pm 2;0,0,0)|[4])^\pm\ , 
\end{align} 
referred to as the chiral (+) and anti-chiral (-) Weyl zero-form modules, respectively, which are $\mathfrak{g}$-irreps with decompositions
\begin{align}
\left.\mathsf{S}^{(\pm)}\right\downarrow_{\mathfrak{g}_C\oplus \mathfrak{g}_R}=\bigoplus_{\tiny\begin{array}{c}J=0,\frac12,1,\frac32,2\\r=4-2J\end{array}}\left.\mathsf{T}^{(\pm)}_{J}\right|_{\mathfrak{g}_C}\otimes \boldsymbol{[r]}|_{\mathfrak{g}_R}\ ,
\end{align}
in terms of 
\begin{itemize}
\item[--] finite-dimensional $\mathfrak{g}_R$-unirreps $\boldsymbol{[r]}$ extending to $G_R$-unirreps; and
\item[--] infinite-dimensional $\mathfrak{g}_C$-irreps 
\begin{align}
\mathsf{T}^{(\varepsilon)}_{J}\equiv \mathsf{T}(\Delta^{(\varepsilon)}_{J};J,J,\pm J)^\varepsilon\ ,\qquad \Delta^{(\pm)}_{J}=\pm(J+2)\ ,
\end{align}
with $H_C$-decompositions  
\begin{align}
\left.\mathsf{T}^{(\varepsilon)}_J\right\downarrow_{H_C}=\bigoplus_{k=0}^\infty \boldsymbol{\tau}^{(\varepsilon)}_{J;k}\ ,\qquad 
\boldsymbol{\tau}^{(\varepsilon)}_{J;k}\cong \left(\pm(J+k+2);J+k,J,J,\varepsilon J\right)\ ,\qquad \varepsilon=\pm\ ,
\end{align}
where $(\Delta;s_1,s_2,\varepsilon s_3)$ with $\Delta\in \mathbb{R}$, and $s_1\geqslant s_2\geqslant s_3$ with $s_1,s_2,s_3\in\{0,\tfrac12,1,\dots\}$ denote chiral $Spin(1,5)$-tensors of conformal weight $\Delta$, i.e., the conformal weights are bounded from below and above in $\mathsf{S}^{(+)}$ and $\mathsf{S}^{(-)}$, respectively. 
\end{itemize}
Thus, the linearised Weyl zero-form $\boldsymbol{c}$, which is regular as an element in the $G$-module $\mathsf{S}$, admits conformally dual chiral and anti-chiral restrictions
\begin{align}
\boldsymbol{c}^{(\varepsilon)}:= t^{(\varepsilon)}(\boldsymbol{c})\ ,
\end{align}
consisting of $H$-tensorial component fields defined locally on $\boldsymbol{M}$; thus, in a given spectrum
\begin{align}
\boldsymbol{c}^{(\varepsilon)}=\sum_\xi \boldsymbol{c}^{(\varepsilon)}(\xi)\ ,
\end{align}
consisting of components that are 

-- bounded in all of $\boldsymbol{M}$ if $\mathsf{S}(\xi)$ is unitarizable; and 

-- bounded away from singular subspaces of $\boldsymbol{M}$ of strictly positive codimensions if $\mathsf{S}(\xi)$ is non-unitarizable.

\noindent For example, if $\mathsf{S}(\xi)$ is a real six-manifold with the topology of conformal Minkowski spacetime ${\rm CMink}^{1,5}$ or its compacted extension $S^1\times S^5$, $\boldsymbol{c}^{(\pm)}$ are bounded and unbounded, respectively, when expanded in modes from the exotic supergravity multiplet and the self-dual string $G$-module.

The intertwining property (iv) enters the second-quantised conformal duality transformation defined in Section \ref{Sec:4.2}; for the transformation of $\boldsymbol{c}^{(\varepsilon)}$, see Eqs. \eqref{4.16},  \eqref{4.20}, and \eqref{4.21}.

\subsection{Superconformally covariant derivatives}

Acting on forms valued in a first-quantised  $\mathfrak{g}$-module $\mathsf{M}_{\mathfrak{g}}$,
the superconformally covariant derivative 
\begin{align}
\boldsymbol{D}_{\mathsf{M}_{\mathfrak{g}}}:\Omega_{[p]}(\boldsymbol{M'})\otimes \mathsf{M}_{\mathfrak{g}}\to \Omega_{[p+1]}(\boldsymbol{M}')\otimes \mathsf{M}_{\mathfrak{g}}\ ,
\end{align}
is defined by
\begin{align}
\boldsymbol{D}_{\mathsf{M}_{\mathfrak{g}}}:= d+ \rho_{\mathsf{M}_{\mathfrak{g}}}(\boldsymbol{\Omega}_{\mathfrak{g}})\ ,\qquad \rho_{\mathsf{M}_{\mathfrak{g}}}:\mathfrak{g}\to {\rm End}\left(\mathsf{M}_{\mathfrak{g}}\right)\ ,
\end{align}
written as $\boldsymbol{D}_{\mathfrak{g}}= d+ \boldsymbol{\Omega}_{\mathfrak{g}}$ in contexts free from ambiguity.
Under the structure group,
\begin{align}
\mathfrak{g}\downarrow_{{H}}=\mathfrak{h}\oplus \mathfrak{t}_{(1/2)}\oplus \mathfrak{t}_{(1)}\ ,\qquad
\left.\mathfrak{t}_{(\Delta)}\right\downarrow_{H}= \mathfrak{t}_{-\Delta}\oplus\mathfrak{t}_{\Delta}\ ,\qquad \Delta=\tfrac12, 1\ ,
\end{align}
where $\mathfrak{t}_{\Delta}$ has first-quantised conformal weight $\Delta$, and 
\begin{align}
{\rm Ad}_{\kappa_\sigma}|_{\mathfrak{h}}&= \left(-{\rm Id}_{\mathfrak{so}(1,1)}\right)\oplus{\rm Id}_{\mathfrak{so}(1,5)}\oplus {\rm Id}_{\mathfrak{usp}(8)}\ ,\\
 {\rm Ad}_{\kappa_\sigma}\mathfrak{t}_{\Delta}&= \mathfrak{t}_{-\Delta}\ .
\end{align}
Correspondingly, we split  
\begin{align}
\boldsymbol{\Omega}_{\mathfrak{g}}|_{\mathfrak{h}}= \boldsymbol{\Omega}_{\mathfrak{h}}+ \sum_{\Delta} \boldsymbol{e}_{\Delta}\ ,
\end{align}
summing over $\Delta=\pm\frac12,\pm1$ and using $H$-covariant one-forms
\begin{align}
\boldsymbol{\Omega}_{\mathfrak{h}}\in \Omega_{[1]}(\boldsymbol{M}')\otimes \mathfrak{h}\ ,\qquad  \boldsymbol{e}_{\Delta}\in \Omega_{[1]}(\boldsymbol{M}')\otimes \mathfrak{t}_{-\Delta}\ ,\qquad \Delta=\pm\tfrac12,\pm 1\ ,
\end{align}
expanded into $H$-tensorial components using the conventions 
\begin{align}\label{3.24}
\boldsymbol{\Omega}_{\mathfrak{h}}&=i\left(M_\alpha{}^\beta \Omega_\beta{}^\alpha+D \sigma+\frac{1}2 N^{IJ}\Omega_{IJ}\right)\ ,\\
\label{3.22}
\boldsymbol{e}_{+1}&=-\frac{i}2  T_{\alpha\beta}E^{\alpha\beta} \ ,\qquad \boldsymbol{e}_{+1/2}= i Q_{\alpha}{}^{I}F^{\alpha}{}_{ I}\ ,\\
\boldsymbol{e}_{-1}&=-\frac{i}2 K^{\alpha\beta}\widetilde{E}_{\alpha\beta}\ ,\qquad \boldsymbol{e}_{-1/2}=i  S^{\alpha I}\widetilde{F}_{\alpha I}\ .
\end{align}
The conformal duality transformations reverse conformal weights and exchange chiral and anti-chiral indices of $Spin(1,5)$ spinors, viz.,
\begin{align}
&{\rm Ad}_{\kappa_\sigma}(M_\alpha{}^\beta,D,N^{IJ},  T_{\alpha\beta},Q_{\alpha}{}^{I}, S^{\alpha I},K^{\alpha\beta})=\\&=(\delta_{\alpha\alpha'}\delta^{\beta\beta'}M^{\alpha'}{}_{\beta'},-D,N^{IJ},  \delta_{\alpha\alpha'}\delta_{\beta\beta'}K^{\alpha'\beta'},\delta_{\alpha\alpha'}S^{\alpha' I}, -\delta^{\alpha\alpha'}Q_{\alpha'}{}^{I},\delta^{\alpha\alpha'}\delta^{\beta\beta'}T_{\alpha'\beta'})\ ,
\end{align}
using
\begin{align}
M^\alpha{}_\beta\equiv -M_\beta{}^\alpha\ ;
\end{align}
the bi-spinorial frame and dual frame fields can be dualised using the $Spin(1,5)$ covariant epsilon tensor, viz.,
\begin{align}
E_{\alpha\beta}:=\frac12\epsilon_{\alpha\beta\gamma\delta}E^{\gamma\delta}\ ,\qquad  \widetilde{E}^{\alpha\beta}:=\frac12\epsilon^{\alpha\beta\gamma\delta}\widetilde{E}_{\gamma\delta}\ ,\qquad \epsilon^{\alpha\beta\gamma\delta}\epsilon_{\alpha\beta\gamma\delta}=4!\ .
\end{align} 
The component forms are thus coordinates of $\mathcal{M}$ with second-quantised conformal weights 
\begin{align}\label{3.25}
\hspace{-.5cm}\Delta_{\mathcal{M}}(E^{\alpha\beta})=-1\ ,\qquad
\Delta_{\mathcal{M}}(F^{\alpha I})=-\tfrac12\ ,\qquad\Delta_{\mathcal{M}}(\widetilde{F}_{\alpha I})=+\tfrac12\ ,\qquad \Delta_{\mathcal{M}}(\widetilde{E}_{\alpha\beta})=+1\ .
\end{align}

\subsection{Structure-group covariant derivatives}

Acting on forms valued in a first-quantised $H$-module $\mathsf{M}_{H}$,
the $H$-covariant derivative 
\begin{align}
\boldsymbol{\nabla}_{\mathsf{M}_{H}}:\Omega_{[p]}(\boldsymbol{M}')\otimes \mathsf{M}_{H}\to \Omega_{[p+1]}(\boldsymbol{M}')\otimes \mathsf{M}_{H}\ ,
\end{align}
is defined by 
\begin{align}
\boldsymbol{\nabla}_{\mathsf{M}_{H}}:= d+ \rho_{\mathsf{M}_{H}}(\boldsymbol{\Omega}_{\mathfrak{h}})\ ,\qquad \rho_{\mathsf{M}_{H}}:H\to {\rm End}\left(\mathsf{M}_{H}\right)\ ,
\end{align}
written as $\boldsymbol{\nabla}= d+ \boldsymbol{\Omega}_{\mathfrak{h}}$ in contexts free from ambiguity.
Decomposing 
\begin{align}
\left.\mathsf{M}_{\mathfrak{g}}\right\downarrow_{H}=\bigoplus_{\vec w} \mathsf{M}_{(\vec w)}\ ,   \end{align}
where ${\vec w}$ are $H$-labels, we have 
\begin{align}
\boldsymbol{D}_{\mathsf{M}_{\mathfrak{g}}}=\sum_{\vec w} \nabla_{(\vec w)}+\sum_{\vec w,\vec w'}\sigma_{(\vec w)}^{(\vec w')}\left(\sum_{\Delta}\boldsymbol{e}_{\Delta}\right)\ ,\qquad \nabla_{\vec w}\equiv \nabla_{\mathsf{M}_{(\vec w)}}\ ,
\end{align}
summing over $\Delta=\pm \tfrac12,\pm 1$, written as $\boldsymbol{D}_{\mathfrak{g}}=\boldsymbol{\nabla}+\sum_{\Delta} \boldsymbol{e}_\Delta$ in ambiguity-free contexts, and where 
\begin{align}
\sigma_{(\vec w)}^{(\vec w')}: \mathsf{M}_{(\vec w')}\to \mathsf{M}_{(\vec w)}\ ,
\end{align}
using $H$-covariant cocycles of $\rho_{\mathsf{M}_{\mathfrak{g}}}$ which are active when $ \mathfrak{t}_{\Delta}\otimes \mathsf{M}_{(\vec w')}$ contains $\mathsf{M}_{(\vec w)}$.

\subsection{Basis for superconformal tensors}

We equip the fundamental representation $\boldsymbol{(1)}$ of $\pi(G)\equiv OSp(8^\ast|8)$ and its dual $\boldsymbol{(1)}^\ast$, which are thus unfaithful representations of $G$, with bases elements $\boldsymbol{\tau}^A$ and $\boldsymbol{\tau}^\ast_A$ of vanishing AKSZ degrees and distinct fermion numbers, viz., 
\begin{align}
\boldsymbol{\tau}^\ast_A(\boldsymbol{\tau}^B)= \delta_A^B\ ,\qquad 
(-1)^{{\rm fer}(\boldsymbol{\tau}^A)}=(-1)^{{\rm fer}(\boldsymbol{\tau}^\ast_A)}\equiv (-1)^A\ ,
\end{align}
and $\mathfrak{g}$ with a corresponding tensorial basis $L^{AB}$, viz., 
\begin{align}
[L^{AB},L^{CD}]&=4i\zeta^{[B|[C|}L^{A]|D]}\ ,\qquad L^{AB}=-(-1)^{AB}L^{BA}\ ,\\
L^{AB} \boldsymbol{\tau}^C&=2i\zeta^{[B|C}\boldsymbol{\tau}^{|A]}\ ,\qquad \boldsymbol{\tau}^\ast_A L_{BC}=2i\zeta_{A[B}\boldsymbol{\tau}^\ast_{C]}\ ,
\end{align}
where the $\pi(G)$-invariant tensor $\zeta^{AB}$ obeys
\begin{align}
\zeta^{AB}=(-1)^{AB}\zeta^{BA}\ ,\qquad \zeta^{AB}\zeta_{AC}=\delta^B_C\ ,
\end{align}
and fundamental indices are dualised using the conventions 
\begin{align}
V^A=\zeta^{AB}V_B\ ,\qquad V_A=V^B\zeta_{BA}\quad \Rightarrow\quad V^A W_A=(-1)^A V_A W^A\ .
\end{align}
Expanding $\boldsymbol{\Omega}_{\mathfrak{g}}$ and $\boldsymbol{\psi}_{(1)}\in \Omega(\boldsymbol{M}')\otimes (1)|_G$ as
\begin{align}
\boldsymbol{\Omega}_{\mathfrak{g}}:=\frac{i}2 L^{AB}\Omega_{BA}\ ,\qquad \Omega^{AB}=-(-1)^{AB}\Omega^{BA}\ ,\qquad 
\boldsymbol{\psi}_{(1)}=\boldsymbol{\tau}^A \Psi_A\ ,
\end{align}
yields 
\begin{align}
\boldsymbol{R}_{\mathfrak{g}}&:=d\boldsymbol{\Omega}_{\mathfrak{g}}+\boldsymbol{\Omega}_{\mathfrak{g}}\wedge \boldsymbol{\Omega}_{\mathfrak{g}}=: \frac{i}2 L^{AB}R_{BA}\ ,\qquad R_{AB}= d\Omega_{AB}+\Omega_{A}{}^C\wedge \Omega_{CB}\ ,\\
\boldsymbol{D}_{\mathfrak{g}}\boldsymbol{\psi}_{(1)}&:= d\boldsymbol{\psi}_{(1)}+\boldsymbol{\Omega}_{\mathfrak{g}}\wedge \boldsymbol{\psi}_{(1)} =:\boldsymbol{\tau}^A D\Psi_A\ ,\qquad D\Psi_A=d\Psi_A+\Omega_A{}^B\wedge \Psi_B\ .
\end{align}
The resulting component forms of the Bianchi identities read
\begin{align}
DD\Psi_A\equiv R_A{}^B \wedge \Psi_B\ ,\qquad DR_{AB}\equiv 0\ .
\end{align}
Differential forms valued in rank-$R$ supertensors, viz.,
\begin{align}
\boldsymbol{\psi}_{(1|\cdots|1)}= \boldsymbol{\tau}^{A_1}\otimes \cdots \otimes \boldsymbol{\tau}^{A_R}  \Psi_{A_R|\dots|A_1}\in \Omega(\boldsymbol{M}')\otimes (1)_{\pi(G)}\otimes\cdots \otimes (1)_{\pi(G)}\ ,
\end{align}
are differentiated using the convention
\begin{align}
\boldsymbol{D}_{\mathfrak{g}}\boldsymbol{\psi}_{(1|\cdots|1)}:= d\boldsymbol{\psi}_{(1|\cdots|1)}+\boldsymbol{\Omega}_{\mathfrak{g}}\wedge\boldsymbol{\psi}_{(1|\cdots|1)}=: \boldsymbol{\tau}^{A_1}\otimes \cdots \otimes \boldsymbol{\tau}^{A_R}  D\Psi_{A_R|\dots|A_1}\ ,
\end{align}
where 
\begin{align}
D\Psi_{A_1|\dots|A_R}={}& d\Psi_{A_1|\dots|A_R}+\Omega_{A_1}{}^B \wedge\Psi_{B|A_2|\dots|A_R}+(-1)^{(A_2+B)A_1}\Omega_{A_2}{}^B \wedge\Psi_{A_1|B|\dots|A_R}\\&+\cdots+(-1)^{(A_R+B)(A_1+\cdots+A_{R-1})}\Omega_{A_R}{}^B \wedge\Psi_{A_1|\dots|A_{R-1}|B}\ .
\end{align}

\subsection{Basis for structure-group tensors} 

Under the maximal bosonic subalgebra $\mathfrak{g}_C\oplus \mathfrak{g}_R$, we decompose
\begin{align}
\boldsymbol{\tau}^A=(\boldsymbol{\tau}^{\underline\alpha},\boldsymbol{\tau}^I)\ ,\qquad \zeta^{AB}= \left[\begin{array}{cc} \underline{C}^{\underline{\alpha\beta}}&0\\0& -i\eta^{IJ}\end{array}\right]\ ,
\end{align}
where $\underline\alpha,\underline\beta=1,\dots,8$, $I,J=1,\dots,8$,
and 
\begin{align}
L^{AB}= \left[\begin{array}{cc} M^{\underline{\alpha\beta}}& iQ^{\underline{\alpha} J}\\ -iQ^{\underline{\beta} I}& -i N^{IJ}\end{array}\right]\ ,\qquad \Omega_{AB}=\left[\begin{array}{cc} \Omega_{\underline{\alpha\beta}}&i E_{\underline{\alpha}J}\\[2pt]-iE_{\underline{\beta}I}& i\eta_{IJ}\end{array}\right]\ .
\end{align}
It follows that
\begin{align}
M^{\underline{\alpha\beta}} \boldsymbol{\tau}^{\underline\gamma}= 2i \underline{C}^{\underline{\beta}[\underline{\gamma}}\boldsymbol{\tau}^{\underline{\alpha}]}\ ,\qquad N^{IJ}\boldsymbol{\tau}^K= 2i\eta^{(J|K}\boldsymbol{\tau}^{|I)}\ .
\end{align}
Under $H$, the further decompositions
\begin{align} 
\boldsymbol{\tau}^{\underline\alpha}=\left(\left(\boldsymbol{\tau}^{(-)}_{-1/2}\right)^\alpha,\left(\boldsymbol{\tau}^{(+)}_{+1/2}\right)_\alpha\right)\ ,\qquad \underline{C}^{\underline{\alpha\beta}}=\left[\begin{array}{cc} 0& \delta^\alpha_\beta\\ 
\delta_\alpha^\beta& 0\end{array}\right]\ ,
\end{align}
and
\begin{align}
M^{\underline{\alpha\beta}}&=\left[\begin{array}{cc} K^{\alpha\beta}& M^\alpha{}_\beta-\frac12 \delta_\beta^\alpha D\\[2pt] M_\alpha{}^\beta+\frac12 \delta_\alpha^\beta D& T_{\alpha\beta}\end{array}\right]\ ,\qquad M^\alpha{}_\beta=-M_\beta{}^\alpha\ ,\\
Q^{\underline\alpha I}&=(S^{\alpha I},Q_{\alpha}{}^I)\ ,
\end{align}
where $\alpha,\beta=1,\dots,4$, yields
\begin{align}
M^\alpha{}_\beta \boldsymbol{\tau}^\gamma&=i\left(\delta_\beta^\gamma \boldsymbol{\tau}^\alpha-\frac14 \delta^\alpha_\beta \boldsymbol{\tau}^\gamma\right)\ ,\qquad D\boldsymbol{\tau}^\alpha=-\frac{i}{2} \boldsymbol{\tau}^\alpha\ ,\\
M_\alpha{}^\beta \boldsymbol{\tau}_\gamma&=i\left(\delta^\beta_\gamma \boldsymbol{\tau}_\alpha-\frac14 \delta_\alpha^\beta \boldsymbol{\tau}_\gamma\right)\ ,\qquad \, \, D\boldsymbol{\tau}_\alpha=\frac{i}{2} \boldsymbol{\tau}_\alpha\ .
\end{align}
Correspondingly, we decompose 
\begin{align}
\Omega_{\underline{\alpha\beta}}&=\left[\begin{array}{cc} \widetilde{E}_{\alpha\beta}& \Omega_\alpha{}^\beta+\frac12\delta_\alpha^\beta \sigma\\[2pt]\Omega^\alpha{}_\beta-\frac12\delta_\beta^\alpha \sigma& E^{\alpha\beta}\end{array}\right]\ ,  \qquad \Omega^\alpha{}_\beta=- \Omega_\beta{}^\alpha\ ,\\
E_{\underline{\alpha} I}&=(\widetilde{F}_{\alpha I},{F}^{\alpha}{}_I)\ ,
\end{align}
i.e,
\begin{align}
\Omega_{A}{}^{B}&=\left[\begin{array}{cc|c} \Omega_\alpha{}^\beta +\frac12\delta_\alpha^\beta \sigma&  \widetilde{E}_{\alpha\beta} & \widetilde{F}_{\alpha}{}^{J}\\[2pt]
E^{\alpha\beta}&\Omega^\alpha{}_\beta-\frac12\delta_\beta^\alpha\sigma&{F}^{\alpha J}\\\hline\\[-12pt]
-i{F}^{\beta}{}_{I}&-i\widetilde{F}_{\beta I}& \Omega_{I}{}^{J}\end{array}\right]\ ,\\
\Omega_{AB}&=\left[\begin{array}{cc|c} \widetilde{E}_{\alpha\beta} & \Omega_\alpha{}^\beta +\frac12\delta_\alpha^\beta \sigma& i \widetilde{F}_{\alpha J}\\[2pt]
\Omega^\alpha{}_\beta-\frac12\delta_\beta^\alpha\sigma&E^{\alpha\beta}&i{F}^\alpha{}_J\\\hline\\[-12pt]
-i\widetilde{F}_{\beta I}&-i{F}^\beta{}_I&i \Omega_{IJ}\end{array}\right]\ .
\end{align}
In this basis, the superconformal curvature tensor 
\begin{align}
R_{AB}=\left[\begin{array}{cc|c} \widetilde{R}_{\alpha\beta} & R_\alpha{}^\beta +\frac12\delta_\alpha^\beta R& i \widetilde{R}_{\alpha J}\\[2pt]
R^\alpha{}_\beta-\frac12\delta_\beta^\alpha R & R^{\alpha\beta}&i{R}^\alpha{}_J\\\hline\\[-12pt]
-i\widetilde{R}_{\beta I}&-i{R}^\beta{}_I&i R_{IJ}\end{array}\right]\ ,\qquad R^\alpha{}_\beta=-R_\beta{}^\alpha\ ,
\end{align}
where 
\begin{align}
R_\alpha{}^\beta&={\rm Rie}_\alpha{}^\beta+\widetilde{E}_{\alpha\gamma}\wedge E^{\gamma\beta}-i\widetilde{F}_{\alpha}{}^I\wedge F^\beta{}_I\ ,\\
R^\alpha{}_\beta&={\rm Rie}^\alpha{}_\beta+{E}^{\alpha\gamma}\wedge \widetilde{E}_{\gamma\beta}-i{F}^{\alpha I}\wedge \widetilde{F}_{\beta I}\ ,\\ 
R_{IJ}&=F_{IJ}-i\left(\widetilde{F}_{\alpha I}\wedge F^\alpha{}_J+F^\alpha{}_I\wedge \widetilde{F}_{\alpha J}\right)\ ,\\
R&=R^{(\sigma)}+\frac12\widetilde{E}_{\alpha\beta}\wedge E^{\beta\alpha}-\frac{i}2 \widetilde{F}_\alpha{}^I\wedge F^\alpha{}_I\ ,\\
\widetilde{R}_{\alpha\beta}&=\nabla \widetilde{E}_{\alpha\beta}-i \widetilde{F}_\alpha{}^I \wedge \widetilde{F}_{\beta I}\ ,\\
{R}^{\alpha\beta}&=\nabla E^{\alpha\beta}-iF^{\alpha I}\wedge {F}^{\beta}{}_{I}\ ,\\
\widetilde{R}_{\alpha I}&=\nabla \widetilde{F}_{\alpha I}+\widetilde{E}_{\alpha\beta}\wedge {F}^\beta{}_I\ ,\\
R^\alpha{}_I&=\nabla F^\alpha{}_I+E^{\alpha\beta}\wedge \widetilde{F}_{\beta I}\ ,
\end{align}
are built from $H$-covariant frames, curvatures
\begin{align}
{\rm Rie}_{\alpha}{}^\beta&=-{\rm Rie}^\beta{}_\alpha=d\Omega_\alpha{}^\beta+\Omega_\alpha{}^\gamma\wedge \Omega_\gamma{}^\beta\ ,\qquad 
R^{(\sigma)}=d\sigma\ ,\\
F_{IJ}&=d\Omega_{IJ}+\Omega_{I}{}^K\wedge \Omega_{KJ}\ , 
\end{align}
and covariant derivatives
\begin{align}
\nabla \widetilde{E}_{\alpha\beta}&=d\widetilde{E}_{\alpha\beta}+2\Omega_{[\alpha|}{}^\gamma\wedge \widetilde{E}_{\gamma|\beta]}+\Delta_{\mathcal{M}}(\widetilde{E}_{\alpha\beta})\sigma \widetilde{E}_{\alpha\beta}\ ,\\ 
\nabla {E}^{\alpha\beta}&=d{E}^{\alpha\beta}+2\Omega^{[\alpha|}{}_\gamma\wedge {E}^{\gamma|\beta]}+\Delta_{\mathcal{M}}({E}^{\alpha\beta})\sigma {E}^{\alpha\beta}\ ,\\
\nabla \widetilde{F}_{\alpha I}&=d\widetilde{F}_{\alpha I}+\Omega_\alpha{}^\beta \wedge \widetilde{F}_{\beta I}+A_I{}^J\wedge \widetilde{F}_{\alpha J}+\Delta_{\mathcal{M}}(\widetilde{F}_{\alpha I}) \sigma\wedge \widetilde{F}_{\alpha I}\ ,\\
\nabla F^\alpha{}_I&=dF^\alpha{}_I+\Omega^\alpha{}_\beta \wedge F^\beta{}_I+A_I{}^J\wedge F^\alpha{}_J+\Delta_{\mathcal{M}}({F}^{\alpha}{}_{ I})\sigma\wedge F^\alpha{}_I\ .
\end{align}
Likewise, expanding 
\begin{align}
\boldsymbol{\Omega}_{\mathfrak{h}}&= i\left( M_\alpha{}^\beta \Omega_\beta{}^\alpha+ D \sigma +\frac{1}2 N^{IJ}\Omega_{IJ}\right)\ ,
\\
\boldsymbol{\psi}_{(1)}&=\boldsymbol{\tau}^\alpha \left(\Psi^{(+)}_{+1/2}\right)_\alpha+\boldsymbol{\tau}_\alpha\left(\Psi^{(-)}_{-1/2}\right)^\alpha+ \boldsymbol{\tau}^I\Psi_I\ ,
\end{align}
where thus
\begin{align}
\Delta_{\mathcal{M}}(\Psi_\alpha)=\frac12\ ,\qquad \Delta_{\mathcal{M}}(\Psi^\alpha)=-\frac12\ ,\qquad \Delta_{\mathcal{M}}(\Psi_I)=0\ ,\label{3.100}
\end{align}
the ${H}$-covariant derivative
\begin{align}\label{3.109}
\boldsymbol{\nabla} \boldsymbol{\psi}_{(1)}:= d\boldsymbol{\psi}_{(1)}+\boldsymbol{\Omega}_{\mathfrak{h}}\wedge \boldsymbol{\psi}_{(1)}\ ,
\end{align}
decomposes under $H$ into
\begin{align}
\boldsymbol{\nabla} \boldsymbol{\psi}_{(1)}=\boldsymbol{\tau}^\alpha \nabla\Psi_\alpha+\boldsymbol{\tau}_\alpha\nabla\Psi^\alpha+ \boldsymbol{\tau}^I\nabla\Psi_I\ ,
\end{align}
where the $H$-covariant derivatives
\begin{align}
\nabla \Psi_\alpha&= d\Psi_\alpha+\Omega_{\alpha}{}^\beta\wedge \Psi_\beta+\Delta_{\mathcal{M}}(\Psi_\alpha)\sigma \wedge \Psi_\alpha\ ,\\
\nabla \Psi^\alpha&= d\Psi^\alpha+\Omega^\alpha{}_{\beta}\wedge \Psi^\beta+\Delta_{\mathcal{M}}(\Psi^\alpha)\sigma \wedge \Psi^\alpha\ ,\\
\nabla \Psi_I&= d\Psi_I+\Omega_{I}{}^J\wedge \Psi_J\ ,
\end{align}
obeying the Bianchi identities 
\begin{align}
\nabla^2 \Psi_\alpha&\equiv {\rm Rie}_\alpha{}^\beta \wedge \Psi_\beta+\Delta_{\mathcal{M}}(\Psi_\alpha)R^{(\sigma)} \wedge \Psi_\alpha\ , \\
\nabla^2 \Psi^\alpha&\equiv {\rm Rie}^\alpha{}_\beta \wedge \Psi^\beta+\Delta_{\mathcal{M}}(\Psi^\alpha)R^{(\sigma)} \wedge \Psi^\alpha\ ,\\
\nabla^2 \Psi_I&\equiv F_I{}^J\wedge \Psi_J\ .
\end{align}
The $\mathfrak{so}(2,6)$-valued connection
\begin{align}
\boldsymbol{\Omega}_{\mathfrak{so}(2,6)}&:=\frac{i}2 M^{\underline{\alpha\beta}}\Omega_{\underline{\beta\alpha}}=i\left(M_{\alpha}{}^\beta \Omega_\beta{}^\alpha+\sigma D+\frac12 T_{\alpha\beta} E^{\beta\alpha}+\frac12K^{\alpha\beta}\widetilde{E}_{\beta\alpha}\right)\\
&\equiv i\left(\frac12 M^{ab}\Omega_{ba}+\sigma D+T^a E_a+ K^a \widetilde{E}_a\right)\ ,
\end{align}
where thus
\begin{align}
\Omega^{ab}&=-\frac12 (\sigma^{ab})_\alpha{}^\beta \Omega_\beta{}^\alpha\ ,\qquad E^a=\frac14(\sigma^a)_{\alpha\beta}E^{\alpha\beta}\ ,\qquad \widetilde{E}^a=-\frac14 (\tilde\sigma^a)^{\alpha\beta}\widetilde{E}_{\alpha\beta}\ ,\\
\Omega_\alpha{}^\beta&=\frac14 (\sigma_{ab})_\alpha{}^\beta\Omega^{ab}\ ,\qquad E^{\alpha\beta}=-(\tilde\sigma_a)^{\alpha\beta} E^a\ ,\qquad \widetilde{E}_{\alpha\beta}=(\sigma_a)_{\alpha\beta} \widetilde{E}^a\ ,
\end{align}
The $\mathfrak{so}(2,6)$-valued curvature
\begin{align}
\hspace{-1cm}\boldsymbol{R}_{\mathfrak{so}(2,6)}&:= d\boldsymbol{\Omega}_{\mathfrak{so}(2,6)}+\boldsymbol{\Omega}_{\mathfrak{so}(2,6)}\wedge \boldsymbol{\Omega}_{\mathfrak{so}(2,6)}\cr
&=i\left(M_\alpha{}^\beta \left({\rm Rie}_{\beta}{}^\alpha+\widetilde{E}_{\beta\gamma}\wedge E^{\gamma\alpha}\right)+D \left(d\sigma +\frac12\widetilde{E}_{\alpha\beta}\wedge E^{\beta\alpha}\right)\right.\cr
&\left.+\frac12 T_{\alpha\beta} \nabla E^{\beta\alpha}+\frac12K^{\alpha\beta}\nabla \widetilde{E}_{\beta\alpha}\right)\cr
&\equiv i\left(\frac12 M^{ab}\left({\rm Rie}_{ba}-4E_b\wedge \widetilde{E}_a\right) +D \left(d\sigma-2\widetilde{E}^a\wedge {E}_a\right)+T^a \nabla E_a+ K^a \nabla \widetilde{E}_a\right)\ ,
\end{align}
where
\begin{align}
\nabla E^a&=dE^a+\Omega^{ab}\wedge E_b-\sigma\wedge E^a\ ,\qquad \nabla \widetilde{E}^a=d \widetilde{E}+\Omega^{ab}\wedge  \widetilde{E}_b+\sigma\wedge  \widetilde{E}^a\ ,\\
{\rm Rie}^{ab}&=d\Omega^{ab}+\Omega^{ac}\wedge \Omega_c{}^b\ .
\end{align}

\subsection{Weyl zero-form in components}
\label{Sec:3.10}

Introducing chiral polarization tensors $(\Pi^{(\varepsilon)})^{AB}$, $\varepsilon=\pm$, viz., 
\begin{align}
\hspace{-.5cm}(\Pi^{(+)})^{AB} \boldsymbol{\tau}_B&=(\boldsymbol{\tau}^\alpha,0;\boldsymbol{\tau}^I)\ ,\qquad (\Pi^{(-)})_{A}{}^{B} \boldsymbol{\tau}_B=(\boldsymbol{\tau}_\alpha,0;\boldsymbol{\tau}_I)\ ,\\ (\Pi^{(\varepsilon)})_{AB}&=(\Pi^{(-\varepsilon)})_{BA}\ ,
\end{align}
the chiral and anti-chiral conformal modules $\mathsf{S}^{(\varepsilon)}$ are spanned by polarized basis elements 
\begin{align}
\boldsymbol{\tau}^{(\varepsilon)}_{A(k+4),B(k)}=(\Pi^{(\varepsilon)})_{(A_1|}{}^C \boldsymbol{\tau}^{(\varepsilon)}_{C|A(k+3)),B(k)}\ ,\qquad k=0,1,\dots\ ,
\end{align}
obeying Young projections and trace constraints, viz.,
\begin{align}
\boldsymbol{\tau}^{(\varepsilon)}_{(A(k+4),A_{k+5})B(k-1)}=0\ ,\qquad \zeta^{AB}\boldsymbol{\tau}^{(\varepsilon)}_{ABC(k+2),D(k)}\approx 0\ .
\end{align}
The representation matrices 
\begin{align}
&\rho_{\mathsf{S}^{(\varepsilon)}}\!\left(L_{AB}\right)\boldsymbol{\tau}^{(\varepsilon)}_{C(k+4),D(k)}=\nonumber\\
&2i(k+4)(\Pi^{(-\varepsilon)})_{[B|(C_1|}\boldsymbol{\tau}^{(\varepsilon)}_{A]|C(k+3)),D(k)}\nonumber\\&+ 2ik(-1)^{B(C_1+\cdots+C_k)} (\Pi^{(-\varepsilon)})_{[B|((D_1|}\boldsymbol{\tau}^{(\varepsilon)}_{C(k+4)),|A]|D(k-1))}\nonumber\\
&+(-1)^{B(C_1+\cdots+C_{k+4})}\boldsymbol{\tau}^{(\varepsilon)}_{AC(k+4),BD(k)}\nonumber\\
&+(-1)^{A(B+C_1)}\lambda_k (\Pi^{(-\varepsilon)})_{[A|\{C_2|} (\Pi^{(-\varepsilon)})_{B]|C_1}\boldsymbol{\tau}^{(\varepsilon)}_{C(k+2),D(k)\}}\ ,
\end{align}
where the first two terms represent $\mathfrak{h}$, and the last two terms represent $\mathfrak{g}/\mathfrak{h}$ with $\lambda_k$ determined by the closure relations.
Correspondingly, the chiral and anti-chiral Weyl zero-forms
\begin{align}
\boldsymbol{c}^{(\varepsilon)}=\sum_{k=0}^\infty (\boldsymbol{\tau}_{[1]})^{(\varepsilon)A(k+4),B(k)}C^{(\varepsilon)}_{B(k),A(k+4)}\in \Omega_{[0]}(\boldsymbol{M}')\otimes \mathsf{S}^{(\varepsilon)}_{[1]}\ ,
\end{align}
in terms of chiral and anti-chiral components with conformal weights
\begin{align}
\Delta_{\mathcal{M}}\left(C^{(\varepsilon)}_{B(k),A(k+4)}\right)&=\varepsilon \left(\sum_{\xi=1}^{k+4}\Delta_{\mathcal{M}}(\Psi_{A_\xi})+\sum_{\xi=1}^{k}\Delta_{\mathcal{M}}(\Psi_{B_\xi})\right)\ ,
\end{align}
using \eqref{3.100}.
We refer to the components arising in ${\rm ker}\,\rho_{\boldsymbol{(2)}}(T_{\alpha\beta})$ and ${\rm ker}\,\rho_{\boldsymbol{(2)}}(K^{\alpha\beta})$ as primaries and anti-primaries, respectively, which we denote by
\begin{align}
\hspace{-.7cm}\mbox{Primaries:}\quad\!\!& (R^{(+)}_{+4})_{\alpha(4)}\ ,\quad\!\! (\Psi^{(+)}_{+7/2})_{\alpha(3) I}\ ,\quad\!\! (H^{(+)}_{+3})_{\alpha(2) I[2]}\ ,\quad\!\! (\Theta^{(+)}_{+5/2})_{\alpha I[3]}\ ,\quad\!\! (\Phi^{(+)}_{+2})_{I[4]}\ ,\\[3pt]
\hspace{-.7cm}\mbox{Anti-primaries:}\quad\!\!& (\widetilde{R}^{(-)}_{-4})^{\alpha(4)}\ ,\quad\!\! (\widetilde{\Psi}^{(-)}_{-7/2})^{\alpha(3) I}\ ,\quad\!\! (\widetilde{H}^{(-)}_{-3})^{\alpha(2) I[2]}\ ,\quad\!\! (\widetilde{\Theta}^{(-)}_{-5/2})^{\alpha I[3]}\ ,\quad\!\! (\widetilde{\Phi}^{(-)}_{-2})^{I[4]}\ ,
\end{align}
whose super- and subscripts indicate second-quantised chiralities and conformal weights, respectively; for $J\geqslant 1$ conversion to Lorentz tensors and tensor-spinors yields
\begin{align}
\mbox{Primaries with $J\geqslant 1$:}\quad& (R^{(-)}_{+4})_{abc,def}\ ,\quad(\Psi^{(-)}_{+7/2})_{abc,\alpha I}\ ,\quad(H^{(-)}_{+3})_{abc I[2]}\ ,\\
\mbox{Anti-primaries with $J\geqslant 1$ :}\quad &  (\widetilde{R}^{(+)}_{+4})_{abc,def}\ ,\quad(\widetilde{\Psi}^{(+)}_{+7/2})_{abc}{}^{\alpha I}\ ,\quad(\widetilde{H}^{(+)}_{+3})_{abc}{}^{I[2]}\ ,
\end{align}
whose superscripts indicate self-duality $(+)$ and anti-self-duality $(-)$ projections.

\subsection{Two-form potential in components}

\begin{figure}
    \centering
\begin{tikzpicture}[->,>=stealth',node distance=3.2cm,
  thick, main node/.style={circle,draw}]
\node[main node] (1) {${\boldsymbol{\tau}}^{\alpha\beta}$};
\node[main node] (2) [right of=1] {${\boldsymbol{\tau}}^{\alpha I}$};
\node[main node] (3) at (6.4,0.55) {${\boldsymbol{\tau}}^{({\rm z})}$};
\node[main node] (4) at (9.6,0.0) {${\boldsymbol{\tau}}_\alpha{}^{I}$};
\node[main node] (5) [right of=4] {${\boldsymbol{\tau}}_{\alpha\beta}$};
\node[main node] (6) at (6.4,3.2) {${\boldsymbol{\tau}}^\alpha{}_\beta$};  
\node[main node] (8) at (6.4,-0.55) {${\boldsymbol{\tau}}^{({\rm h})}$};
\node[main node] (7) at (6.5,-3.2) {${\boldsymbol{\tau}}^{IJ}$};

\draw [->] (1) to [out=5,in=175]  node[midway, above] {$Q$} (2);
\draw [->,red] (2) to [out=185,in=-5] node[midway, below] {$S$} (1) ;
\draw [->] (2) to [out=10,in=190] node[midway, above] {$Q$} (3);
\draw [->,red] (4) to [out=170,in=-10] node[midway, above] {$S$} (3);
\draw [->] (4) to [out=5,in=175] node[midway, above] {$Q$} (5);
\draw [->,red] (5) to [out=185,in=-5] node[midway, below] {$S$} (4);
\draw [->] (2) to [out=60,in=-155] node[midway, above] {$Q$} (6);
\draw [->] (2) to [out=-60,in=150] node[midway, below] {$Q$} (7);
\draw [->] (6) to [out=-30,in=120] node[midway, above] {$Q$} (4);
\draw [->] (7) to [out=30,in=-120] node[midway, below] {$Q$} (4);
\draw [->,red] (6) to [out=-135,in=45] node[midway, below] {$S$} (2);
\draw [->,red] (7) to [out=135,in=-45] node[midway, above] {$S$} (2);
\draw [->,red] (4) to [out=-135,in=45] node[midway, above] {$S$} (7);
\draw [->,red] (4) to [out=135,in=-45] node[midway, below] {$S$} (6);
\draw [->,blue] (2) to [out=35,in=145] node[midway, above] {$T$} (4);
\draw [->,green] (4) to [out=215,in=-35] node[midway, below] {$K$} (2);
\draw [->,blue] (1) to [out=60,in=-180] node[midway, above] {$T$} (6);
\draw [->,blue] (6) to [out=0,in=120]  node[midway, above] {$T$} (5);
\draw [->,green] (6) to [out=-165,in=35] node[midway, above] {$K$} (1) ;
\draw [->,green] (5) to [out=145,in=-15] node[midway, above] {$K$} (6);
\draw [->,red] (8) to [out=170,in=-10] node[midway,below] {$S$} (2);
\draw [->] (8) to [out=10,in=190] node[midway,below] {$Q$} (4);

\end{tikzpicture}
\caption{The conformal supermultiplet $\widehat{\mathsf{T}}$ consists of a hypercharge, a central element, and the supermultiplet $\boldsymbol{(2)}'$ containing the minimal set of two-forms.}
    \label{fig:1}
\end{figure}
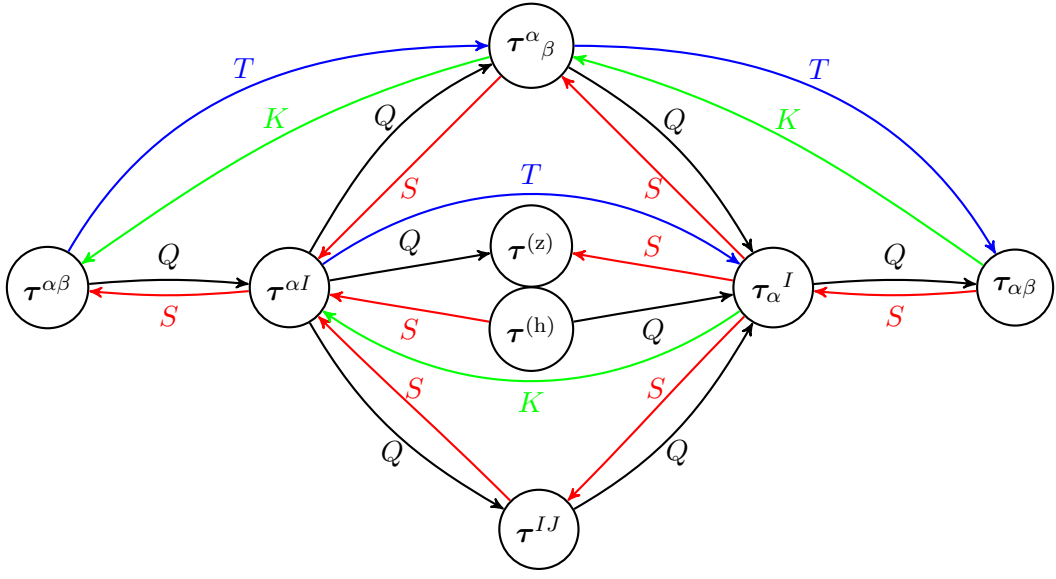

The $\pi(G)$-tensor $\boldsymbol{(2)}$ decomposes under the maximal bosonic subgroup into 
\begin{align}
\left.\boldsymbol{(2)}\right\downarrow_{G_C\times G_R}=\boldsymbol{(2)}|_{G_C}\otimes \boldsymbol{[0]}|_{G_R}\,\oplus\, \boldsymbol{(1)}|_{G_C}\otimes \boldsymbol{[1]}|_{G_R}\,\oplus\, \boldsymbol{(0)}|_{G_C}\otimes \boldsymbol{[2]}|_{G_R}\ ,
\end{align} 
with further $H$-decomposition 
\begin{align}
\hspace{-1cm}\boldsymbol{(2)}\!\!\downarrow_{H}=\left[\begin{array}{ccc}(\boldsymbol{\tau}_{-1})^{\alpha\beta}&\!\!\boldsymbol{\tau}^{\alpha}{}_{\beta}+\tfrac1{16}\delta^\alpha_\beta \left(\btau^{({\rm h})}+\btau^{({\rm z})}\right)& (\boldsymbol{\tau}_{-1/2})^{\alpha J}\\ \!\!\boldsymbol{\tau}_{\alpha}{}^\beta+\tfrac1{16}\delta_\alpha^\beta \left(\btau^{({\rm h})}+\btau^{({\rm z})}\right)& (\boldsymbol{\tau}_{+1})_{\alpha\beta}& (\boldsymbol{\tau}_{+1/2})_{\alpha}{}^J\\ (\boldsymbol{\tau}_{-1/2})^{\beta I}&(\boldsymbol{\tau}_{+1/2})_\beta{}^I& \!\!\!\!\boldsymbol{\tau}^{IJ}+\tfrac{i}{16}\eta^{IJ}\left(\btau^{({\rm z})}-\btau^{({\rm h})}\right)\end{array}\right]\ ,
\end{align}
where subscripts indicate non-vanishing first-quantised conformal weights, 
\begin{align}
\boldsymbol{\tau}_{\alpha}{}^\beta=\boldsymbol{\tau}^{\beta}{}_{\alpha}\ ,\qquad \btau_\alpha{}^\alpha=0=\tau^{I}{}_I \ ,
\end{align}
and $\btau^{({\rm h})}$ is the hypercharge;
in terms of Young diagrams,
\begin{align}
\left.\boldsymbol{(2)}\right\downarrow_{H}\approx \yngBlue{;;}_{-1}\oplus \left(\yngBlue{;}_{-1/2}\otimes \yngRed{;}\right)\oplus\yngBlue{;;,;,;}_0\oplus \yngRed{;,;}_0\oplus \bullet^{({\rm h})}\oplus\left(\yngBlue{;,;,;}_{+1/2}\otimes \yngRed{;}\right)\oplus \yngBlue{;;,;;,;;}_{+1} \ .
\end{align}
Correspondingly, we expand the conformally dual pair of two-forms over $\boldsymbol{(2)}|_G$ as 
\begin{align}\label{expandb}
\boldsymbol{b}^{(\varepsilon)}\!\!\downarrow_{H}&=\frac12\left(\btau^{\alpha\beta} \left((\omega^{(\varepsilon)})^{(+)}_{+1}\right)_{\alpha\beta} +\btau^{\alpha I}\left((\Psi^{(\varepsilon)})^{(+)}_{+1/2}\right)_{\alpha I} \right.\cr
&+\btau_\alpha{}^\beta ((G^{(\varepsilon)})_{0})_\beta{}^\alpha +\btau^{IJ}((B^{(\varepsilon)})_0)_{IJ}+  \btau^{({\rm h})} (b_0)^{({\rm h})}\cr
&\left.+\btau_\alpha{}^I\left((\widetilde{\Psi}^{(\varepsilon)})^{(-)}_{-1/2}\right)^\alpha{}_I+\btau_{\alpha\beta}\left((\widetilde{\omega}^{(\varepsilon)})^{(-)}_{-1}\right)^{\alpha\beta}\right) \ ,
\end{align}
with super and subscripts, respectively, indicating second-quantised spinorial chiralities and conformal weights.
In $\boldsymbol{(2)}$, the non-trivial $\mathfrak{g}$-transformations of the hypercharge are
\begin{align}
Q_{\alpha I}\btau^{({\rm h})}=-4\btau_{\alpha I}\ ,\qquad  S^{\alpha I}\btau^{({\rm h})}=-4\btau^{\alpha I}\ ;
\end{align} 
the remaining representation matrix for $\mathfrak{g}/\mathfrak{h}$ has non-vanishing entries 
\begin{align}\label{Tcommutators}
T_{\alpha\beta} \btau^{\gamma I}&=2i\delta^\gamma_{[\beta}\btau^{\phantom{[}}_{\alpha]}{}^I
\ ,\qquad  
T_{\alpha\beta}\btau^{\gamma\delta}= 4i\delta_{[\beta}^{(\gamma}\btau^{\delta)}{}_{\alpha]}^{\phantom{[}}
\ ,\qquad  
T_{\alpha\beta}\btau^\gamma{}_\delta= 2i\delta^\gamma_{[\beta}\btau^{\phantom{[}}_{\alpha]\delta}\ ,\\ 
\label{Kcommutators}
K^{\alpha\beta}\btau_{\gamma}{}^I&= 2i\delta_\gamma^{[\beta}\btau^{\alpha]I}\ ,\qquad K^{\alpha\beta}\btau_{\gamma\delta}= 4i\delta_{(\gamma}^{[\beta}\btau^{\alpha]}{}_{\delta)}^{\phantom{[}}\ ,\qquad 
K^{\alpha\beta}\btau^\gamma{}_\delta= 2i\delta_\delta^{[\beta}\btau^{\alpha]\gamma}\ , 
\\
\label{Qcommutators1}
Q_{\alpha I}\btau^{\beta J}&=-\delta_\alpha^\beta\btau_I{}^J-i\delta_I^J\btau^\beta{}_\alpha-\frac{i}8\delta^J_I\delta_\alpha^\beta\btau^{({\rm z})}\ , \\
Q_{\alpha I}\btau^{\beta\gamma}&=-2\delta_\alpha^{(\beta}\btau^{\gamma)}{}_I
\ ,\qquad 
Q_{\alpha I}\btau_{\beta}{}^{J}=-i\delta_I^J\btau_{\alpha\beta}\ ,
    \\ \label{Qcommutators2}
Q_{\alpha I}\btau^{JK}&=-2i\delta_I^{[J}\btau_\alpha{}^{K]}-\frac{i}{4}\eta^{JK}\btau_{\alpha I}
\ ,\qquad 
Q_{\alpha I}\btau^{\beta}{}_\gamma= -\delta_\alpha^\beta \btau_{\gamma I}+\frac14\delta_\gamma^\beta\btau_{\alpha I}\ , 
\\
\label{Scommutators1}
S^{\alpha I}\btau_{\beta }{}^J&=-\delta_\beta^\alpha\btau^{IJ}-i\eta^{IJ}\btau^\alpha{}_\beta-\frac{i}8 \eta^{IJ}\delta^\alpha_\beta\btau^{({\rm z})}\ ,\\
S^{\alpha I}\btau^{\beta J}&=-i\eta^{IJ}\btau^{\alpha\beta}\ ,\qquad 
 S^{\alpha I}\btau_{\beta\gamma}=-2\delta^\alpha_{(\beta}\btau_{\gamma)}{}^I, \\ 
S^{\alpha I}\btau^{JK}&=-2i\eta^{I[J}\btau^{\alpha K]}-\frac{i}4\eta^{JK}\btau^{\alpha I}\ ,\qquad  
 S^{\alpha I}\btau^{\beta}{}_\gamma=-\delta_\gamma^\alpha\btau^{\beta I} +\frac14 \delta^\beta_\gamma\btau^{\alpha I}\ ;
\end{align}
among the vanishing entries, one has
\begin{align}
L^{AB}\btau^{({\rm z})}&=0\ .
\end{align}
Thus, $\boldsymbol{(2)}$ contains an ideal $\boldsymbol{(2)}'$ spanned by all generators except $\btau^{({\rm h})}$, 
implying the Casimir identities 
\begin{align}
C_{2r}\left(\mathfrak{g}|\boldsymbol{(2)}\right)=C_{2r}\left(\mathfrak{g}|\boldsymbol{(2)}'\right)=C_{2r}\left(\mathfrak{g}|\btau^{({\rm h})}\right)=0\ ,\qquad r=1,\dots,8\ .
\end{align}
We view 
\begin{align}
\widehat{\mathsf{T}}:=\boldsymbol{(2)}\oplus (\mathbb{C}\otimes \btau^{({\rm z})})
\end{align}
as a six-dimensional analogue of  $\mathfrak{u}(2,2|4)$ decomposed as a $\mathfrak{psu}(2,2|4)$-module with $\btau^{({\rm z})}$ and  $\btau^{({\rm h})}$ corresponding to the central element and hypercharge of $\mathfrak{u}(2,2|4)$, respectively, and $\boldsymbol{(2)}'$ to $\mathfrak{psu}(2,2|4)$.
The kernels and images in $\widehat{\mathsf{T}}$ of the coset elements are given by 
\begin{align}
{\rm ker}\, \rho_{\widehat{\mathsf{T}}}(T)&={\rm span}\,\{\btau_{\alpha\beta},\btau_\alpha{}^I,\btau^{IJ},\btau^{({\rm z})},\btau^{({\rm h})}\}\ ,\qquad  
{\rm im}\, \rho_{\widehat{\mathsf{T}}}(T)={\rm span}\,\{\btau_{\alpha\beta},\btau_\alpha{}^I,\btau^\alpha{}_\beta\}\ ,\\
{\rm ker}\, \rho_{\widehat{\mathsf{T}}}(K)&={\rm span}\,\{\btau^{\alpha\beta},\btau^{\alpha I},\btau^{IJ},\btau^{({\rm z})},\btau^{({\rm h})}\}\ ,\qquad  
{\rm im}\, \rho_{\widehat{\mathsf{T}}}(K)={\rm span}\,\{\btau^{\alpha\beta},\btau^{\alpha I},\btau^\alpha{}_\beta\}\ ,\\
{\rm ker}\, \rho_{\widehat{\mathsf{T}}}(Q)&={\rm span}\,\{\btau_{\alpha\beta},\btau^{({\rm z})}\}\ ,\qquad 
{\rm im}\, \rho_{\widehat{\mathsf{T}}}(Q)={\rm span}\,\{\btau_{\alpha\beta},\btau_\alpha{}^I,\btau^\alpha{}_\beta, \btau^{IJ},\btau^{\alpha I},\btau^{({\rm z})}\}\ ,\\
{\rm ker}\, \rho_{\widehat{\mathsf{T}}}(S)&={\rm span}\,\{\btau^{\alpha\beta},\btau^{({\rm z})}\}\ ,\qquad 
{\rm im}\, \rho_{\widehat{\mathsf{T}}}(S)={\rm span}\,\{\btau_\alpha{}^I,\btau^\alpha{}_\beta, \btau^{IJ},\btau^{\alpha I},\btau^{\alpha\beta},\btau^{({\rm z})}\}\ ;
\end{align}
likewise, upon factoring out $\btau^{(\rm z)}$ and passing to the ideal $\boldsymbol{(2)}'$, the kernels and images of the generators in the coset $\mathfrak{g}/\mathfrak{h}$ are given by 
\begin{align}
{\rm ker}\, \rho_{\boldsymbol{(2)}'}(T)&\approx {\rm span}\,\{\btau_{\alpha\beta},\btau_\alpha{}^I,\btau^{IJ}\}\ ,\qquad  
{\rm im}\, \rho_{\boldsymbol{(2)}'}(T)\approx {\rm span}\,\{\btau_{\alpha\beta},\btau_\alpha{}^I,\btau^\alpha{}_\beta\}\ ,\\
{\rm ker}\, \rho_{\boldsymbol{(2)}'}(K)&\approx {\rm span}\,\{\btau^{\alpha\beta},\btau^{\alpha I},\btau^{IJ}\}\ ,\qquad  
{\rm im}\, \rho_{\boldsymbol{(2)}'}(K)\approx {\rm span}\,\{\btau^{\alpha\beta},\btau^{\alpha I},\btau^\alpha{}_\beta\}\ ,\\
{\rm ker}\, \rho_{\boldsymbol{(2)}'}(Q)&\approx {\rm span}\,\{\btau_{\alpha\beta},\btau^{({\rm z})}\}\ ,\qquad 
{\rm im}\, \rho_{\boldsymbol{(2)}'}(Q)\approx {\rm span}\,\{\btau_{\alpha\beta},\btau_\alpha{}^I,\btau^\alpha{}_\beta, \btau^{IJ}\}\ ,\\
{\rm ker}\, \rho_{\boldsymbol{(2)}'}(S)&\approx {\rm span}\,\{\btau^{\alpha\beta},\btau^{({\rm z})}\}\ ,\qquad 
{\rm im}\, \rho_{\boldsymbol{(2)}'}(S)\approx {\rm span}\,\{\btau_\alpha{}^I,\btau^\alpha{}_\beta, \btau^{IJ},\btau^{\alpha I}\}\ ;
\end{align}

\subsection{Superoscillator realisation and Howe-dual gauging}

Introducing superoscillators $(q^\alpha, p_\alpha, \bar q^\alpha, \bar p_\alpha| \xi^I, \bar\xi^I)$ treated as canonical coordinates of the noncommutative supersymplectic $\mathbb{R}^{16|16}$ with quantum numbers 
\begin{align}
{\rm fer}_{\boldsymbol{\mathcal{A}}}(q^\alpha, p_\alpha, \bar q^\alpha, \bar p_\alpha|\xi^I, \bar\xi^I)&=(0,0,0,0|1,1)\ ,\\ 
\Delta_{\boldsymbol{\mathcal{A}}}(q^\alpha, p_\alpha, \bar q^\alpha, \bar p_\alpha| \xi^I, \bar\xi^I)&=\left(-\tfrac12,\tfrac12,-\tfrac12,\tfrac12|0,0\right)\ ,
\end{align}
and graded commutation rules 
\begin{align}
[q^\alpha,\bar p_\beta]_\star=i\delta_\beta^\alpha\ , \qquad
[\bar q^\alpha,p_\beta]_\star=-i\delta_\beta^\alpha\ ,\qquad  [\xi^I,\bar\xi^J]_\star=\eta^{IJ}\ ,
\end{align}
the generators of $\mathfrak{g}$ can be realised as 
\begin{align}
M_\alpha{}^\beta&= \bar{p}_\alpha q^\beta-  p_\alpha \bar{q}^\beta - \tfrac12\delta_\alpha^\beta D\ ,\qquad
D=\tfrac12\left(q^\alpha\bar p_\alpha-\bar q^\alpha p_\alpha \right)\ ,\\
\label{supertranslations}
T_{\alpha\beta}&= 2
\bar{p}_{[\alpha} p_{\beta]}\ ,\qquad 
Q_{\alpha I}=i(\bar{\xi}_I p_\alpha-\xi_I\bar{p}_\alpha)\ ,\\
\label{superconformaltranslations}
K^{\alpha\beta}&=2\bar{q}^{[\alpha}q^{\beta]}\ ,\qquad 
S^{\alpha I}=i(\bar\xi^I q^\alpha-\xi^I\bar q^\alpha)\ ,\\
N_{IJ}&=-2{i} \bar\xi_{(I}\xi_{J)}\ , 
\end{align}
treated as symbols defined in Weyl order. 
The Howe-dual group is generated by 
\begin{align}
\widetilde{K}_{++}&=\bar q^\alpha\bar p_\alpha+\frac{i}2 \bar\xi^I \bar\xi_I\ ,\qquad 
\widetilde{K}_{-\,-}=q^\alpha p_\alpha+\frac{i}2 \xi^I \xi_I, \\
\widetilde{K}_{+\,-}&=-\frac{i}2\left(q^\alpha \bar p_\alpha+\bar q^\alpha p_\alpha\right) +\frac12 \xi^I \bar\xi_I\ ,
\end{align}
viz., 
\begin{align}
[\widetilde{K}_{+\,-},\widetilde{K}_{\pm,\pm}]_\star=\pm \widetilde{K}_{\pm,\pm}\ , \qquad [\widetilde{K}_{++},\widetilde{K}_{-\,-}]_\star=2\widetilde{K}_{+\,-}\ .
\end{align}
commuting to $\mathfrak{g}$ and assigning the canonical coordinates Howe-dual spins 
\begin{align}
{\rm ad}^\star_{\widetilde{K}_{+\,-}}(q^\alpha, p_\alpha, \bar q^\alpha, \bar p_\alpha, \xi^I, \bar\xi^I)=\left(\frac12,-\frac12,-\frac12,\frac12,\frac12,-\frac12\right)\ .
\end{align}
Under the conformal duality transformation,
\begin{align}
{\rm Ad}^\star_{\kappa_\sigma}(q^{\alpha}, p_\alpha, \bar q^{\alpha}, \bar p_\alpha| \xi^I, \bar\xi^I)= (p'_\alpha,q^{\prime\alpha},   \bar p'_\alpha, \bar q^{\prime\alpha}|\xi^I, \bar\xi^I)\ ,
\end{align}
where the primed spinorial oscillators are defined using a rotated Howe-dual spin-frame \cite{wip:2026}.

The space $\mathsf{B}$ of 
bilinears in superoscillators equipped with 
\begin{align}
\rho_{{\mathsf{B}}}(x)={\rm ad}^\star_x \ ,\qquad x\in \mathfrak{g}\oplus \tilde{\mathfrak{k}}\ ,
\end{align}
has the decomposition
\begin{align}
\mathsf{B}\downarrow_{\mathfrak{g}\oplus \tilde{\mathfrak{k}}}=\bullet|_G\otimes (1)_{\widetilde{K}}\oplus \boldsymbol{(2)}|_G\otimes (1)_{\widetilde{K}}\ ,
\end{align}
i.e., the sum of the  oscillator realisation of $\tilde{\mathfrak{k}}$ with basis $\widetilde{K}_{ij}$, $i,j=1,2$, obeying
\begin{align}
\rho_{\mathsf{B}}(\widetilde{K}_{ij})\widetilde{K}_{kl}=4i\epsilon_{(j(k}\widetilde{K}_{l)|i)}\ ,
\end{align}
and the two-form module with basis  
\begin{align}
\btau^{AB}_{ij}= (-1)^{AB}\btau^{BA}_{ij}=\btau^{AB}_{ji}\ ,\qquad \btau^{AB}_{ij}\zeta_{BA}= 0\ ,
\end{align}
obeying 
\begin{align}
\rho_{\mathsf{B}}\left(L^{AB}\right)\btau^{CD}_{ij}=4i\zeta^{[B|(C|}\btau^{A]|D)}_{ij}\ ,\qquad
\rho_{\mathsf{B}}(\widetilde{K}_{ij})\btau^{AB}_{kl}=4i\epsilon_{(j(k}\btau^{AB}_{l)|i)}\ .
\end{align}
From 
$\zeta^{AB}\zeta_{BA}=0$, it follows that traces cannot be subtracted from $G$-tensors; instead, they can be factored out by embedding the linearised module into a larger operator algebra as the cohomology of a first-quantised BRST operator gauging $\widetilde{K}$. Expanding
\begin{align}
\boldsymbol{b}^{(\varepsilon)}=\tfrac12 (\btau_{[-1]})^{AB}_{ij} (B^{(\varepsilon)})^{ij}_{AB}\ ,
\end{align}
the BRST transformations induce shifts
\begin{align}
\delta \boldsymbol{b}^{(\varepsilon)}= \tfrac12 \widetilde{K}_{ij} (\beta^{(\varepsilon)})^{ij}\ ,\qquad (\beta^{(\varepsilon)})^{ij}\in \Omega_{[2]}(\boldsymbol{M}')\ ,
\end{align} 
i.e., 
\begin{align}
\delta (B^{(\varepsilon)})^{ij}_{AB} =\zeta_{AB} (\beta^{(\varepsilon)})^{ij}\ ,
\end{align}
such that the trace-part of $(B^{(\varepsilon)})^{ij}_{AB}$factors out at the level of the cohomology.
The centrally extended tensor $\widehat{\mathsf{T}}$ can thus be realised using a space of bilinears of fixed Howe-dual spin, e.g.,
\begin{align}
{\boldsymbol{\tau}}^{\alpha I}{}&=\bar q^\alpha \bar\xi^I\ ,\qquad
{\boldsymbol{\tau}}^{\alpha\beta}{}=\bar q^\alpha \bar q^\beta\ ,\qquad {\boldsymbol{\tau}}^{IJ}=\bar\xi^I\bar \xi^J+\frac18 \eta^{IJ}\bar\xi^K\bar\xi_K\ ,\\ {\boldsymbol{\tau}}_{\alpha\beta}&=\bar p_\alpha \bar p_\beta\ ,\qquad {\boldsymbol{\tau}}^\alpha{}_\beta {}= \bar q^\alpha \bar p_\beta-\frac14\delta^\alpha_\beta\bar q^\gamma\bar p_\gamma\ ,\qquad {\boldsymbol{\tau}}_{\alpha}{}^I=\bar p_\alpha \bar\xi^I\, \\ 
{\boldsymbol{\tau}}^{({\rm z})}&=2\bar q^\alpha\bar p_\alpha+{i}\bar\xi^I\bar\xi_I\ , \qquad 
\btau^{({\rm h})}=2\bar q^\alpha\bar p_\alpha -{i}\bar\xi^I\bar\xi_I\ ,
\end{align}
with Howe-dual spin $+1$.

Finally, the Howe-dual zero-form is expanded as
\begin{align}
\boldsymbol{k}=\tfrac12 \widetilde{K}_{ij} k^{ij}\ .
\end{align}

\section{Unfolded system}
\label{unfoldedeom}

In this section, we exhibit the linearised unfolded system, its conformal duality symmetry, and two quadratic invariants, 

\subsection{Equations of motion}

The superconformally covariant linearised equations of motion constitute a universally integrable system with  Cartan curvatures
\begin{align}
\boldsymbol{r}^{\boldsymbol{\Omega}}&:= d\boldsymbol{\Omega}+\boldsymbol{\Omega}\star \boldsymbol{\Omega}\ ,\\\label{3.2}
\boldsymbol{r}^{\boldsymbol{c}}&:= \boldsymbol{D}\boldsymbol{c}\ ,\\
\boldsymbol{r}^{\boldsymbol{k}}&:= \boldsymbol{D}\boldsymbol{k}\ ,\\
\boldsymbol{r}^{\boldsymbol{b}^{(\varepsilon)}}&:= \boldsymbol{D}\boldsymbol{b}^{(\varepsilon)}+\boldsymbol{\Sigma}^{(\varepsilon)}(\boldsymbol{c}^{(\varepsilon)})\ ,
\end{align}
where 
\begin{align}
\boldsymbol{D}\boldsymbol{c}=d\boldsymbol{c}+\boldsymbol{\Omega}\star \boldsymbol{c}\ ,\qquad \boldsymbol{D}\boldsymbol{k}=d\boldsymbol{k}+[\boldsymbol{\Omega}, \boldsymbol{k}]_\star\ ,\qquad \boldsymbol{D}\boldsymbol{b}^{(\varepsilon)}=d\boldsymbol{b}^{(\varepsilon)}+[\boldsymbol{\Omega},\boldsymbol{b}^{(\varepsilon)}]_\star\ ,
\end{align}
and $\boldsymbol{\Sigma}^{(\varepsilon)}$ are composite three-forms linear in $\boldsymbol{k}$ and trilinear in $(\boldsymbol{e}_{-\varepsilon/2},\boldsymbol{e}_{-\varepsilon})$ providing $H$-covariant linear maps
\begin{align}\label{3.4}
\boldsymbol{\Sigma}^{({\varepsilon})}: \Omega_{[p]}(\boldsymbol{M}')\otimes \mathsf{S}^{({\varepsilon})}_{[1]}\to \Omega_{[3+p]}(\boldsymbol{M}')\otimes \left(\boldsymbol{(2)}|_G\otimes \boldsymbol{(1)}_{{\widetilde{K}}}\right)_{[-1]}\ ,\qquad p=0,1,\dots\ ,
\end{align}
with co-kernel and image given by 
\begin{align}\label{3.4}
\boldsymbol{\Sigma}^{({\varepsilon})}& :\  \Omega_{[p]}(\boldsymbol{M}')\otimes \left({\rm ker}\,\rho_{\mathsf{S}^{(\varepsilon)}}(\mathfrak{t}_{-\varepsilon})\right)_{[+1]}\ \to\ \Omega_{[3+p]}(\boldsymbol{M}')\otimes \left({\rm ker}\,\rho_{\boldsymbol{(2)}}(\mathfrak{t}_{-\varepsilon})\otimes \boldsymbol{(1)}|_{{\widetilde{K}}}\right)_{[-1]}\ ,
\end{align}
where 
\begin{align}
{\rm ker}\,\rho_{\mathsf{S}^{(\varepsilon)}}\left( \mathfrak{t}_{-\varepsilon}\right) &=\bigoplus_{J=0,1/2,1,3/2,2}\left.\boldsymbol{\tau}^{(\varepsilon)}_{J;0}\right|_{\mathfrak{h}}\ ,\\
{\rm ker}\,\rho_{\boldsymbol{(2)}}\left( \mathfrak{t}_{1}\right)&=\yngRed{;,;}_0 \oplus \bullet^{({\rm h})}\oplus \left(\yngBlue{;,;,;}_{+1/2}\otimes \yngRed{;}\right)\oplus \yngBlue{;;,;;,;;}_{+1} \ ,\\
{\rm ker}\,\rho_{\boldsymbol{(2)}}\left( \mathfrak{t}_{-1}\right)&=   \yngBlue{;;}_{-1}\oplus \left(\yngBlue{;}_{-1/2}\otimes \yngRed{;}\right)\oplus \yngRed{;,;}_0\oplus \bullet^{({\rm h})}\ ,
\end{align}
i.e., $\boldsymbol{\Sigma}^{(+)}$ and  $\boldsymbol{\Sigma}^{(-)}$ glue primaries and anti-primaries, respectively, to two-forms in $\boldsymbol{b}^{(+)}$ and $\boldsymbol{b}^{(-)}$
whose components have non-negative and non-positive conformal weights.
The cocycles have expansions
\begin{align}
\boldsymbol{\Sigma}^{(\varepsilon)}=k^{ij} \boldsymbol{\Sigma}^{(\varepsilon)}_{ij}\ ,\qquad \boldsymbol{\Sigma}^{(\varepsilon)}_{ij}= \btau^{AB}_{ij} {\Sigma}^{(\varepsilon)}_{AB}\ ,
\end{align}
where the linear maps
\begin{align}
{\Sigma}^{(\varepsilon)}_{AB}: \Omega_{[p]}(\boldsymbol{M}')\otimes \mathsf{S}^{({\varepsilon})}_{[1]}\to \Omega_{[3+p]}(\boldsymbol{M}')\ ,
\end{align}
are thus trilinear in  $(\boldsymbol{e}_{-\varepsilon/2},\boldsymbol{e}_{-\varepsilon})$.
In components, the unfolded  system reads 
\begin{align}
&R_{AB}\approx 0\ ,\qquad 
D(C^{(\varepsilon)})_{A(4+k),B(k)}\approx 0\ ,\qquad Dk_{ij}\approx 0\ ,\cr 
&D(B^{(\varepsilon)})^{ij}_{AB}+k^{ij} (\Sigma^{(\varepsilon)})_{AB}{}^{C(4)} C^{(\varepsilon)}_{C(4)}\approx 0\ ,
\end{align}
where $\Sigma^{(\varepsilon)}_{AB}{}^{C(4)}$ are $H$-covariant trilinear functions of $(E^{\alpha\beta}, F^{\alpha I})$ for $\varepsilon=+$ and $(\widetilde{E}_{\alpha\beta}, \widetilde{F}_{\alpha I})$ for $\varepsilon=-$.
The universal Cartan integrability of the unfolded system is equivalent to the cocycle condition $[\boldsymbol{D},\boldsymbol{\Sigma}^{({\varepsilon})}]\approx 0$, i.e., 
\begin{align}\label{3.5}
\boldsymbol{D}\left(\boldsymbol{\Sigma}^{({\varepsilon})}\left(\boldsymbol{c}^{({\varepsilon})}\right)\right)\equiv (d+\rho_{\boldsymbol{(2)}}(\boldsymbol{\Omega}))\left(\boldsymbol{\Sigma}^{({\varepsilon})}\left(\boldsymbol{c}^{({\varepsilon})}\right)\right)\approx 0\ ,
\end{align}
where $\approx$ denotes equality modulo Cartan curvatures, shown below, first in super-Poincaré backgrounds and then in superconformal backgrounds.

\subsection{Conformal duality symmetry}
\label{Sec:4.2}

While the superoscillator realisation only plays an auxiliary role in determining the linearised system's $Q$-structure, the conformal duality map induces a second-quantised conformal duality symmetry group $\mathbb{Z}_4(\gamma_\sigma)$ with generator 
\begin{align}
\gamma_\sigma:\mathcal{M}\to \mathcal{M}\ ,\qquad (\gamma_\sigma)^4={\rm Id}_{\mathcal{M}}\ ,
\end{align}
defined by 
\begin{align}
\gamma_\sigma(\boldsymbol{\Omega})&:={\rm Ad}^\star_{\kappa_\sigma} \boldsymbol{\Omega}\ ,\\\label{4.16}
\gamma_\sigma(\boldsymbol{c})&:=i(\kappa_\sigma)^{-1}\star \boldsymbol{c}\ ,\\
\gamma_\sigma(\boldsymbol{k})&:=\boldsymbol{k}\ ,\\
\gamma_\sigma(\boldsymbol{b}^{(\varepsilon)})&:=i{\rm Ad}^\star_{\kappa_\sigma}\boldsymbol{b}^{(-\varepsilon)}\ ,
\end{align}
Since $t^{(\varepsilon)}$ acts on the first-quantised algebra, it follows that 
\begin{align}
\gamma_\sigma\circ t^{(\varepsilon)}=t^{(\varepsilon)}\circ \gamma_\sigma\ ;
\end{align}
hence, using also the properties of the domains of $t^{(\varepsilon)}$, one has
\begin{align}\label{4.20}
\gamma_\sigma(\boldsymbol{c}^{(\varepsilon)})&=\gamma_\sigma(t^{(\varepsilon)}(\boldsymbol{c}))=t^{(\varepsilon)}(\gamma_\sigma(\boldsymbol{c}))=t^{(\varepsilon)}(i(\kappa_\sigma)^{-1}\star \boldsymbol{c})= \\\label{4.21}&=i(\kappa_\sigma)^{-1}\star t^{(-\varepsilon)}(\boldsymbol{c})=i(\kappa_\sigma)^{-1}\star \boldsymbol{c}^{(-\varepsilon)}\ .
\end{align}
Moreover, we assume that  $\boldsymbol{\Sigma}^{(\varepsilon)}$ are normalised such that
\begin{align}
\gamma_\sigma(\boldsymbol{\Sigma}^{(\varepsilon)})= {\rm Ad}^\star_{\kappa_\sigma}\boldsymbol{\Sigma}^{(-\varepsilon)}\star \kappa_\sigma \ ,
\end{align}
where thus ${\rm Ad}^\star_{\kappa_\sigma}\boldsymbol{\Sigma}^{(-\varepsilon)}\equiv k^{ij} {\rm Ad}^\star_{\kappa_\sigma}(\btau^{AB}_{ij}) {\Sigma}^{(-\varepsilon)}_{AB}\star \kappa_\sigma$, i.e., 
\begin{align}
\gamma_\sigma\left(\boldsymbol{\Sigma}^{(\varepsilon)}(\boldsymbol{c}^{(\varepsilon)})\right)= i{\rm Ad}^\star_{\kappa_\sigma}\left(\boldsymbol{\Sigma}^{(-\varepsilon)}(\boldsymbol{c}^{(-\varepsilon)})\right)\ .
\end{align}
We also have ${\rm Ad}^\star_{\kappa_\sigma} \boldsymbol{k}=\boldsymbol{k}$.
Thus, the linearised unfolded system is conformal-duality covariant\footnote{$(\gamma_\sigma)^2$ is the discrete symmetry transformation amounting to flipping the sign of all fermionic component fields.}.

\subsection{Zero-form charge and exotic topological invariant}
\label{Sec:4.3}

The linearised unfolded system admits two quadratic forms that are on-shell gauge invariant:
\begin{itemize}
\item[i)] the zero-form charge
\begin{align}
I_0:= \oint_{\boldsymbol{M}_0} \left\langle \boldsymbol{c}^{(-)}\right|\left. \boldsymbol{c}^{(+)}\right\rangle_{\mathsf{S}}\ , 
\end{align}
where $\boldsymbol{M}_0$ is a point and $\left\langle \cdot| \cdot \right\rangle_{\mathsf{S}}$ denotes the $G$-invariant Hermitian, i.e., symmetric sesquilinear, form on $\mathsf{S}$; and
\item[ii)] the six-form charge
\begin{align}
I_6\approx \oint_{\boldsymbol{M}_6} \left\langle \boldsymbol{\Sigma}^{(-)}(\boldsymbol{c}^{(-)}), \boldsymbol{\Sigma}^{(+)}(\boldsymbol{c}^{(+)})\right\rangle_{\boldsymbol{(2)}\otimes \boldsymbol{(1)}}\ , 
\end{align}
where $\boldsymbol{M}_6$ is a orientable closed six-manifold and $\left\langle \cdot, \cdot \right\rangle_{\boldsymbol{(2)}\otimes \boldsymbol{(1)}}$ denotes the $G\times \widetilde{K}$-invariant bilinear form on $\boldsymbol{(2)}|_{G}\otimes \boldsymbol{(1)}|_{\widetilde{K}}$.
\end{itemize}
$I_0$ and $I_6$ are  

-- globally well-defined, i.e., their integrands are locally $H$-invariant; and

-- functions of the homology classes of $\boldsymbol{M}_0$ and $\boldsymbol{M}_6$ , i.e., their integrands are on-shell de Rham closed; and

-- $\gamma_\sigma$-covariant, i.e., 
\begin{align}
\gamma_\sigma(I_0)=\overline{I_0}\ ,\qquad \gamma_\sigma(I_6)={I_6}\ .
\end{align}
Both $I_0$ and $I_6$ are natural candidates for two-point functions of pairs of Weyl zero-forms valued in $G$-modules.
In the case of $I_6$, its evaluation in super-Poincar\'e backgrounds requires a limiting procedure whereby $\widetilde E_a$ is set equal to $\Lambda E_a$ after which one sends $\Lambda\to 0$ in $\Lambda^{-3}I_6$, which we leave for a separate study.

The six-form charge is the on-shell value of the topological invariant
\begin{align}
C&:=\oint_{\boldsymbol{M}_6} \left\langle \boldsymbol{D}\boldsymbol{b}^{(-)}, \boldsymbol{D}\boldsymbol{b}^{(+)}\right\rangle_{\boldsymbol{(2)}|_{G}\otimes \boldsymbol{(1)}|_{\widetilde{K}}}=\oint_{\boldsymbol{M}_6}  (\boldsymbol{D}\boldsymbol{b}^{(-)})_{ij}^{AB} \wedge (\boldsymbol{D}\boldsymbol{b}^{(+)})^{ij}_{BA}\ ,
\end{align}
which is conformal-duality invariant.

\section{Super-Poincar\'e covariant three-form cocycle}
\label{Sec:5}

In this section, we show the existence of a Cartan integrable $H$-covariant cocycle gluing the Weyl zero-form to the two-form potentials in a chiral super-Poincar\'e background built from 
\begin{itemize}
\item[i)] conformal chiral primaries;
\item[ii)] a set of nearly anti-chiral three-form building blocks in traceless R-symmetry tensors with salient features;
\item[iii)] the elements in the two-form modules $\boldsymbol{(2)}^{(+)}$ and $\boldsymbol{(2)}^{(-)}$ the  kernel of $K^{\alpha\beta}$ and $T_{\alpha\beta}$, respectively.
\end{itemize}

\subsection{Super-Poincar\'e background}

We refer to backgrounds in which 
\begin{align}
\boldsymbol{e}_{\varepsilon/2}+\boldsymbol{e}_{\varepsilon}\stackrel{!}{=}0\ ,\qquad \Omega_{\tilde{\mathfrak{k}}}=0\ ,
\end{align}
as chiral and anti-chiral super-Poincar\'e backgrounds for $\varepsilon=-$ and $\varepsilon=+$, respectively.
We choose
\begin{align}
\boldsymbol{\Omega}\stackrel{!}{=}\boldsymbol{\Omega}_+\equiv \boldsymbol{\Omega}_{\boldsymbol{h}}+\boldsymbol{e}_{+1/2}+\boldsymbol{e}_{+1}\ ,
\end{align}
activating the cocycle $\boldsymbol{\Sigma}^{(+)}$ glueing chiral primary tensors of positive conformal weight to two-forms $\boldsymbol{b}^{(+)}$.
We let 
\begin{align}
\boldsymbol{D}_+:=d+\boldsymbol{\Omega}_+\ ,
\end{align}
and use the normalisations
\begin{align}
\boldsymbol{\Omega}_+=i\left(\Omega_\beta{}^\alpha M_\alpha{}^\beta+D\sigma +E^{\alpha\beta}T_{\alpha\beta}+ F^{\alpha I} Q_{\alpha I}\right)\ ,
\end{align}
for which the nilpotency of $\boldsymbol{D}_+$  amounts to the background field equations 
\begin{align}
{\rm Rie}_\alpha{}^\beta&\approx 0\ ,\qquad R^{(\sigma)}\approx0\ ,\qquad 
\nabla E^{\alpha\beta} \approx \frac{i}{2}\eta^{IJ}F_I^{\alpha}\wedge F_J^{\beta}\ ,\qquad \nabla F^{I\alpha}\approx 0\ ,
\end{align}
where $\nabla$ is the $H$-covariant derivative defined in \eqref{3.109} which is nilpotent by the first two equations.

\subsection{Poincar\'e background}
\label{poincareBackground}

Let us first consider a chiral Poincaré background with nilpotent
\begin{align}
\boldsymbol{D}_+:=\nabla +\boldsymbol{e}_{+1}\ ,
\end{align}
and a two-form 
\begin{align}
\boldsymbol{b}^{(-)}&=\frac12\left(\btau^{\alpha\beta} \omega_{\alpha\beta} +\btau^{\alpha I}\Psi_{\alpha I} +\btau_\alpha{}^\beta  h_\beta{}^\alpha\right. \cr &\left. +\btau^{IJ} B_{IJ}+  \btau^{({\rm h})} B^{({\rm h})}+\btau_\alpha{}^I \widetilde{\Psi}^\alpha{}_I+\btau_{\alpha\beta}\widetilde{\omega}^{\alpha\beta}\right) ,
\end{align}
with equation of motion  
\begin{align}\label{bosoniczerocurvature}
\boldsymbol{D}_+\boldsymbol{b}^{(-)}+\boldsymbol{\Sigma}^{(+)}\approx 0\ ,
\end{align}
where the three-form cocycle is assumed to be of the form
\begin{align}
\boldsymbol{\Sigma}^{(+)}:= \frac12(\Sigma_{[3]})^{\alpha\beta}\left(\btau^{\gamma\delta} R_{\alpha\beta|\gamma\delta}+\btau^{\gamma I}\Psi_{\alpha\beta|\gamma I} +\btau^{IJ}H_{\alpha\beta IJ} \right),
\end{align}
i.e., glueing chiral zero-forms  $(R_{\alpha\beta|\gamma\delta},\Psi_{\alpha\beta|\gamma I}, H_{\alpha\beta|IJ})$ into the kernel of $K^{\alpha\beta}$ in $\boldsymbol{(2)}^{(-)}$ via the anti-chiral three-form 
\begin{align}\label{5.8}
(\Sigma_{[3]})^{\alpha\beta}:=\frac14  (\tilde \sigma_{abc})^{\alpha\beta} E^a\wedge E^b\wedge E^c=2E^{\alpha\gamma_1}\wedge E_{\gamma_1\gamma_2}\wedge E^{\gamma_2\beta}\ .
\end{align}
Expanding the two-form as in \eqref{expandb}, and using
\begin{align}
\boldsymbol{e}_{+1} {\boldsymbol\tau}^{\alpha\beta} &=4E^{(\alpha|\gamma}\boldsymbol{\tau}_\gamma{}^{|\beta)}
\ ,\qquad 
\boldsymbol{e}_{+1}{\boldsymbol\tau}_\alpha{}^\beta=2E^{\beta\gamma}\boldsymbol{\tau}_{\gamma\alpha}
\ ,\qquad  \boldsymbol{e}_{+1}\btau^{\alpha I}=2E^{\alpha\beta}\btau_\beta{}^I, \\
\boldsymbol{e}_{+1}{\boldsymbol\tau}_{\alpha\beta}&=0\ ,\qquad  
\boldsymbol{e}_{+1}{\boldsymbol\tau}^{IJ}=0\ ,\qquad \boldsymbol{e}_{+1}\btau_\alpha{}^I=0\ ,
\end{align}
one has 
\begin{align}
\boldsymbol{D}_+\boldsymbol{b}^{(-)}&=\frac12\left(\btau^{\alpha\beta} \nabla\omega_{\alpha\beta} +\btau^{\alpha I}\nabla\Psi_{\alpha I} +\btau_\alpha{}^\beta \nabla h_\beta{}^\alpha \right.\cr &+\btau^{IJ}\nabla B_{IJ}+  \btau^{({\rm h})}dB^{({\rm h})}+\btau_\alpha{}^I \nabla\widetilde{\Psi}^\alpha{}_I+\btau_{\alpha\beta}\nabla\widetilde{\omega}^{\alpha\beta} \cr  
&\left.-4\btau_\alpha{}^\beta E^{\alpha\gamma}\wedge \omega_{\beta\gamma}-2\btau_{\alpha\beta}E^{\beta\gamma}\wedge h_\gamma{}^\alpha-2\btau_\alpha{}^I E^{\alpha\beta}\Psi_{\beta I}\right)\ ,
\end{align}
hence, in components,
\begin{align}
\nabla\omega_{\alpha\beta}+ (\Sigma_{[3]})^{\gamma\delta} R_{\gamma\delta|\alpha\beta}&\approx0\ ,\\
\nabla\psi_{\alpha I}+(\Sigma_{[3]})^{\beta\gamma}\psi_{\beta\gamma|\alpha I}&\approx0\ ,\\
\nabla h_\alpha{}^\beta -4E^{\beta\gamma}\wedge\omega_{\alpha \gamma}&\approx0\ ,\\
\nabla B_{IJ} +(\Sigma_{[3]})^{\alpha\beta}H_{\alpha\beta IJ}&\approx0\ ,\\
\nabla B^{({\rm h})}&\approx0\ ,\\
\nabla\widetilde{\psi}^\alpha{}_I -E^{\alpha\beta}\wedge\psi_{\beta I}&\approx0\ ,\\
\nabla \widetilde{\omega}^{\alpha\beta}-2 E^{\gamma(\alpha}\wedge h_{\gamma}{}^{\beta)}&\approx0\ .
\end{align}
The Cartan curvature of $\widetilde{\omega}^{\alpha\beta}$ obeys the Bianchi identity trivially. 
From 
\begin{align}
(\Sigma_{[3]})^{\alpha\beta}\wedge E^{\rho\lambda}=\epsilon^{\rho\lambda\sigma(\alpha}(\Sigma_{[4]})_\sigma{}^{\beta)}\ ,\qquad  (\Sigma_{[4]})_\alpha{}^\beta:=\frac{1}{4!}(\sigma_{abcd})_\alpha{}^\beta E^a\wedge E^b\wedge E^c\wedge E^d\ ,
\end{align}
it follows that the Cartan curvatures of $h_\beta{}^\alpha$ and $\widetilde{\Psi}^{\alpha I}$ obey Bianchi identities iff
\begin{align}
\epsilon^{\alpha\gamma\delta\lambda}(\Sigma_{[4]})_\delta{}^\rho R_{\lambda\rho|\gamma\beta}\approx 0\ ,\qquad  \epsilon^{\alpha\beta\gamma\delta}(\Sigma_{[4]})_\gamma{}^\lambda \Psi_{\delta\lambda|\beta I}\approx 0\ ,
\end{align}
respectively, which in view of $\yngBlue{;;}\otimes\yngBlue{;;}=\yngBlue{;;;;}\oplus\yngBlue{;;;,;}\oplus\yngBlue{;;,;;}$ and $\yngBlue{;;}\otimes\yngBlue{;}=\yngBlue{;;;}\oplus\yngBlue{;;,;}$ are equivalent to
\begin{align}
R_{\alpha\beta|\gamma\delta}\in\yngBlue{;;;;}\ ,\qquad \Psi_{\alpha\beta|\gamma I}\in\yngBlue{;;;}\otimes\yngRed{;}\ .
\end{align}
Similarly, the Bianchi identities for the Cartan curvatures of $\omega_{\alpha\beta}$, $\Psi_{\alpha I}$, and $B_{IJ}$, are equivalent to 
\begin{align}
\nabla R_{\alpha(4)}&\approx E^{\beta\gamma}R_{\alpha(4)\beta,\gamma}\ ,\qquad R_{\alpha(5),\beta}\in\yngBlue{;;;;;,;}\ ,\\
\nabla\Psi_{\alpha(3) I}&\approx E^{\beta\gamma}\Psi_{\alpha(3)\beta,\gamma I}\ ,\qquad \Psi_{\alpha(4),\beta I}\in\yngBlue{;;;;,;}\otimes\yngRed{;}\ ,\\
\nabla H_{\alpha(2) IJ}&\approx E^{\beta\gamma}H_{\alpha(2)\beta,\gamma IJ}\ ,\qquad H_{\alpha(3),\beta IJ}\in\yngBlue{;;;,;}\otimes\yngRed{;,;}\ ,
\end{align}
leaving $\eta^{IJ}H_{\alpha\beta IJ}$ and $\eta^{IJ}R_{\alpha\beta\gamma,\delta IJ}$ algebraically unconstrained, as such constraints require supersymmetry. 
On real six-manifolds with invertible $E^{\alpha\beta}$, the exotic graviton, gravitini and two-form potentials of the metric-like formulation arise in $h_\alpha{}^\beta$, $\Psi_{\alpha I}$, and $B_{IJ}$, respectively, upon elimination of auxiliary fields and gauge fixing, with $\omega_{\alpha\beta}$ being an exotic Lorentz connection, and $\widetilde{\omega}^{\alpha\beta}$ and $\widetilde{\Psi}^\alpha{}_I$ being flat, auxiliary two-forms.
Finally, the hypercharged two-form $B^{({\rm h})}$ remains a decoupled, topological field (at the linearised level).

\subsection{Three-form building blocks}

Proceeding as for the $(2,0)$-multiplet \cite{TwoCommaZero}, we introduce a set of nearly anti-chiral three-forms in traceless R-symmetry tensors consisting of \eqref{5.8} and 
\begin{align}
(\Sigma_{[3]})_{\alpha}{}^{\beta|\gamma I}&:=(\Sigma_{[2]})_{\alpha}{}^{\beta}\wedge F^{\gamma I}\ ,\\ 
(\Sigma_{[3]})^{\alpha,\beta|\gamma,\delta| IJ}&:=E^{\alpha\beta}\wedge(\Sigma_{[2]})^{\gamma,\delta|IJ}\ ,\\
(\Sigma_{[3]})_\alpha{}^{IJK}&:=\epsilon_{\alpha\beta\gamma\delta}F^{\beta I}\wedge F^{\gamma J}\wedge F^{\delta K}+\frac12\epsilon_{\alpha\beta\gamma\delta}\eta^{[IJ} F^{\beta K]}\wedge F^{\gamma L}\wedge F^{\delta}{}_{L}\ ,
\end{align}
where
\begin{align}
(\Sigma_{[2]})_\alpha{}^\beta&:=E_{\alpha\gamma}\wedge E^{\gamma\beta}\ ,\\
(\Sigma_{[2]})^{\alpha,\beta|IJ}&:=\frac12\left(F^{\alpha I}\wedge F^{\beta J}-F^{\alpha J}\wedge F^{\beta I}+\frac14\eta^{IJ}F^{\alpha K}\wedge F^{\beta}{}_K \right)\ .
\end{align}
In a super-Poincaré background, they obey 
\begin{align} \label{nablaSigma1}
\nabla(\Sigma_{[3]})^{\alpha\beta}&\approx 3i(\Sigma_{[3]})_\gamma{}^{(\alpha|\gamma I}\wedge F^{\beta)}{}_I\ , \\
\label{nablaSigma2}
\nabla (\Sigma_{[3]})_\alpha{}^{\beta|\gamma I}&\approx i\left( (\Sigma_{[3]})_\alpha{}^{\beta|IJ}-\frac14\delta_\alpha{}^\beta(\Sigma_{[3]})_\rho{}^{\rho|IJ}\right)\wedge F^{\gamma}{}_J\ ,\\
\label{nablaSigma3}
\nabla (\Sigma_{[3]})_\alpha{}^{IJK}&\approx0\ .
\end{align}
There are also the following algebraic relations:
\begin{align}\label{1stalgebraic}
& E_{\alpha\beta}\wedge E^{\gamma\delta}=-2\delta^{[\gamma}_{[\alpha}(\Sigma_{[2]})_{\beta]}{}^{\delta]}\ ,\\ \label{2ndalgebraic}
& E^{\alpha\beta}\wedge E^{\gamma\delta}=\epsilon^{\alpha\beta\lambda[\gamma}(\Sigma_{[2]})_\lambda{}^{\delta]}\ ,\\ 
&(\Sigma_{[3]})^{\alpha\beta}\wedge E^{\rho\lambda}=\epsilon^{\rho\lambda\sigma(\alpha}(\Sigma_{[4]})_\sigma{}^{\beta)}\ , \\ \label{3rdalgebraic}
&(\Sigma_{[2]})_\alpha{}^{\beta}\wedge E^{\rho\lambda}=\frac13\delta_\alpha^{[\rho}(\Sigma_{[3]})^{\lambda]\beta}  + 
\frac1{12}\epsilon^{\beta\rho\lambda\sigma} (\Sigma_{[3]})_{\sigma\alpha}\ ,
\end{align}
where
\begin{align}
(\Sigma_{[3]})_{\alpha\beta}&:=2E_{\alpha\gamma_1}\wedge E^{\gamma_1\gamma_2}\wedge E_{\gamma_2\beta}\ ,\\ 
(\Sigma_{[4]})_{\alpha}{}^\beta&:=\frac{1}{4!}(\sigma_{abcd})_\alpha{}^\beta E^a\wedge E^b\wedge E^c\wedge E^d\ .
\end{align} 

It will be useful to compute from \eqref{Tcommutators}, \eqref{Qcommutators1}, and \eqref{Qcommutators2}:  
\begin{align}
    &[\boldsymbol{e}_{+1}+\boldsymbol{e}_{+1/2},\btau^{\alpha\beta}]_\star= -4E^{\gamma(\alpha}\btau^{\beta)}{}_\gamma-iF^{(\alpha I}\btau^{\beta)}{}_I, \\
    &[\boldsymbol{e}_{+1}+\boldsymbol{e}_{+1/2},\btau^{\alpha I}]_\star= -2E^{\alpha\beta}\btau_\beta{}^I-iF^{\alpha J}\btau_J{}^I +F^{\beta I}\btau^\alpha{}_\beta+F^{\alpha I}\btau^{({\rm z})}, \\
    &[\boldsymbol{e}_{+1}+\boldsymbol{e}_{+1/2},\btau^{IJ}]_\star= 2F^{\alpha[I}\btau_\alpha{}^{J]}+\frac{i}4\eta^{IJ}F^{\alpha K}\btau_{\alpha K}\ ,\\
    &[\boldsymbol{e}_{+1}+\boldsymbol{e}_{+1/2},\btau^\alpha{}_\beta]_\star=2E^{\alpha\gamma}\btau_{\gamma\beta}-iF^{\alpha I}\btau_{\beta I}+\frac{i}4\delta^\alpha_\beta F^{\gamma I}\btau_{\gamma I}\ ,\\
    &[\boldsymbol{e}_{+1}+\boldsymbol{e}_{+1/2},\btau_{\alpha I}]_\star=F^{\beta}{}_I\btau_{\beta\alpha}\ ,\\
    &[\boldsymbol{e}_{+1}+\boldsymbol{e}_{+1/2},\btau_{\alpha\beta}]_\star=0\ .
\end{align}
For later use, we will also compute 
\begin{align}\label{tauFtilde1}
    &[\boldsymbol{e}_{-1/2},\btau^{\alpha\beta}]_\star=0\ ,\\
    \label{tauFtilde2}
    &[\boldsymbol{e}_{-1/2},\btau^{\alpha I}]_\star=\tilde F_{\beta}{}^I\btau^{\beta\alpha}\ ,\\
    \label{tauFtilde3}
    &[\boldsymbol{e}_{-1/2},\btau^{IJ}]_\star=2\tilde F_\alpha{}^{[I}\btau^{\alpha J]}-\frac{1}{4}\eta^{IJ}\tilde F_{\alpha K}\btau^{\alpha K},
\end{align}

\subsection{Cocycle valued in finite-dimensional \texorpdfstring{$\mathfrak{g}$}{g}-module}

Let us consider a two-form 
\begin{align}
    \boldsymbol{b}=  B_\mathcal{A} {\btau^\mathcal{A}}\ ,
\end{align}
valued in a $\mathfrak{g}$-module $\mathsf{M}$ with basis $ {\boldsymbol{\tau}^{\mathcal{A}}}$. Elements in the image of $\mathfrak{t}_{+1}$ will be $ {\mathsf{t}^\mathcal{A}}$  and elements in the difference $\mathsf{M}\setminus (\mathfrak{t}_{+1}\mathsf{M})$ will be $ {\mathsf{s}^\mathcal{A}}$. 
The Cartan curvature 
\begin{align}
    \boldsymbol{R}:= \boldsymbol{D}_+\boldsymbol{b}+\boldsymbol{\Sigma}\ .
\end{align}
A non-trivial cocycle will be 
\begin{align}
\boldsymbol{\Sigma}=\left(\Sigma^{\alpha\beta}R_{\alpha\beta|\mathcal{A}}+\Sigma_\alpha{}^\beta\wedge F^{\gamma I} \Psi^{\alpha}_{\beta|\gamma I|\mathcal{A}}+E^{\alpha\beta}\wedge \Sigma_{(2)}^{\gamma,\delta|IJ} H_{\alpha,\beta|\gamma,\delta|IJ|\mathcal{A}}+ 
\Sigma^{IJK}_\alpha \Theta^{\alpha}_{IJK|\mathcal{A}}\right)  {\mathsf{s}^\mathcal{A}}\ .
\end{align}
Then we compute
\begin{align}
\boldsymbol{D_+}\boldsymbol{\Sigma}&= \left(\nabla\Sigma^{\alpha\beta}R_{\alpha\beta|\mathcal{A}}+\nabla\Sigma_\alpha{}^\beta\wedge F^{\gamma I} \Psi^{\alpha}_{\beta|\gamma I|\mathcal{A}}+\nabla E^{\alpha\beta}\wedge \Sigma_{(2)}^{\gamma,\delta|IJ} H_{\alpha,\beta|\gamma,\delta|IJ|\mathcal{A}}\right. \cr 
    &- \left.\Sigma^{\alpha\beta}\nabla R_{\alpha\beta|\mathcal{A}}-\Sigma_\alpha{}^\beta\wedge F^{\gamma I} \nabla\Psi^{\alpha}_{\beta|\gamma I|\mathcal{A}}-E^{\alpha\beta}\wedge \Sigma_{(2)}^{\gamma,\delta|IJ} \nabla H_{\alpha,\beta|\gamma,\delta|IJ|\mathcal{A}}\right.\cr&-\left. 
    \Sigma^{IJK}_\alpha \nabla\Theta^{\alpha}_{IJK|\mathcal{A}}\right)  {\mathsf{s}^\mathcal{A}} - \left(\Sigma^{\alpha\beta} R_{\alpha\beta|\mathcal{A}}+ \Sigma_\alpha{}^\beta\wedge F^{\gamma I} \Psi^{\alpha}_{\beta|\gamma I|\mathcal{A}}\right.\cr &+E^{\alpha\beta}\wedge \Sigma_{(2)}^{\gamma,\delta|IJ} H_{\alpha,\beta|\gamma,\delta|IJ|\mathcal{A}}+\left. 
    \Sigma^{IJK}_\alpha \Theta^{\alpha}_{IJK|\mathcal{A}}\right) \boldsymbol{E}^+ {\mathsf{s}^\mathcal{A}}\ .
\end{align}
In the last line we use 
\begin{align}
    \boldsymbol{E}^+ {\mathsf{s}^\mathcal{A}} = 
    iE^{\alpha\beta} (T^{\mathcal{A}}{}_\mathcal{B})_{\alpha\beta}  {\mathsf{t}^\mathcal{B}}+iF^{\alpha I} (T^{\mathcal{A}}{}_\mathcal{B})_{\alpha I} {\mathsf{t}^\mathcal{B}} + 
    iF^{\alpha I} (S^{\mathcal{A}}{}_\mathcal{B})_{\alpha I} {\mathsf{s}^\mathcal{B}},
\end{align}
since $\mathfrak{q}^+(\mathsf{M}\setminus (\mathfrak{t}_{+1}\mathsf{M}))$ may be a generic element of $\mathsf{M}$. The coefficients $T$ and $S$ are representation matrices for the module $\mathsf{M}$. 

The terms with $ {\mathsf{t}^\mathcal{B}}$ are 
\begin{align}\label{tequations}
    0&\approx i\left( -\Sigma^{\alpha\beta}\wedge E^{\gamma\delta} 
    R_{\alpha\beta|\mathcal{A}}(T^{\mathcal{A}}{}_\mathcal{B})_{\gamma\delta} +\Sigma_\alpha{}^\beta\wedge E^{\delta\rho}\wedge F^{\gamma I} \Psi^{\alpha}_{\beta|\gamma I|\mathcal{A}} (T^{\mathcal{A}}{}_\mathcal{B})_{\delta\rho}\right.\cr &\left. -E^{\alpha\beta}\wedge E^{\rho\lambda}\wedge\Sigma^{\gamma,\delta|IJ} H_{\alpha,\beta|\gamma,\delta|IJ|\mathcal{A}} (T^{\mathcal{A}}{}_\mathcal{B})_{\rho\lambda} -\Sigma_\alpha^{IJK}\wedge E^{\beta\gamma}\Theta^{\alpha}_{IJK|\mathcal{A}}
    (T^{\mathcal{A}}{}_\mathcal{B})_{\beta\gamma}\right. \cr 
    &\left.-\Sigma^{\alpha\beta}\wedge F^{\gamma I} R_{\alpha\beta|\mathcal{A}} (T^{\mathcal{A}}{}_\mathcal{B})_{\gamma I} -\Sigma_\alpha{}^\beta\wedge F^{\gamma I}\wedge F^{\delta J} \Psi^{\alpha}_{\beta|\gamma I|\mathcal{A}}
    (T^{\mathcal{A}}{}_\mathcal{B})_{\delta J}\right.\cr 
    &-\left.
    E^{\alpha\beta}\wedge \Sigma^{\gamma,\delta|IJ}\wedge F^{\rho K}H_{\alpha,\beta|\gamma,\delta|IJ|\mathcal{A}}
    (T^{\mathcal{A}}{}_\mathcal{B})_{\rho K}-
    \Sigma_\alpha^{IJK}\wedge F^{\beta L}\Theta^{\alpha}_{IJK|\mathcal{A}}(T^{\mathcal{A}}{}_\mathcal{B})_{\beta L}
    \right) {\mathsf{t}^\mathcal{B}} 
\end{align}
Each term with the same power of $E$ and $F$ should cancel independently. From this, we get 
\begin{align}\label{teq1}
    &\epsilon^{\gamma\delta\sigma(\alpha}\Sigma_{(4)\sigma}{}^{\beta)} R_{\alpha\beta|\mathcal{A}}(T^{\mathcal{A}}{}_\mathcal{B})_{\gamma\delta} \approx 0\ ,\\
    \label{teq2}
    &\epsilon^{\beta\delta\rho\sigma} \Sigma_{\sigma\alpha}\Psi^{\alpha}_{\beta|\gamma I|\mathcal{A}} (T^{\mathcal{A}}{}_\mathcal{B})_{\delta\rho}\approx0\ ,\\
    \label{teq3}
    &\frac13\delta_\alpha^{[\delta} \Sigma^{\rho]\beta}\wedge F^{\gamma I}\Psi^{\alpha}_{\beta|\gamma I|\mathcal{A}} (T^{\mathcal{A}}{}_\mathcal{B})_{\delta\rho}-\Sigma^{\alpha\beta}\wedge F^{\gamma I} R_{\alpha\beta|\mathcal{A}}(T^\mathcal{A}{}_\mathcal{B})_{\gamma I}\approx 0\ ,\\
    \label{teq4}
    &\epsilon^{\alpha\beta\sigma[\rho} \Sigma_\sigma{}^{\lambda]}\wedge F^{\gamma I}\wedge F^{\delta J} H_{\alpha,\beta|\gamma,\delta|IJ|\mathcal{A}}(T^\mathcal{A}{}_\mathcal{B})_{\rho\lambda} - \Sigma_\alpha{}^\beta\wedge F^{\gamma I}\wedge F^{\delta J} \Psi^{\alpha}_{\beta|\gamma I|\mathcal{A}}
    (T^{\mathcal{A}}{}_\mathcal{B})_{\delta J}\approx0\ ,\\
    \label{teq5}
    &E^{\alpha\beta}\wedge F^{\gamma I}\wedge F^{\delta J}\wedge F^{\rho K}\epsilon_{\lambda\gamma\delta\rho} \Theta^{\lambda}_{IJK|\mathcal{A}}(T^\mathcal{A}{}_\mathcal{B})_{\alpha\beta}\cr &-E^{\alpha\beta}\wedge F^{\gamma I}\wedge F^{\delta J}\wedge F^{\rho K} H_{\alpha,\beta|\gamma,\delta|IJ|\mathcal{A}}(T^\mathcal{A}{}_\mathcal{B})_{\rho K}\approx0\ , \\ 
    \label{teq6}
    &\Sigma_\alpha^{IJK}\wedge F^{\beta L}\Theta^{\alpha}_{IJK|\mathcal{A}}(T^{\mathcal{A}}{}_\mathcal{B})_{\beta L}\approx0\ ,
\end{align}

Now we proceed with the terms with $ {\mathsf{s}^\mathcal{B}}$. First we write 
\begin{align}
    &\nabla R_{\alpha\beta|\mathcal{A}}\approx E^{\gamma\delta} \left(C^R\right)_{\alpha\beta|\mathcal{A}|\gamma,\delta}+ F^{\gamma I}\C{R}_{\alpha\beta|\mathcal{A}|\gamma I},\cr 
    &\nabla \Psi^\alpha_{\beta|\gamma I|\mathcal{A}}\approx E^{\delta\lambda} \C{\Psi}^{\alpha}_{\beta|\gamma I|\mathcal{A}|\delta,\lambda}+F^{\delta J} \C{\Psi}^{\alpha}_{\beta|\gamma I|\mathcal{A}|\delta J} ,\cr
    &\nabla H_{\alpha,\beta|\gamma,\delta|IJ|\mathcal{A}}\approx E^{\lambda\rho}\C{H}_{\alpha,\beta|\gamma,\delta|IJ|\mathcal{A}|\lambda\rho}+ F^{\lambda K}\C{H}_{\alpha,\beta|\gamma,\delta|IJ|\mathcal{A}|\lambda K},\cr    &\nabla\Theta^\alpha_{IJK|\mathcal{A}}\approx E^{\beta\gamma} \C{\Theta}^\alpha_{IJK|\mathcal{A}|\beta,\gamma} +F^{\beta L} \C{\Theta}^\alpha_{IJK|\mathcal{A}|\beta L}\ .
\end{align}

Using that, we get five different on-shell equations with different powers of $E$ and $F$
\begin{align}\label{sequations}
    &\epsilon^{\gamma\delta\sigma(\alpha}\Sigma_{(4)\sigma}{}^{\beta)}\C{R}_{\alpha\beta|\mathcal{A}|\gamma,\delta}\approx0\ ,\cr 
    &-\Sigma^{\alpha\beta}\wedge F^{\gamma I}\C{R}_{\alpha\beta|\mathcal{A}|\gamma I}+\left( \frac13\delta_\alpha^{[\rho}\Sigma^{\lambda]\beta}  + 
    \frac1{12}\epsilon^{\beta\rho\lambda\sigma} \Sigma_{\sigma\alpha} \right)\wedge F^{\delta J}\C{\Psi}^\alpha_{\beta|\gamma I|\mathcal{A}|\rho,\lambda}\cr &-i\Sigma^{\alpha\beta}\wedge F^{\gamma I}R_{\alpha\beta|\mathcal{B}}(S^\mathcal{B}{}_\mathcal{A})_{\gamma I}\approx0\ ,\cr 
    & -\frac{i}{2}\Sigma_\gamma{}^\alpha\wedge F^{\gamma I}\wedge F^{\beta J} R_{\alpha\beta|\mathcal{A}}\eta_{IJ} -\Sigma_\alpha{}^\beta\wedge F^{\gamma I}\wedge F^{\delta J}\C{\Psi}^\alpha_{\beta|\gamma I|\mathcal{A}|\delta J}\cr &-
    \epsilon^{\alpha\beta\sigma[\rho} \Sigma_\sigma{}^{\lambda]}\wedge F^{\gamma I}\wedge F^{\delta J}\C{H}_{\alpha,\beta|\gamma,\delta|IJ|\mathcal{A}|\rho,\lambda}-i\Sigma_\alpha{}^\beta \wedge F^{\gamma I}\wedge F^{\delta J}\Psi^\alpha_{\beta|\gamma I|\mathcal{B}}(S^\mathcal{B}{}_\mathcal{A})_{\delta J}\approx0\ ,\cr 
    &E^{\rho\beta}\wedge F^{\gamma I}\wedge F^{\delta J}\wedge F^{\lambda K}\left( \frac{i}{4}\epsilon_{\alpha\rho\gamma\delta}\eta_{IJ} \Psi^\alpha_{\beta|\lambda K|\mathcal{A}} -\frac{i}{4}\epsilon_{\alpha\rho\beta\gamma}\eta_{IJ}\Psi^\alpha_{\delta|\lambda K|\mathcal{A}}-\C{H}_{\rho,\beta|\gamma,\delta|IJ|\mathcal{A}|\lambda K} \right.\cr 
    &\left. -\epsilon_{\alpha\gamma\delta\lambda}\C{\Theta}^\alpha_{IJK| \mathcal{A}|\rho,\beta} + iH_{\rho,\beta|\gamma,\delta|IJ|\mathcal{B}}(S^\mathcal{B}{}_\mathcal{A})_{\lambda K} \right)\approx0\ ,\cr 
    &F^{\alpha I}\wedge F^{\beta J}\wedge F^{\gamma K}\wedge F^{\delta  L}
    \left( -\frac{i}{2}\eta_{IJ} H_{\alpha,\beta|\gamma,\delta|KL|\mathcal{A}}  -\epsilon_{\lambda\beta\gamma\delta} \C{\Theta}^\lambda_{JKL|\mathcal{A}|\alpha I}\right.\cr
    &\left. -i\epsilon_{\lambda\beta\gamma\delta} \Theta^\lambda_{JKL|\mathcal{B}}(S^\mathcal{B}{}_\mathcal{A})_{\alpha I}\right)\approx 0\ .
\end{align}
Some terms of the above equations can be solved before solving the equations 
restricting the representations \eqref{tequations}. The first line is solved with 
\begin{align}
    \C{R}_{\alpha\beta|\mathcal{A}|\gamma,\delta}:= \C{R}_{\alpha\beta\gamma,\delta|\mathcal{A}}\in \yngBlue{;;;,;}\otimes\mathsf{M}\ .
\end{align}
The term with $\Sigma_{\sigma\alpha}$ in the second line implies 
\begin{align}
    \C{\Psi}^\alpha_{\beta|\gamma I|\mathcal{A}|\rho,\lambda}:= \C{\Psi}^\alpha_{\beta\rho,\lambda|\gamma I|\mathcal{A}}\in \yngBlue{;;,;}\otimes\yngBlue{;}^\yngBlue{!\bluefill;}\otimes\yngRed{;} \otimes\mathsf{M}\ .
\end{align}

\subsection{Exotic cocycle}

In the exotic case, we need the representation matrices $(T^{\mathcal{A}}{}_\mathcal{B})$ and $(S^{\mathcal{A}}{}_\mathcal{B})$ of the module $\mathsf{T}$ to go further. The 
relevant ones for the cocycle are 
\begin{align}\label{eq:module-matrices}
    &(T^{\rho\sigma}{}_\lambda{}^\omega)_{\gamma\delta}=-4i\,\delta_\omega^{(\rho}\delta_{[\gamma}^{\sigma)}\delta^\lambda_{\delta]}\ ,\qquad  
    (T^{\rho I}\;{}^\omega{}_J)_{\gamma\delta}=-2i\,\delta_{[\gamma}^{\rho}\delta_{\delta]}^{\omega}\delta^I_{J},\cr 
    &(T^{\rho I}{}_\sigma{}^\omega)_{\lambda J}=-i\,\delta^I_J\left(\delta_\lambda^\sigma\delta_\omega^\rho -\frac14\delta^\omega_\sigma \delta_\lambda^\rho\right)\ ,\qquad  
    (T^{IJ}\;{}^\omega{}_K)_{\lambda L}=2i\,\delta^\omega_\lambda\left( \delta^{[I}_K\delta^{J]}_L-\frac18\eta^{IJ}\eta_{KL}\right),\cr 
    & (S^{\rho\sigma}{}_{\omega I})_{\lambda J}=-2\,\delta_\omega^{(\rho}\delta_{\lambda}^{\sigma)}\eta_{IJ}\ ,\qquad  
    (S^{\rho I}{}_{JK})_{\lambda L}=\delta_\lambda^{\rho}\left(\delta^I_{[J}\eta_{K]L}+\frac18\delta^{I}_L\eta_{JK}\right).
\end{align}

The equation \eqref{teq1} implies 
\begin{align}
    R_{\alpha\beta|\gamma\delta}=R_{\alpha\beta\gamma\delta}\in\yngBlue{;;;;}\otimes\redBullet\ ,\\
    R_{\alpha\beta|\gamma I}=\Psi_{\alpha\beta\gamma I}\in\yngBlue{;;;}\otimes\yngRed{;}\ ,
\end{align}
also, by construction
\begin{align}
    R_{\alpha\beta|IJ}=H_{\alpha\beta IJ} \in\yngBlue{;;}\otimes\yngRed{;,;}_{\,\eta}.
\end{align}

The equation \eqref{teq2}  implies 
\begin{align}
    \Psi^\alpha_{\beta|\lambda I|\gamma\delta}\in \yngBlue{;;;}\otimes \yngBlue{;}^\yngBlue{!\bluefill;}\otimes \yngRed{;}\ ,\qquad  
    \Psi^\alpha_{\beta|\lambda I|\gamma J}\in \yngBlue{;;}\otimes \yngBlue{;}^\yngBlue{!\bluefill;}\otimes \yngRed{;}\otimes \yngRed{;}\ .
\end{align}

From equations \eqref{teq3} and \eqref{teq4} we get
\begin{align}
    &\Psi^\alpha_{\beta|\gamma I|\rho_1\rho_2}=\frac{3}{2} 
    \delta^\alpha_\gamma \Psi_{\beta\rho_1\rho_2 I}- \frac34\delta^\alpha_{(\rho_1}\Psi_{\rho_2)\beta\gamma I},\\ 
    &\Psi^\alpha_{\beta|\gamma I|\delta J}=6\delta^\alpha_\gamma H_{\beta\delta IJ} -2\delta^\alpha_\delta H_{\beta\gamma IJ},\\
    &\Psi^\alpha{}_{\beta|\gamma I|JK}=
    \delta^\alpha_\gamma\Theta_{\beta IJK}
    -\frac14\delta^\alpha_\beta\Theta_{\gamma IJK} \in
    \yngBlue{;}\otimes\yngRed{;,;,;}_{\,\eta}.
\end{align}

The equation \eqref{teq4} also implies
\begin{align}
    &H_{\alpha,\beta|\gamma,\delta|IJ|\sigma_1\sigma_2}= 
    \frac34 \epsilon_{\alpha\beta\gamma\delta}H_{\sigma_1\sigma_2 IJ}+\epsilon_{\alpha\beta(\sigma_1[\gamma}H_{\delta]\sigma_2)IJ},\\
    &H_{\alpha,\beta|\gamma,\delta|IJ|\sigma_1 K}=\frac12 \epsilon_{\alpha\beta\gamma\delta}\Theta_{\sigma_1 IJK}+\frac12 \epsilon_{\sigma_1 \alpha\beta [\gamma}\Theta_{\delta] IJK}. 
\end{align}

From equation \eqref{teq6} we obtain 
\begin{align}
    &\Theta^\alpha_{IJK|\beta L}=-2\delta^\alpha_\beta \Phi_{IJKL}\in \blueBullet\otimes\yngRed{;,;,;,;}_\eta,\\
    &\Theta^\alpha_{IJK|LM}=0.    
\end{align}
From this result and equation \eqref{teq5} we finally obtain
\begin{align}
    &H_{\alpha,\beta|\gamma,\delta|IJ|KL}= 3\epsilon_{\alpha\beta\gamma\delta}\Phi_{IJKL},\\
    &\Theta^\alpha_{IJK|\beta\gamma}=-\frac{1}{12} \delta^\alpha_{(\beta}\Theta_{\gamma)|IJK}.
\end{align}

The cocycle for the module $\mathsf T$ can be written in terms of five fields-strengths, as expected
\begin{align}\label{finalcocycle}
    \boldsymbol{\Sigma}^{(+)}&= 
    \Sigma^{\alpha\beta}\left( R_{\alpha\beta\gamma\delta}\btau^{\gamma\delta}+ \Psi_{\alpha\beta\gamma I}\btau^{\gamma I} + 
    H_{\alpha\beta IJ}\btau^{IJ}
    \right) \cr
    &+\Sigma_\alpha{}^\beta\wedge F^{\gamma I} \left( 
    \left[\frac32\delta^\alpha_\gamma \Psi_{\beta\sigma_1\sigma_2 I}-\frac34 \delta^\alpha_{\sigma_1}\Psi_{\sigma_2\beta\gamma I} \right]\btau^{\sigma_1\sigma_2} 
    +\left[6\delta_\gamma^\alpha H_{\beta\delta IJ}-2\delta^\alpha_\delta H_{\beta\gamma IJ} \right]\btau^{\delta J}\right.\cr
    &\left.
    +\delta_\gamma^\alpha\Theta_{\beta IJK}\btau^{JK}\right) \cr
    &+E^{\alpha\beta}\wedge \Sigma^{\gamma,\delta|IJ}\left( 
    \left[ \frac34\epsilon_{\alpha\beta\gamma\delta} H_{\sigma_1\sigma_2 IJ}
    +\epsilon_{\alpha\beta\sigma_1\gamma}H_{\delta\sigma_2 IJ} \right]\btau^{\sigma_1\sigma_2}\right.\cr 
    &\left.+\left[ \frac12\epsilon_{\alpha\beta\gamma\delta} \Theta_{\sigma_1 IJK}-\frac12 \epsilon_{\sigma_1\alpha\beta\gamma}\Theta_{\delta IJK} \right]\btau^{\sigma_1 K}+3 \epsilon_{\alpha\beta\gamma\delta}\Phi_{IJKL}\btau^{KL} \right)\cr
    &+\Sigma_\alpha{}^{IJK}\left( 
    -\frac{1}{12}\delta^\alpha_\beta \Theta_{\gamma IJK}\btau^{\beta\gamma} -2\delta^\alpha_\beta \Phi_{IJKL}\btau^{\beta L}
    \right) .
\end{align}

It is also useful to see the expression factorized in field-strengths 
\begin{align}
\label{finalcocycle-factorized}
\boldsymbol{\Sigma}^{(+)}
={}&
R_{\alpha\beta\gamma\delta}\,
\Sigma^{\alpha\beta}\btau^{\gamma\delta}
\cr
&+
\Psi_{\alpha\beta\gamma I}
\left(
\Sigma^{\alpha\beta}\btau^{\gamma I}
-\frac32
\Sigma_\rho{}^{\alpha\mid\rho I}
\btau^{\beta\gamma}
+\frac34
\Sigma_\rho{}^{\beta\mid\gamma I}
\btau^{\rho\alpha}
\right)
\cr
&+
H_{\alpha\beta IJ}
\left(
\Sigma^{\alpha\beta}\btau^{IJ}
+6
\Sigma_\rho{}^{\alpha\mid\rho I}
\btau^{\beta J}
-2
\Sigma_\rho{}^{\alpha\mid\beta I}
\btau^{\rho J}
\right.
\cr
&\hspace{3.4cm}\left.
+\frac32
\Sigma^{IJ}\btau^{\alpha\beta}
+2
\Sigma_\lambda{}^{\alpha\mid IJ}
\btau^{\lambda\beta}
\right)
\cr
&+
\Theta_{\alpha IJK}
\left(
-\Sigma_\rho{}^{\alpha\mid\rho I}
\btau^{JK}
+\Sigma^{IJ}\btau^{\alpha K}
+\Sigma_\lambda{}^{\alpha\mid IJ}
\btau^{\lambda K}
+\frac1{12}
\Sigma_\rho{}^{IJK}
\btau^{\rho\alpha}
\right)
\cr
&+
\Phi_{IJKL}
\left(
6\Sigma^{IJ}\btau^{KL}
-2\Sigma_\alpha{}^{IJK}
\btau^{\alpha L}
\right),
\end{align}
where we used
\begin{align}
    &\Sigma^{IJ}
    :=E_{\alpha\beta}\wedge\Sigma^{\alpha,\beta\mid IJ},
    \\
    &\Sigma_\alpha{}^{\beta\mid IJ}:=
    E_{\alpha\gamma}\wedge\Sigma^{\gamma,\beta\mid IJ},
\end{align}
and the identities
\begin{align}
    &\epsilon_{\rho\sigma\lambda\kappa}
    \Sigma^{\rho,\sigma\mid\lambda,\kappa\mid IJ}
    =2\Sigma^{IJ},
    \\
    &\epsilon_{\rho\sigma\lambda\kappa}
    \Sigma^{\rho,\sigma\mid\kappa,\alpha\mid IJ}
    =2\Sigma_\lambda{}^{\alpha\mid IJ},
    \\
    &\Sigma_\rho{}^{\rho\mid IJ}
    =-\Sigma^{IJ}.
\end{align}

Inserting \eqref{finalcocycle} in \eqref{sequations} and after some algebraic simplifications we finally obtain
\begin{align}
    &\nabla R_{\alpha(4)}\approx E^{\beta\gamma}C^R_{\alpha(4)\beta,\gamma} +F^{\beta I} \left(-\frac{5}{16} C^\Psi_{\alpha(4),\beta I} \right)\\
    &\nabla \Psi_{\alpha(3)I}\approx E^{\beta\gamma} C^\Psi_{\alpha(3)\beta,\gamma|I}+F^{\beta J} 
    \left( \frac34 C^H_{\alpha(3),\beta|IJ}+ 2i\eta_{IJ}R_{\alpha(3)\beta}\right)\\
    &\nabla H_{\alpha\beta IJ}\approx E^{\gamma\delta} C^H_{\alpha\beta\gamma,\delta|IJ}+ F^{\gamma K} 
    \left(-\frac14 C^\Theta_{\alpha\beta,\gamma IJK}
   -i\eta_{K[I}\Psi_{\alpha\beta\gamma J]}
   -\frac{i}{8}\eta_{IJ}\Psi_{\alpha\beta\gamma K}\right)\\
    &\nabla \Theta_{\alpha I[3]}\approx E^{\beta\gamma} C^\Theta_{\alpha\beta,\gamma|I[3]}+ F^{\beta J} 
    \left( 12 C^\Phi_{\alpha,\beta|I[3]J}
   +9i\eta_{JI}H_{\alpha\beta I[2]}
   -3i\eta_{I[2]}H_{\alpha\beta I J}\right)\\
    &\nabla \Phi_{I[4]}\approx E^{\alpha\beta} C^\Phi_{\alpha,\beta| I[4]}+ F^{\alpha J} 
    \left( -\frac{2i}{3}\eta_{JI}\Theta_{\alpha I[3]}
   -\frac{i}{2}\eta_{I[2]}\Theta_{\alpha I[2]J} \right).    
\end{align}

\subsection{Bi-grading of exotic cocycle}

The component expression for the three-form cocycle appears complicated because
the Lorentz contractions, the $USp(8)$ trace subtractions, and the numerical
coefficients of the different embeddings are all displayed simultaneously.
There is, however, a simple grading underlying the entire expression.  The five
primary field strengths form a single sequence
\begin{align}
    \mathcal W_r
    :=
    C_{\alpha(4-r)I[r]},
    \qquad
    r=0,\ldots,4,
\end{align}
where $I[r]$ denotes $r$ antisymmetric and symplectic-traceless $USp(8)$
indices.  Explicitly,
\begin{align}
    \mathcal W_0&=R_{\alpha(4)},&
    \mathcal W_1&=\Psi_{\alpha(3)I},&
    \mathcal W_2&=H_{\alpha(2)I[2]},&
    \mathcal W_3&=\Theta_{\alpha I[3]},&
    \mathcal W_4&=\Phi_{I[4]}.
\end{align}
Thus, increasing $r$ by one replaces one symmetric chiral spinor index by one
antisymmetric $USp(8)$ index.  In the oscillator description this is simply
the fermionic oscillator degree,
\begin{align}
    \mathcal W_r
    \quad\longleftrightarrow\quad
    \bar q^{\,4-r}\bar\xi^{\,r}.
\end{align}
The five apparently different field strengths are therefore the five
$H$-components of a single graded-symmetric rank-four object
$C_{A(4)}$.

We grade the possible three-forms by the number $s$ of fermionic frames:
\begin{align}
    \Omega_s^{(3)}
    \sim E^{3-s}F^s,
    \qquad
    s=0,\ldots,3.
\end{align}
Here $\Omega_s^{(3)}$ denotes the corresponding projected three-form module,
rather than a single form.  In terms of the basis used above,
\begin{align}
    \Omega_0^{(3)}
    &\sim \Sigma^{\alpha\beta},
    &
    \Omega_1^{(3)}
    &\sim \Sigma_\alpha{}^\beta\wedge F^{\gamma I},
    \cr
    \Omega_2^{(3)}
    &\sim E^{\alpha\beta}\wedge
    \Sigma^{\gamma,\delta|IJ},
    &
    \Omega_3^{(3)}
    &\sim \Sigma_\alpha{}^{IJK}.
\end{align}
We similarly grade the part of the two-form module entering the cocycle by
the number $t$ of odd superindices,
\begin{align}
    \mathsf T_0&=\left\langle\btau^{\alpha\beta}\right\rangle,
    &
    \mathsf T_1&=\left\langle\btau^{\alpha I}\right\rangle,
    &
    \mathsf T_2&=\left\langle\btau^{IJ}\right\rangle.
\end{align}
These are precisely the three components of the supertraceless,
graded-symmetric rank-two tensor $\btau^{AB}$.

The possible couplings obey the selection rule
\begin{align}\label{eq:cocycle-level-selection}
    r=s+t.
\end{align}
Consequently, the field strength appearing in the cell labelled by $(s,t)$
is always $\mathcal W_{s+t}$.  The complete pattern is
\begin{align}\label{eq:cocycle-hankel-pattern}
\begin{array}{c|ccc}
 & \mathsf T_0 &
   \mathsf T_1 &
   \mathsf T_2
\\[2pt] \hline
\Omega_0^{(3)}
 & R & \Psi & H
\\
\Omega_1^{(3)}
 & \Psi & H & \Theta
\\
\Omega_2^{(3)}
 & H & \Theta & \Phi
\\
\Omega_3^{(3)}
 & \Theta & \Phi & 0
\end{array}.
\end{align}
The matrix is constant along its anti-diagonals.  The vanishing lower-right
entry is not an additional constraint: it would require a field strength
$\mathcal W_5$, which is absent from the $(4,0)$ multiplet.

The same rule follows directly from conformal weights.  In the chiral patch
under consideration, the three gradings carry
\begin{align}
    \Delta(\mathcal W_r)
    &=4-\frac{r}{2},
    &
    \Delta(\Omega_s^{(3)})
    &=-3+\frac{s}{2},
    &
    \Delta(\mathsf T_t)
    &=-1+\frac{t}{2}.
\end{align}
It follows that
\begin{align}
    \Delta\left(
      \Omega_s^{(3)}
      \otimes\mathsf T_t
      \otimes\mathcal W_r
    \right)
    =
    \frac{s+t-r}{2}.
\end{align}
Conformal-weight conservation therefore gives $r=s+t$.  On the dual chiral
patch all conformal weights are reversed, but the same selection rule
follows.

The content of the cocycle can now be written as
\begin{align}\label{eq:cocycle-intertwiner-decomposition}
    \boldsymbol\Sigma^{(+)}
    =
    \sum_{\substack{0\leq s\leq3\\
                    0\leq t\leq2\\
                    s+t\leq4}}
    \mathcal I_{s,t}\left(\mathcal W_{s+t}\right),
\end{align}
where
\begin{align}
    \mathcal I_{s,t}
    \in
    \operatorname{Hom}_{H}
    \left(
       \mathcal W_{s+t},
       \Omega_s^{(3)}\otimes\mathsf T_t
    \right)
\end{align}
is a $H$-equivariant intertwiner.  Schematically, the sum of these
intertwiners reconstructs the single graded-superindex expression
\begin{align}
    \boldsymbol\Sigma^{(+)}
    :=
    \btau^{AB}\,
    \Sigma_{AB}{}^{C(4)}\,
    C_{C(4)}.
\end{align}
The long component expression is simply the
$Spin(1,5)\times SO(1,1)\times USp(8)$ decomposition of this object.

\section{Superconformally covariant unfolding} 

\label{Sec:6}

The super-Poincar\'e analysis of the previous section fixes the chiral three-form
cocycle and its first fermionic descendants on a chiral super-Poincar\'e  patch.  We
now restore the conformal superframe and ask how these data assemble into a
single superconformally covariant unfolded system.  The new
$\widetilde F$-components of the background connection act by special
supersymmetry.  Requiring the chiral cocycle to remain closed determines their action
on the five zero-form primaries and completes the corresponding
superconformal transformation laws.

Applying the conformal duality transformation to the chiral cocycle produces the anti-chiral cocycle as described in Section \ref{Sec:4.2}.

We also construct the superoscillator realization of the conformally dual chiral
and anti-chiral Weyl zero-forms.  
The Howe-dual constraints select the
$(4,0)$ field content, while the oscillator-level expansion organises the
primaries and their spacetime and fermionic descendants.  This provides a
compact superconformal packaging of the component equations derived above,
makes the passage between the two conformal patches explicit without
introducing additional propagating degrees of freedom, and confirms the construction in Section \ref{Sec:5}.

\subsection{Three-form cocycle}

In a superconformal background the covariant derivative of $F^{\alpha I}$ no longer vanishes 
\begin{align}
    \nabla F^{\alpha I}\approx - E^{\alpha\beta}\wedge 
    {\widetilde F}_\beta{}^I.
\end{align}
The three-forms in \eqref{finalcocycle-factorized} now satisfy
\begin{align}
    &\nabla\Sigma^{\alpha\beta}\big|_{\widetilde{F}}
    =0,\cr
    &\nabla\Sigma_\alpha{}^{\beta\mid\gamma I}\big|_{\widetilde{F}}=
    -\Sigma_\alpha{}^\beta
    \wedge E^{\gamma\rho}
    \wedge\widetilde F_\rho{}^I ,
    &\\
    &\Xi^{\alpha,\beta\mid IJ}
    :=\nabla\Sigma^{\alpha,\beta\mid IJ}\big|_{\widetilde{F}}
    =
    E^{\alpha\rho}\wedge F^{\beta[I}\wedge\widetilde F_\rho{}^{J]}
    -E^{\beta\rho}\wedge F^{\alpha[I} \wedge\widetilde F_\rho{}^{J]}
    \\
    &\quad
    +\frac18\eta^{IJ}
    \left(
    E^{\alpha\rho}\wedge F^{\beta K}
    -E^{\beta\rho}\wedge F^{\alpha K}
    \right)\wedge\widetilde F_{\rho K},
    \\
    &\nabla\Sigma^{IJ}\big|_{\widetilde{F}}=
    -E_{\alpha\beta}\wedge
    \Xi^{\alpha,\beta\mid IJ},
    \\
    &\Sigma_\lambda{}^{\alpha\mid IJ}\big|_{\widetilde{F}}
    =
    -E_{\lambda\beta}\wedge
    \Xi^{\beta,\alpha\mid IJ},
    \\
    &\nabla\Sigma_\alpha{}^{IJK}\big|_{\widetilde{F}}
    =
    6\Sigma_\alpha{}^{\rho\mid[IJ}
    \wedge\widetilde F_\rho{}^{K]}
    +3\Sigma^{[IJ}
    \wedge\widetilde F_\alpha{}^{K]}
    \\
    &\quad
    +2\eta^{[IJ}
    \Sigma_\alpha{}^{\rho\mid K]L}
    \wedge\widetilde F_{\rho L}
    +\eta^{[IJ}\Sigma^{K]L}
    \wedge\widetilde F_{\alpha L}.
\end{align}

Together with \eqref{tauFtilde1}, \eqref{tauFtilde2} and \eqref{tauFtilde3}, the cocycle $\boldsymbol{\Sigma}$ is closed in the superconformal background if
\begin{align}
\left.
\nabla R_{\alpha\beta\gamma\delta}
\right|_{\widetilde F}
&\approx
\frac54\,
\widetilde F_{(\alpha}{}^I
\Psi_{\beta\gamma\delta)I},
\\
\left.
\nabla\Psi_{\alpha\beta\gamma I}
\right|_{\widetilde F}
&\approx
-4\,
\widetilde F_{(\alpha}{}^J
H_{\beta\gamma)IJ},
\\
\left.
\nabla H_{\alpha\beta IJ}
\right|_{\widetilde F}
&\approx
\frac12\,
\widetilde F_{(\alpha}{}^K
\Theta_{\beta)IJK},
\\
\left.
\nabla\Theta_{\alpha IJK}
\right|_{\widetilde F}
&\approx
12\,
\widetilde F_\alpha{}^L
\Phi_{IJKL},
\\
\left.
\nabla\Phi_{IJKL}
\right|_{\widetilde F}
&\approx 0 .
\end{align}
As discussed in \cite{TwoCommaZero}, the components $F^{\alpha I}$ and $\widetilde F_\alpha{}^I$ are identified with the supersymmetry and superconformal transformations of the field-strengths respectively. The last equation then implies $\Phi_{IJKL}$ is a superconformal primary, as expected.

\subsection{Chiral and anti-chiral Weyl zero-forms}

The zero-form master field $\boldsymbol{c}(q,\bar q,\bar\xi)$ is a power series in its variables subject to the Howe dual constraints
\begin{align}
\widetilde{K}_{ij} \boldsymbol{c} = 0\ .
\end{align}
The constraint from $\widetilde K_{+-}$ implies that the powers in each monomial $q^a\bar q^b \bar\xi^c$ (suppressing indices) satisfy $-a+b+c=4$. 
To construct the zero-form master field, we start from the five conformal
primaries and dress each of them with the unique Howe-dual singlet polynomial
compatible with its number of spinor indices.  The operator
$\widetilde X=\widetilde K_{++}^{(\mathrm f)}
\widetilde K_{--}^{(\mathrm b)}$ replaces one $\bar q$ by one $q$ and inserts
the $USp(8)$-invariant fermionic bilinear
$U:=\bar\xi^I\bar\xi_I$.  Since
$-i\widetilde K_{++}^{(\mathrm f)}=U/2$, the term containing $k$ insertions
has coefficient $1/(2^k k!)$.  The series for a field with $m$ spinor indices
terminates at $k=m$, and the sum of the five resulting polynomials gives the
Howe-dual singlet master field.

\begin{align}
\boldsymbol{c}(q,\bar q,\bar\xi)
={}&
R_{\alpha\beta\gamma\delta}
\Bigg[
 \bar q^\alpha\bar q^\beta\bar q^\gamma\bar q^\delta
 +\frac12
 q^{(\alpha}\bar q^\beta\bar q^\gamma\bar q^{\delta)}U
 \nonumber\\
&\hspace{25mm}
 +\frac18
 q^{(\alpha}q^\beta\bar q^\gamma\bar q^{\delta)}U^2
 +\frac1{48}
 q^{(\alpha}q^\beta q^\gamma\bar q^{\delta)}U^3
 +\frac1{384}
 q^\alpha q^\beta q^\gamma q^\delta U^4
\Bigg]
\nonumber\\[2mm]
&+
\Psi_{\alpha\beta\gamma I}\bar\xi^I
\Bigg[
 \bar q^\alpha\bar q^\beta\bar q^\gamma
 +\frac12
 q^{(\alpha}\bar q^\beta\bar q^{\gamma)}U
 \nonumber\\
&\hspace{37mm}
 +\frac18
 q^{(\alpha}q^\beta\bar q^{\gamma)}U^2
 +\frac1{48}
 q^\alpha q^\beta q^\gamma U^3
\Bigg]
\nonumber\\[2mm]
&+
H_{\alpha\beta IJ}\bar\xi^I\bar\xi^J
\Bigg[
 \bar q^\alpha\bar q^\beta
 +\frac12 q^{(\alpha}\bar q^{\beta)}U
 +\frac18 q^\alpha q^\beta U^2
\Bigg]
\nonumber\\[2mm]
&+
\Theta_{\alpha IJK}
\bar\xi^I\bar\xi^J\bar\xi^K
\Bigg[
 \bar q^\alpha
 +\frac12 q^\alpha U
\Bigg] +
\Phi_{IJKL}
\bar\xi^I\bar\xi^J\bar\xi^K\bar\xi^L+\cdots ,
\label{eq:zero-form-master-field}
\end{align}

Here parentheses denote unit-weight symmetrisation of the spinor indices.
The ordering of the Grassmann-odd zero-forms and the fermionic oscillators is
kept exactly as displayed.  Equivalently, the construction can be written
compactly as
\begin{align}
\boldsymbol{c}
={}&
\sum_{m=0}^{4}
C_{\alpha(m)I[4-m]}\,
\bar\xi^{I_1}\cdots\bar\xi^{I_{4-m}}
\sum_{k=0}^{m}
\frac{1}{2^k k!}
\left(q^k\bar q^{\,m-k}\right)^{\alpha(m)}
U^k ,
\end{align}
where
\begin{align}
C_{\alpha(4)}&=R_{\alpha(4)},&
C_{\alpha(3)I}&=\Psi_{\alpha(3)I},&
C_{\alpha(2)IJ}&=H_{\alpha(2)IJ},&
C_{\alpha IJK}&=\Theta_{\alpha IJK},&
C_{IJKL}&=\Phi_{IJKL}.
\end{align}

The polynomial written above is the level-zero master field.  The different
powers of $U:=\bar\xi^I\bar\xi_I$
are not spacetime descendants.  Together, they form the Howe-dual completion
of the same conformal primary.  The first unfolded descendants appear at level
$\ell=1$, where each conformal module acquires one additional pair of spinor
indices.

For a primary with $m$ symmetric spinor indices, the first descendant is
\begin{equation}
X_{\alpha(m+1),\beta\mid I[4-m]},
\end{equation}
with Young symmetry
\begin{align}
X_{\alpha(m+1),\beta\mid I[4-m]}
&=
X_{(\alpha_1\cdots\alpha_{m+1}),\beta\mid I[4-m]},
\\
X_{(\alpha_1\cdots\alpha_{m+1},\beta)\mid I[4-m]}
&=0.
\end{align}
Thus, the five independent level-one tensors are
\begin{equation}
R_{\alpha(5),\beta},
\qquad
\Psi_{\alpha(4),\beta I},
\qquad
H_{\alpha(3),\beta IJ},
\qquad
\Theta_{\alpha(2),\beta IJK},
\qquad
\Phi_{\alpha,\beta\mid IJKL}.
\end{equation}
In the last case, the Young condition reduces to
\begin{equation}
\Phi_{\alpha,\beta\mid IJKL}
=
-\Phi_{\beta,\alpha\mid IJKL}.
\end{equation}
Their conformal weights are, respectively,
\begin{equation}
5,
\qquad
\frac92,
\qquad
4,
\qquad
\frac72,
\qquad
3.
\end{equation}

In oscillator language, the first descendant level is
\begin{align}
\boldsymbol C^{(1)}
={}&
R_{\alpha(5),\beta}\,
C_4(\widetilde X)\star
\bar q^{\alpha(5)}q^\beta
\nonumber\\
&+
\Psi_{\alpha(4),\beta I}\bar\xi^I\,
C_3(\widetilde X)\star
\bar q^{\alpha(4)}q^\beta
\nonumber\\
&+
H_{\alpha(3),\beta IJ}\bar\xi^I\bar\xi^J\,
C_2(\widetilde X)\star
\bar q^{\alpha(3)}q^\beta
\nonumber\\
&+
\Theta_{\alpha(2),\beta IJK}
\bar\xi^I\bar\xi^J\bar\xi^K\,
C_1(\widetilde X)\star
\bar q^{\alpha(2)}q^\beta
\nonumber\\
&+
\Phi_{\alpha,\beta\mid IJKL}
\bar\xi^I\bar\xi^J\bar\xi^K\bar\xi^L\,
\bar q^\alpha q^\beta .
\end{align}

The same dressing polynomial $C_m(\widetilde X)$ is used as at level zero.
Its explicit coefficients nevertheless change because
$\widetilde K_{--}^{(\mathrm b)}$ now acts on $m+1$, rather than $m$, factors
of $\bar q$.  Explicitly,
\begin{align}
C_m(\widetilde X)\star
\bar q^{\alpha(m+1)}q^\beta
={}&
\sum_{j=0}^{m}
a^{(1)}_{m,j}
\left(q^j\bar q^{\,m+1-j}\right)^{\alpha(m+1)}
q^\beta U^j,
\\
a^{(1)}_{m,j}
={}&
\frac{m+1}{m+1-j}\,
\frac{1}{2^j j!}.
\end{align}
The five coefficient sequences are therefore
\begin{align}
R:\qquad&
1,\quad
\frac58,\quad
\frac5{24},\quad
\frac5{96},\quad
\frac5{384},
\\
\Psi:\qquad&
1,\quad
\frac23,\quad
\frac14,\quad
\frac1{12},
\\
H:\qquad&
1,\quad
\frac34,\quad
\frac38,
\\
\Theta:\qquad&
1,\quad
1,
\\
\Phi:\qquad&
1.
\end{align}

Equivalently, the five oscillator dressing polynomials are
\begin{align}
\mathcal P_R^{(1)\,\alpha(5)\mid\beta}
={}&
\Bigg[
\bar q^{\alpha(5)}
+\frac58
\left(q\bar q^4\right)^{\alpha(5)}U
+\frac5{24}
\left(q^2\bar q^3\right)^{\alpha(5)}U^2
\nonumber\\
&\hspace{15mm}
+\frac5{96}
\left(q^3\bar q^2\right)^{\alpha(5)}U^3
+\frac5{384}
\left(q^4\bar q\right)^{\alpha(5)}U^4
\Bigg]q^\beta,
\\[2mm]
\mathcal P_\Psi^{(1)\,\alpha(4)\mid\beta}
={}&
\bar\xi^I
\Bigg[
\bar q^{\alpha(4)}
+\frac23
\left(q\bar q^3\right)^{\alpha(4)}U
+\frac14
\left(q^2\bar q^2\right)^{\alpha(4)}U^2
\nonumber\\
&\hspace{25mm}
+\frac1{12}
\left(q^3\bar q\right)^{\alpha(4)}U^3
\Bigg]q^\beta,
\\[2mm]
\mathcal P_H^{(1)\,\alpha(3)\mid\beta}
={}&
\bar\xi^I\bar\xi^J
\Bigg[
\bar q^{\alpha(3)}
+\frac34
\left(q\bar q^2\right)^{\alpha(3)}U
+\frac38
\left(q^2\bar q\right)^{\alpha(3)}U^2
\Bigg]q^\beta,
\\[2mm]
\mathcal P_\Theta^{(1)\,\alpha(2)\mid\beta}
={}&
\bar\xi^I\bar\xi^J\bar\xi^K
\Bigg[
\bar q^{\alpha(2)}
+\left(q\bar q\right)^{\alpha(2)}U
\Bigg]q^\beta,
\\[2mm]
\mathcal P_\Phi^{(1)\,\alpha\mid\beta}
={}&
\bar\xi^I\bar\xi^J\bar\xi^K\bar\xi^L
\bar q^\alpha q^\beta .
\end{align}

The first unfolded equations take the form
\begin{align}
\nabla R_{\alpha\beta\gamma\delta}
\approx{}&
E^{\rho\sigma}
R_{\alpha\beta\gamma\delta\rho,\sigma}
-\frac5{16}F^{\epsilon I}
\Psi_{\alpha\beta\gamma\delta,\epsilon I},
\\
\nabla\Psi_{\alpha\beta\gamma I}
\approx{}&
E^{\rho\sigma}
\Psi_{\alpha\beta\gamma\rho,\sigma I}
\nonumber\\
&+
F^{\delta J}
\left(
\frac43H_{\alpha\beta\gamma,\delta IJ}
+2i\eta_{IJ}R_{\alpha\beta\gamma\delta}
\right),
\\
\nabla H_{\alpha\beta IJ}
\approx{}&
E^{\rho\sigma}
H_{\alpha\beta\rho,\sigma IJ}
\nonumber\\
&+
F^{\gamma K}
\left(
-\frac14\Theta_{\alpha\beta,\gamma IJK}
-i\eta_{K[I}\Psi_{\alpha\beta\gamma J]}
-\frac{i}{8}\eta_{IJ}\Psi_{\alpha\beta\gamma K}
\right),
\\
\nabla\Theta_{\alpha IJK}
\approx{}&
E^{\rho\sigma}
\Theta_{\alpha\rho,\sigma IJK}
\nonumber\\
&+
F^{\beta L}
\left(
12\Phi_{\alpha,\beta\mid IJKL}
+9i\eta_{L[I}H_{\alpha\beta JK]}
-3i\eta_{[IJ}H_{\alpha\beta K]L}
\right),
\\
\nabla\Phi_{IJKL}
\approx{}&
E^{\rho\sigma}
\Phi_{\rho,\sigma\mid IJKL}
\nonumber\\
&+
F^{\alpha M}
\left(
-\frac{2i}{3}\eta_{M[I}\Theta_{\alpha JKL]}
-\frac{i}{2}\eta_{[IJ}\Theta_{\alpha KL]M}
\right).
\end{align}

Equivalently, defining the fermionic derivative by
\begin{equation}
\left.\nabla X\right|_F
:=
F^{\alpha I}\mathcal D_{\alpha I}X,
\end{equation}
the fermionic descendant chain is
\begin{align}
\mathcal D_{\epsilon I}R_{\alpha\beta\gamma\delta}
&=
-\frac5{16}
\Psi_{\alpha\beta\gamma\delta,\epsilon I},
\\
\mathcal D_{\delta J}\Psi_{\alpha\beta\gamma I}
&=
\frac43H_{\alpha\beta\gamma,\delta IJ}
+2i\eta_{IJ}R_{\alpha\beta\gamma\delta},
\\
\mathcal D_{\gamma K}H_{\alpha\beta IJ}
&=
-\frac14\Theta_{\alpha\beta,\gamma IJK}
-i\eta_{K[I}\Psi_{\alpha\beta\gamma J]}
-\frac{i}{8}\eta_{IJ}\Psi_{\alpha\beta\gamma K},
\\
\mathcal D_{\beta L}\Theta_{\alpha IJK}
&=
12\Phi_{\alpha,\beta\mid IJKL}
+9i\eta_{L[I}H_{\alpha\beta JK]}
-3i\eta_{[IJ}H_{\alpha\beta K]L},
\\
\mathcal D_{\alpha M}\Phi_{IJKL}
&=
-\frac{2i}{3}\eta_{M[I}\Theta_{\alpha JKL]}
-\frac{i}{2}\eta_{[IJ}\Theta_{\alpha KL]M}.
\end{align}

The $E$-components are the genuinely new level-one coordinates of the
unfolded module.  The $F$-components introduce no additional independent
zero-forms: supersymmetry fixes them in terms of the hook descendant in the
adjacent tower and the lower primaries required by the $USp(8)$ trace
completion.  On spacetime, these hook tensors are covariant derivatives of
the original fields rather than new propagating degrees of freedom.

\section{Conclusion and prospects}
\label{Sec:7}

We have unfolded Hull's six-dimensional $(4,0)$ gravity at the linearised level. The Cartan integrable system encodes 
\begin{itemize}
\item[--] background superconformal symmetries including conformal duality transformations; and 
\item[--] the abelian gauge structure and corresponding Bianchi identities.
\end{itemize}
The system consists of 
\begin{itemize}
\item[--] a background connection valued in the direct sum of the superconformal algebra and a Howe-dual algebra;
\item[--] a Weyl zero-form encoding the system's local
degrees of freedom into an extended supersingleton;
\item[--] conformally dual chiral and anti-chiral projections of the Weyl zero-form containing the on-shell curvatures and matter fields as conformal chiral primaries and anti-chiral anti-primaries;
\item[--] a conformally dual pair of two-form potentials; 
\item[--] a Howe-dual zero-form valued in the Howe-dual Lie algebra.
\end{itemize}
The main result is 
\begin{itemize}
\item[--]  the explicit form of a conformally dual pair of universally Cartan integrable three-form cocycles built from the background connection, the conformal primaries and anti-primaries, and the Howe-dual zero-form.
\end{itemize}
On super-Poincar\'e backgrounds, the anti-chiral cocycle vanishes, i.e., the anti-chiral two-form is pure gauge, and the chiral cocycle is determined uniquely by Cartan integrability.
To the latter end, the chiral cocycle is expanded in a special set of three-forms, viz.,
\begin{align}
\Sigma^{\alpha\beta}\ ,\qquad \Sigma_\alpha{}^\beta\wedge F^{\gamma I}\ ,\qquad E^{\alpha\beta}\wedge \Sigma_{(2)}^{\gamma,\delta|IJ}\ ,\qquad 
\Sigma^{IJK}_\alpha\ ,
\end{align}
forming a basis for the space of three-forms valued in traceless antisymmetric $R$-symmetry tensors.
Expanding the integrability condition over $E^{4-r}F^r$, $r=0,1,\dots,4$, it splits into $t$-equations, i.e., algebraic constraints determining how the concycle contains the conformal chiral primaries\footnote{In the $F^4$-sector, the condition restores the full $USp(8)$ trace completion.} 
\begin{align}
R_{\alpha\beta\gamma\delta},
\qquad
\Psi_{\alpha\beta\gamma I},
\qquad
H_{\alpha\beta IJ},
\qquad
\Theta_{\alpha IJK},
\qquad
\Phi_{IJKL}\ ;
\end{align}
and $s$-equations, i.e., differential constraints determining the primaries' first fermionic descendants. 
In the Wess--Zumino gauge, the constraints on the chiral Weyl zero-form and two-form, respectively, imply the linearised $(4,0)$ superspace constraints on chiral primaries
\cite{Koller:1982cs,Howe:1983fr,Chiodaroli:2011pp} and gauge potentials, i.e., the exotic graviton and gravitini, and the self-dual two-form potentials.

The superconformal completion is achieved by 

-- organising the on-shell curvatures on the super-Poincar\'e background into a superconformal chiral Weyl zero-form; and

-- introducing the anti-chiral Weyl zero-form to make the system conformal-duality covariant.

\noindent The chiral and anti-chiral Weyl zero-forms belong to Lie algebra modules that are dual in the sense that one is the space of linear maps on the other.
They do not encode separate local degrees of freedom.
Rather, they are complementary projections of a unique Weyl zero-form valued in a group module, alias, the extended supersingleton, providing overlapping coordinate charts
related by conformal duality symmetry.
Zero-form sectors of unfolded systems organised this way arise in strictly massless limits of massive theories \cite{BMVI}; for related mixtures of
singleton sectors in Coxeter higher-spin theories, see
\cite{Tarusov:2025sre}. 

A higher-spin gravity model with a cosmological constant has zero-form charges built from a single Weyl zero-form valued in a twisted-adjoint Lie algebra module that is its own dual \cite{fibre}.
The twisted-adjoint module has two distinct Poincar\'e limits, yielding a strictly massless Weyl zero-form and its dual.
In this limit, the quadratic zero-form charge of free higher-spin gravity becomes a quadratic zero-form of the strictly massless theory given by the pairing of the Weyl zero-form and its dual.
The corresponding quadratic zero-form charge for the superconformal $(4,0)$-theory is given in Section \ref{Sec:4.3}.

The conformal duality symmetry raises a global issue concerning singularity resolution.
Consider first a commutative 
sigma-model with field $f:X\to S^2$ and potential 
\begin{align}
V(f)=z(f)+\frac{1}{z(f)}
\end{align}
using the standard complex coordinate.
Thinking of the inversion $I:z\to 1/z$ as the analogue of the conformal duality transformation, $V$ is $I$-invariant, and $I$-transitions do not regularize singularities. 
The $(4,0)$ analogue is a configuration for which $\boldsymbol c^{(+)}$ diverges while $\kappa_\sigma\star\boldsymbol c^{(+)}$ remains finite.
If both conformal frames are finite, then both cocycles are finite and the chiral cocycle 
inherits the singularity. A transition function inside the original
structure group $H$ does not remove it.
However, near this singularity, one
may replace the $H$-covariant expansion by a basis adapted to another stabilizer, 
$H'$ say, leading to a finite symbol for $\boldsymbol{c}$.  Four-dimensional higher-spin solutions \cite{2011} provide a concrete precedent for this mechanism: $H$-tensorial
components that are singular in a spacetime expansion can correspond to
well-defined elements of the holomorphic metaplectic group algebra in an
adapted ordering or fibre basis
\cite{2017,COMST,meta,Iazeolla:2022singularities}.  
While this by no means shows that singularities can be resolved in the $(4,0)$ system, it does show
that regularity should ultimately be formulated for the master field and its
admissible transition functions, rather than inferred from a single local
tensor expansion.

Starting from the $Q$-structure of the linearised unfolded system, the interaction problem amounts to finding admissible deformation classes and testing their obstructions.
An associative star-product resolution on a noncommutative correspondence
space would provide additional control, as it does in Vasiliev-type higher-spin models
\cite{Vasiliev:1999ba,Bekaert:2004qos,Didenko:2014dwa}, though it is not implied by the present linear construction,
nor a prerequisite for identifying a classically consistent deformation \cite{Alexandrov:1995kv}.

Cogent tests that such a deformation must pass to be identified with Hull's interacting theory include 

-- reproduction of the
nonlinear couplings and duality structure of five-dimensional maximal supergravity upon circle reduction 
\cite{Hull:2000zn,Hull:2000rr};

-- deformations of the chiral self-duality constraints and the $(4,0)$ algebra.

\noindent They must also admit compatible reality conditions, patching data and boundary conditions, which are valid tests at the level of field
equations and their global solution spaces.  

As an intermediate step towards a fully nonlinear theory, one may seek a $(4,0)$ ``square'' of the self-dual string of the $(2,0)$ tensor multiplet \cite{Howe:1997ue}, which provides an unfolded system with a finite-dimensional zero-form sector, c.f., the unfolded Kerr black hole \cite{Didenko:2008va}.
Another such step would be to fill the niche between the $(4,0)$ theory and the (seemingly straightforward) $OSp(16|4)$ extension of the three-dimensional model of \cite{FSG2}, occupied by four-dimensional superconformal gravity based on $SU(2,2|8)$, with 32 Poincare supercharges, 32 conformal supercharges, and $SU(8)_R$-symmetry, which one may think of as a conformally unbroken phase of ${\cal N}=8$ supergravity extended by a dual Weyl zero-form treated as matter and topological conformal ${\cal N}=8$ higher-spin supergravity.
To this end, it would be natural to seek an extension of Vasiliev's four-dimensional formalism from the spinor oscillator realisation of $Sp(4;\mathbb{R})$ to $Sp(8;\mathbb{R})$.

In summary, while the presented work does not establish the existence of an
interacting $(4,0)$ theory, it provides the linear differential graded system in which that question can be posed sharply.  Hull's circle-reduction
proposal supplies the principal physical test, while the unfolded
formulation supplies the algebraic mechanism for constructing or obstructing
nonlinear deformations.

\paragraph{Acknowledgments.} We have enjoyed conversations with C. Arias, R. Aros, F. Caro, V.E. Didenko, D. Rovere, F. Silva, D. Tempo, O. Valdivia, and M. Valenzuela, and we are particularly thankful to Felipe Diaz for useful discussions on AKSZ sigma models and the nature of cocycles.
We have also benefitted from assisted computations and critical reviews by GPT-5.6 Sol.
This material is based upon work partially supported by the Swedish Research Council under grant no. 2021-06594 while B.C.V. was in residence at Institut Mittag-Leffler in Djursholm, Sweden during the Scientific Program ``Cohomological Aspects of Quantum Field Theory'' in 2025. P.S. expresses his gratitude to the CECs in Valdivia for hospitality during the initial stages of this work and for the support received from UNAP -- VRII Consolida grant ``Higher-spin inspired IR modifications of 3d gravity''; ANID grants Regular N°1250672 and N°1252053; and the MATH-AMSUD project SGP 24-MATH-12 funded by ANID and Minist\`ere de l'Europe et des Affaires \'Etrang\`eres.

\begin{appendix}
\section{Spinor Conventions}\label{App:spinors}

Starting from Clifford algebras over complex and real numbers, following \cite{Sezgin:2023hkc}, we arrive at our conventions for gamma matrices in eight, seven, and six dimensions.

\subsection{Complex Clifford algebras}

Let $Q:V\to \mathbb{C}$ be a quadratic form on a finite-dimensional, complex vector space $V$, i.e., $Q(x)=B(x,x)$ with $B:V\otimes V\to \mathbb{C}$ a bilinear form on $V$.
The Clifford algebra $\boldsymbol{\mathcal{C}}(V)$ is the associative algebra over the complex numbers with unit $\mathbf{1}$ generated by $e_x$, $x\in V$, obeying $(e_x)^2=Q(x)\mathbf{1}$.
Let $(x,y):=\frac12(Q(x+y)-Q(x)-Q(y))=\frac12(B(x,y)+B(y,x))$ be the (symmetric) inner product generated by $Q$.
It follows that 
\begin{align}
e_x e_y+e_ye_x=2(x,y)  \mathbf{1}\ .
\end{align}
Let $n:={\rm dim}_{\mathbb{C}}(V)$, and $x^i$, $i=1,\dots,n$, be an orthogonal basis for $V$, and set $\gamma^i:= e_{x^i}$, such that
\begin{align}
\gamma^i\gamma^j+\gamma^j\gamma^i=2\delta^{ij} \mathbf{1}\ .
\end{align}
It follows that $\mathbf{1}$ and $\gamma^{i[p]}:= \gamma^{[i_1}\cdots \gamma^{i_p]}$, $p=1,\dots,n$, form a basis for $\boldsymbol{\mathcal{C}}(V)$, hence ${\rm dim}(\boldsymbol{\mathcal{C}}(V))=2^n$.
Let $\boldsymbol{\mathcal{C}}_{[0]}(V)$ be the subalgebra of $\boldsymbol{\mathcal{C}}(V)$ spanned by $\mathbf{1}$ and $\gamma^{i[2k]}$, $k=1,\dots,[n/2]$; it follows that ${\rm dim}(\boldsymbol{\mathcal{C}}_{[0]}(V))=2^{n-1}$.
Let $\gamma:=i^{n(n-1)/2}\gamma^1\cdots \gamma^n$ and $P^\sigma:=\frac12(1+\sigma \gamma)$, $\sigma=\pm 1$, which obey $\gamma^2=1$ and $P^\sigma P^{\sigma'}=\delta^{\sigma\sigma'}P^\sigma$.
If $n$ is even, then the center $Z(\boldsymbol{\mathcal{C}}(V))=\mathbb{C}\otimes \mathbf{1}$, i.e., $\boldsymbol{\mathcal{C}}(V)$ is central simple; hence, by the Artin-Wedderburn theorem, 
\begin{align}
\mbox{$n$ even:}\quad {}&\boldsymbol{\mathcal{C}}(V)\cong{\rm Mat}_{2^{n/2}}(\mathbb{C})\ ,\\
{}&\boldsymbol{\mathcal{C}}_{[0]}(V)=\boldsymbol{\mathcal{C}}^+_{[0]}(V)\oplus \boldsymbol{\mathcal{C}}^-_{[0]}(V)\ ,\qquad \boldsymbol{\mathcal{C}}^\pm_{[0]}(V)\cong{\rm Mat}_{2^{(n-2)/2}}(\mathbb{C})\ ,
\end{align}
where $\boldsymbol{\mathcal{C}}^\sigma_{[0]}(V):=P^\sigma  \boldsymbol{\mathcal{C}}_{[0]}(V)$.
If $n$ is odd, then $\gamma$ is central, and $Z(\boldsymbol{\mathcal{C}}(V))=\mathbb{C}\otimes (P^+\oplus P^-)$; hence 
\begin{align}
\mbox{$n$ odd:}\quad {}&\boldsymbol{\mathcal{C}}(V)\cong  \boldsymbol{\mathcal{C}}^+(V)\oplus \boldsymbol{\mathcal{C}}^-(V)\ ,\qquad  \boldsymbol{\mathcal{C}}^\pm(V)\cong{\rm Mat}_{2^{(n-1)/2}}(\mathbb{C})\ ,\\ 
{}&\boldsymbol{\mathcal{C}}_{[0]}(V)\cong{\rm Mat}_{2^{(n-1)/2}}(\mathbb{C})\ ,
\end{align}
where $\boldsymbol{\mathcal{C}}^\sigma(V):=P^\sigma  \boldsymbol{\mathcal{C}}(V)$.
Thus, letting $\boldsymbol{\mathcal{C}}({\rm dim}(V);\mathbb{C}):=\boldsymbol{\mathcal{C}}(V)$, one has the (non-canonical) isomorphisms 
\begin{align}
\boldsymbol{\mathcal{C}}(2m;\mathbb{C})\cong \boldsymbol{\mathcal{C}}_{[0]}(2m+1;\mathbb{C})\ ,\qquad \boldsymbol{\mathcal{C}}(2m+1;\mathbb{C})\cong \boldsymbol{\mathcal{C}}_{[0]}^\pm(2m+2;\mathbb{C})\ .
\end{align}
It follows that $\boldsymbol{\mathcal{C}}(V)$ and $\boldsymbol{\mathcal{C}}_{[0]}(V)$, respectively, admit irreducible representations $D_n$ and $W_n$, viz.,  
\begin{align}
D_{2m}\cong& \mathbb{C}^{2^m}\ ,\qquad\qquad\qquad\qquad\ \  W^\pm_{2m}\cong P^\pm D_{2m}\cong \mathbb{C}^{2^{m-1}}\ ,\\
D^\pm_{2m+1}&= P^\pm D_{2m+1}\cong \mathbb{C}^{2^m}\ ,\qquad W_{2m+1}\cong \mathbb{C}^{2^m}\ ,
\end{align}
whose elements are referred to as Dirac and Weyl spinors.

\subsection{Real forms}

The real Clifford algebra $\boldsymbol{\mathcal{C}}_{t,s}$ of signature $(t,s)$ is the unital, associative algebra with nontrivial generators $\gamma^a$, $a=1,\dots,t+s$, obeying 
\begin{align}
\{\gamma^a,\gamma^b\}=2\eta^{ab} ,\qquad \eta^{ab}={\rm diag}(\underbrace{-,\dots,-}_{t},\underbrace{+,\dots,+}_{s})\ ,
\end{align} 
which admits a representation in the space of Dirac spinors equipped with its positive definite, Hermitian form $(\cdot,\cdot)_D$, viz.,
\begin{align}
S:= (D,(\cdot,\cdot)_D)\ ,\qquad D\cong\mathbb{C}^{2^{\left\lfloor(t+s)/2\right\rfloor}}\ ,
\end{align}
in terms of matrices $\Gamma^a$, referred to as Dirac gamma matrices, obeying
\begin{align}
(\Gamma^a)^\dagger=\left\{\begin{array}{l}
   -\Gamma^a\ ,\qquad a=1,\dots, t\ ,  \\
 +\Gamma^a\ ,\qquad a=t+1,\dots,t+s\ ,
\end{array}\right. 
\end{align}
that is,
\begin{align}
(\Gamma^a)^\dagger=(-1)^t A\Gamma^a A^{-1}\ ,\qquad A=\Gamma^1\cdots \Gamma^{t}\ ,
\end{align}
where $A$, referred to as the Dirac conjugation matrix, obeys
\begin{align}
A^\dagger=(-1)^{\frac12 (t+1)t}A\ ,\qquad A^2= (-1)^{\frac12 (t+1)t}\ .  
\end{align}
The Dirac conjugation map
\begin{align}
\bar\dagger:x\in S\mapsto \bar x := x^\dagger A\in S^\ast\ ,
\end{align}
such that $\bar\dagger( \Gamma^{ab} x)= -\bar x \Gamma^{ab}$. 
It also follows that 
\begin{align}
(\Gamma^a)^\ast=\eta B\Gamma^a B^{-1}\ ,\qquad \eta\in\{\pm 1\}\ ;
\end{align}
hence $B B^\ast$ is central. We choose $B$ to be unitary. Moreover, from
\begin{align}
A^\ast=\eta^t B A B^{-1}
\end{align}
it follows that
\begin{align}
(\Gamma^a)^T=(-1)^t \eta C\Gamma^a C^{-1}\ ,\qquad C:=BA\ , 
\end{align}
referred to as the charge conjugation matrix.
Hence $C(C^{-1})^T$ is central, i.e.,
\begin{align}
C^T=\epsilon_{C} C\ ,\qquad \epsilon_{C}\in \{\pm 1\}\ ,
\end{align}
which, when combined with
\begin{align}
A^T=\eta^t(-1)^{\frac12t(t+1)} C A C^{-1}\ ,
\end{align}
yields
\begin{align}
B^T=\epsilon B\ ,\qquad \epsilon=\epsilon_C \eta^t (-1)^{\frac12 t(t+1)}\ .
\end{align}
It follows that 
\begin{align}
(C\Gamma^{a_1}\cdots \Gamma^{a_n})^T=\epsilon\eta^{t+n}(-1)^{\frac12(t-n)(t-n+1)}C\Gamma^{a_1}\cdots \Gamma^{a_n}\ .
\end{align}
Since $A$ and $B$ are unitary, $C=BA$ is unitary as well. The transposition properties of $B$ and $C$ therefore imply
\begin{align}
B B^\ast=\epsilon\ ,\qquad C C^\ast=\epsilon_C\ ,
\end{align}
and, equivalently,
\begin{align}
B B^\ast=\eta^t(-1)^{\frac12t(t+1)}C C^\ast=\epsilon\ .
\end{align}
The possible reality conditions are
\begin{align}
s-t=0,1,2\ {\rm mod}\ 8\ :{}&\ (\epsilon,\eta)=(+,+)\ ,\qquad x^\ast=Bx\ ,\\
s-t=0,6,7\ {\rm mod}\ 8\ :{}&\ (\epsilon,\eta)=(+,-)\ ,\qquad x^\ast=Bx\ ,\\
s-t=4,5,6\ {\rm mod}\ 8\ :{}&\ (\epsilon,\eta)=(-,+)\ ,\qquad (x_i)^\ast=\Omega^{ij}Bx_j\ ,\\
s-t=2,3,4\ {\rm mod}\ 8\ :{}&\ (\epsilon,\eta)=(-,-)\ ,\qquad (x_i)^\ast=\Omega^{ij}Bx_j\ ,\\
\end{align}
where $\Omega^{ij}$ is a real symplectic matrix. The first two lines give Majorana and pseudo-Majorana conditions, respectively, while the last two give symplectic Majorana and pseudo-symplectic Majorana conditions. The eightfold pattern is the Clifford-algebra manifestation of Bott periodicity.
The chirality matrix
\begin{align}
\Gamma:= i^{t+\frac12(s+t)(s+t-1)}\Gamma^1\cdots \Gamma^{s+t}\ ,
\end{align}
obeys
\begin{align}
\Gamma^2=1\ ,\qquad \Gamma^\dagger=\Gamma\ ,    
\end{align}
and is central, hence proportional to $1$ in $S$, iff $s+t$ is odd.
If $s+t$ is even, the Weyl spinors are
\begin{align}
x^{(\pm)}:=\frac12(1\pm \Gamma)x\ ,
\end{align}
The reality condition preserves each Weyl subspace if and only if $s-t=0\ {\rm mod}\ 4$. Thus Majorana--Weyl conditions occur for $s-t=0\ {\rm mod}\ 8$, while symplectic Majorana--Weyl conditions occur for $s-t=4\ {\rm mod}\ 8$.

\subsection{Eight-dimensional cases} 

We let $t+s=8$, and represent $\boldsymbol{\mathcal{C}}_{t,s}$ using a set of $16\times 16$ Dirac gamma matrices $\widehat{\Gamma}^{\hat a}$, $\hat{a}=1,\dots, 8$, with corresponding conjugation and chirality matrices $(\widehat{A},\widehat{B},\widehat{C},\widehat{\Gamma})$.
Letting $\hat{\alpha}=1,\dots,16$ index Dirac spinors $\hat x$, one has
\begin{align}
\widehat C^{\hat{\alpha}\hat{\beta}}=\widehat C^{\hat{\beta}\hat{\alpha}}\ ,
\end{align}
and we can choose $\widehat C\widehat{\Gamma}^{\hat a}$ to be symmetric, i.e., we take
\begin{align}
\eta=(-1)^t \ ,\qquad (\widehat C\,\widehat \Gamma^{\hat a})^{\hat{\alpha}\hat{\beta}}=(\widehat C\,\widehat \Gamma^{\hat a})^{\hat{\beta}\hat{\alpha}}\ .
\end{align}
Choosing also
\begin{align}
(\widehat C_{\hat{\alpha}\hat{\beta}})^\dagger=\widehat C^{\hat{\alpha}\hat{\beta}}\ , 
\end{align}
it follows that 
\begin{align}
((\widehat A\widehat C)_{\hat{\alpha}\hat{\beta}})^\dagger=(-1)^{\frac12 (t+1)t} (\widehat C\widehat A)^{\hat\beta\hat\alpha}=(-1)^{t} (\widehat C\widehat A)^{\hat{\alpha}\hat{\beta}}\ .
\end{align}
Hence, 
\begin{align}
((\hat{x}_{\hat\alpha})^\dagger \widehat{A}_{\hat{\alpha}\hat{\beta}})^\dagger \widehat{A}_{\hat\beta\hat\gamma}=(-1)^{\frac12 (t-1)t} \hat{x}_{\hat\gamma}\ ,
\end{align}
i.e., the Clifford algebra of signature $(t,8-t)$ and $\eta=(-1)^t$ admits Majorana spinors for $t=0,1,4$, corresponding, respectively, to $(s-t,\eta)=(0,+), (6,-)$, $(0,+)$, and symplectic Majorana spinors for $t=2,3$, corresponding, respectively, to $(s-t,\eta)=(4,+),(2,-)$.

\subsection{Gamma matrices in signature \texorpdfstring{$(2,6)$}{(2,6)}}

We represent $\boldsymbol{\mathcal{C}}_{2,6}$ using $16\times 16$ gamma matrices 
\begin{align}
\widehat \Gamma^{\hat a}\ ,\qquad \hat{a}=0,1,\dots ,6,0'\ ,
\end{align}
obeying
\begin{align}
\widehat \Gamma^{(\hat a} \widehat \Gamma^{\hat b)}=\eta^{\hat a\hat b}={\rm diag}(-1,1,\dots,1,-1)\ ,
\end{align}
and use Dirac conjugation and chirality matrices
\begin{align}\label{A.38}
\widehat{A}=\widehat{\Gamma}^{0'0}\ ,\qquad \widehat{\Gamma}=\widehat\Gamma^{01\cdots 60'}\ ,
\end{align}
obeying
\begin{align}
\widehat \Gamma^2=1\ ,\qquad \widehat{A}^2=-1\ .
\end{align}
We choose the transposition and hermitian conjugation properties
\begin{align}
\widehat{C}^T= \widehat{C}\ ,\qquad (\widehat{C}\widehat{\Gamma}^{\hat{a}})^T=\widehat{C}\widehat{\Gamma}^{\hat{a}}\ ,\qquad \widehat C^\dagger=\widehat C^{-1}\ ,
\end{align}
i.e., we take $(t,s)=(2,6)$ and $(\epsilon,\eta)=(-,+)$, implying
\begin{align}
(\widehat C\widehat A)^{T}=-\widehat C\widehat A\ ,\qquad  (\widehat C\widehat \Gamma)^{T}=\widehat C\widehat \Gamma\ ,\qquad 
\widehat A^\dagger=-\widehat A\ ,\qquad\widehat \Gamma^\dagger=\widehat \Gamma\ .
\end{align}
In components, 
\begin{align}
(\widehat C^{\hat\alpha\hat\beta})^\ast&=(\widehat C^{-1})_{\hat\alpha\hat\beta}\ ,\qquad
((\widehat C\widehat A)^{\hat\alpha\hat\beta})^\ast=(\widehat A\widehat C^{-1})_{\hat\alpha\hat\beta}\ ,\qquad ((\widehat C\widehat \Gamma)^{\hat\alpha\hat\beta})^\ast=(\widehat \Gamma\widehat C^{-1})_{\hat\alpha\hat\beta}\ ,\\
((\widehat C\widehat\Gamma^{\hat a\hat b})^{\hat\alpha\hat\beta})^\ast&=-(\widehat A\widehat\Gamma^{\hat a\hat b}\widehat A\widehat C^{-1})_{\hat\alpha\hat\beta}\ .
\end{align}
The Weyl-spinor projectors 
\begin{align}
\widehat{\Pi}^{(\pm)}:=\frac12(1\pm \widehat \Gamma)\ ,
\end{align}
and we choose the Weyl basis in which 
\begin{align}\label{3.48}
\widehat C^{\hat\alpha\hat\beta}={}&\left[\begin{array}{cc}
\widehat C^{\underline{\alpha\beta}}&0 \\0&\widehat C^{\underline{\dot\alpha\dot\beta}} 
\end{array}\right]\ \ ,\qquad  (\widehat\Gamma^{\hat a})_{\hat \alpha}{}^{\hat\beta}=\left[\begin{array}{cc}
 0& (\widehat\Gamma^{\hat a})_{\underline\alpha}{}^{\dot{\underline\beta}}   \\ (\widehat\Gamma^{\hat a})_{\dot{\underline\alpha}}{}^{\underline\beta} 
     & 0
\end{array}\right]\ ,\qquad \underline{\alpha}, \underline{\dot\alpha}=1,\dots,8\ ,
\end{align} 
with symmetry properties 
\begin{align}
\widehat{C}^{\underline{\alpha\beta}}= \widehat{C}^{\underline{\beta\alpha}}\ ,\qquad \widehat{C}^{\underline{\dot\alpha\dot\beta}}=\widehat{C}^{\underline{\dot\beta\dot\alpha}}\ ,\qquad (\widehat C\widehat\Gamma^{\hat a})^{\underline{\alpha\dot\beta}}=(\widehat C\widehat\Gamma^{\hat a})^{\underline{\dot\beta\alpha}}\ ;
\end{align}
in this basis, 
\begin{align}
\widehat\Gamma_{\hat \alpha}{}^{\hat\beta}=\left[\begin{array}{cc}
\delta_{\underline\alpha}^{{\underline\beta}}  &0 \\ 0&-\delta_{\dot{\underline\alpha}}^{\dot{\underline\beta}}
\end{array}\right]\ .
\end{align}

\subsection{Gamma matrices in signature \texorpdfstring{$(1,6)$}{(1,6)}}

We represent $\boldsymbol{\mathcal{C}}_{1,6}$ using the $8\times 8$ gamma matrices
\begin{align}
\underline{\Gamma}^{\underline{a}}= \widehat{\Pi}^{(+)} \widehat\Gamma^{0'\underline{a}}\ ,\qquad \underline{a}=0,1,\dots,6\ ,
\end{align}
such that 
\begin{align}
\underline{\Gamma}^{01\dots 6}=-\widehat{\Pi}^{(+)}\ ,
\end{align}
and choose 
\begin{align}
\underline{C}= \widehat C\widehat{\Pi}^{(+)}\ ,\qquad  \underline{A}=\widehat{\Pi}^{(+)} \widehat A=\underline{\Gamma}^0\ ,\qquad 
\underline{A}^2=-1\ ,
\end{align}
with the transposition and hermitian conjugation properties
\begin{align}
\underline{C}^{T}&=\underline{C}\ ,\qquad (\underline{C}\,\underline{\Gamma}^{{\underline{a}}})^T=-\underline{C}\,\underline{\Gamma}^{{\underline{a}}}\ ,\qquad (\underline{C}\,\underline{A})^T=    -\underline{C}\,\underline{A}\ ,\\
\underline{A}^\dagger&=-\underline{A}\qquad \underline{C}^\dagger=\underline{C}^{-1}\ ,
\end{align}
i.e., $(t,s)=(1,6)$ and $(\epsilon,\eta)=(-1,+1)$;
in components, 
\begin{align}
((\underline{C}\,\underline{A})^{\underline{\alpha\beta}})^\ast={}&(\underline{A}\,\underline{C}^{-1})_{\underline{\alpha\beta}}\ ,\qquad ( \underline{C}^{\underline{\alpha\beta}})^\ast= (\underline{C}^{-1})_{\underline{\alpha\beta}}\ .    
\end{align}

\subsection{Gamma matrices in signature \texorpdfstring{$(1,5)$}{(1,5)}}

We represent $\boldsymbol{\mathcal{C}}_{1,5}$ using the $8\times 8$ gamma matrices 
\begin{align}
\Gamma^a=\underline{\Gamma}^a\ , \qquad a=0,1,\dots,5\ ,
\end{align}
and choose
\begin{align}
C&=\underline{C}\ ,\qquad A=\underline{A}=\Gamma^0\ ,\qquad \Gamma=\Gamma^{01\dots 5}=-\underline{\Gamma}^6\ ,\\
\qquad A^2&=-1\ ,\qquad \Gamma^2=1\ ,
\end{align}
with transposition and hermitian conjugation properties
\begin{align}
C^T&=C\ ,\qquad (C\Gamma^a)^T=- C\Gamma^a \ ,\qquad 
(CA)^T=-CA\ ,\qquad (C\Gamma)^T=-C\Gamma\ ,\\
A^\dagger&=-A\ ,\qquad C^\dagger=C^{-1}\ ,\qquad
 \Gamma^\dagger=\Gamma\ , 
\end{align}
i.e., $(t,s)=(1,5)$ and $(\epsilon,\eta)=(-1,+1)$; 
in components, 
\begin{align}
( {C}^{\underline{\alpha\beta}})^\ast= ({C}^{-1})_{\underline{\alpha\beta}}\ ,\qquad (({C}{A})^{\underline{\alpha\beta}})^\ast={}&({A}{C}^{-1})_{\underline{\alpha\beta}}\ ,\qquad (({C}{\Gamma})^{\underline{\alpha\beta}})^\ast=-({\Gamma}{C}^{-1})_{\underline{\alpha\beta}}\ ,\qquad     
\end{align}

\subsection{Six-dimensional chiral notation}

We use the Weyl basis 
\begin{align}\label{A.60}
(\Gamma^a)_{\underline{\alpha}}{}^{\underline{\beta}} =\left[\begin{array}{cc}
    0&(\sigma^a)_{\alpha\beta} \\
(\tilde{\sigma}^a)^{\alpha\beta}&0
\end{array}\right]\ ,\qquad C^{\underline{\alpha\beta}}=\left[\begin{array}{cc}
    0 &\delta^\alpha_\beta  \\
    \delta^\beta_\alpha &0 
\end{array}\right]\ ,
\end{align}
in which
\begin{align}
\Gamma_{\underline{\alpha}}{}^{\underline{\beta}} =\left[\begin{array}{cc}
    \delta_\alpha^\beta &  0\\
    0 &-\delta_\beta^\alpha 
\end{array}\right]\ ,\qquad 
\qquad \qquad A=\left[\begin{array}{cc}
    0&(\sigma^0)_{\alpha\beta} \\
(\tilde{\sigma}^0)^{\alpha\beta}&0
\end{array}\right]\ ,
\end{align}
where the $4\times 4$ sigma matrices obey
\begin{align}
    (\sigma^a)_{\alpha\beta}(\tilde\sigma^b)^{\beta\gamma}+ (\sigma^b)_{\alpha\beta}(\tilde\sigma^a)^{\beta\gamma}=2\delta_\alpha^\gamma\eta^{ab}\ ,
\end{align}
the symmetry and reality conditions
\begin{align}
(\tilde{\sigma}^a)^{\alpha\beta}={}&-(\tilde{\sigma}^a)^{\beta\alpha}\ ,\qquad (\sigma^a)_{\alpha\beta}=-(\sigma^a)_{\beta\alpha}\ ,\\
((\tilde{\sigma}^a)^{\alpha\beta})^\ast={}&-(\sigma^0 \tilde{\sigma}^a\sigma^0)_{\alpha\beta}\ ,\qquad ((\sigma^a)_{\alpha\beta})^\ast=-(\tilde{\sigma}^0 \sigma^a\tilde{\sigma}^0)^{\alpha\beta}\ ,
\end{align}
and the chirality condition
\begin{align}\label{A.62}
(\sigma^{[a} \tilde\sigma^b\sigma^c\tilde\sigma^d\sigma^e\tilde\sigma^{f]})_\alpha{}^\beta=\epsilon^{abcdef}\delta_\alpha^\beta \ , \qquad \epsilon^{01\dots 5}=+1\ ,
\end{align}
upon chosing an orientation of the Lorentz frame with volume form
\begin{align}
\mathcal{E}_{[6]}:= -\frac{1}{6!} \epsilon_{abcdef} E^a\wedge \cdots \wedge E^f= E^0\wedge \cdots \wedge E^5\ .
\end{align}
The real form implies that 
\begin{align}
((\tilde{\sigma}^0)^{\alpha\beta})^\ast=(\sigma^0)_{\alpha\beta}\ ,\qquad ((\sigma^0)_{\alpha\beta})^\ast=(\tilde{\sigma}^0)^{\alpha\beta}\ ,\label{3.53}
\end{align}
and that\footnote{The Lorentz invariance implies that there exist $\lambda,\tilde \lambda\in\mathbf{C}$ such that $\epsilon^{\alpha\beta\gamma\delta}(\sigma_a)_{\gamma\delta}=\lambda(\tilde\sigma_a)^{\alpha\beta}$ and $\epsilon_{\alpha\beta\gamma\delta}(\tilde\sigma_a)^{\gamma\delta}=\tilde\lambda(\sigma_a)_{\alpha\beta}$; iteration yields $\lambda\tilde\lambda=4$. Setting $a=0$ and complex conjugating yields $\tilde \lambda = \bar\lambda$, which can also be shown by combining direct complex conjugation, which yields $\lambda^\ast = {\rm det}( \tilde\sigma^0) \lambda = ({\rm Pf} (\tilde\sigma^0))^2 \lambda$, with $\epsilon_{\alpha\beta\gamma\delta}(\tilde\sigma_a)^{\alpha\beta}(\tilde\sigma_a)^{\gamma\delta}=-4\tilde\lambda \eta^{ab}$, which yields ${\rm Pf} (\tilde\sigma^0)=\tilde\lambda/2$, and finally using $\lambda\tilde\lambda=4$. It follows that $\lambda=2\eta$ and $\tilde\lambda=2\bar\eta$.} 
\begin{align}\label{raiseandlowersigma}
\epsilon^{\alpha\beta\gamma\delta}(\sigma_a)_{\gamma\delta}=2\eta(\tilde\sigma_a)^{\alpha\beta}\ , \qquad \epsilon_{\alpha\beta\gamma\delta}(\tilde\sigma_a)^{\gamma\delta}=2\bar\eta(\sigma_a)_{\alpha\beta}\ ,
\end{align}
where $|\eta|=1$ and $\epsilon^{\alpha\beta\gamma\delta}\epsilon_{\alpha\beta\gamma\delta}:=4!$; we choose the overall phase of the sigma matrices such that
\begin{align}
\eta=1\ .
\end{align}
One also has the Fierz identity
\begin{align}
(\sigma^a)_{\alpha\beta}(\tilde\sigma_a)^{\gamma\delta}=-4\delta_{\alpha\beta}^{\gamma\delta}\ ,
\end{align}
which, for the choice of $\eta$ made above, implies
\begin{align}
(\sigma^a)_{\alpha\beta}(\sigma_a)_{\gamma\delta}=-2\epsilon_{\alpha\beta\gamma\delta}\ , \qquad (\tilde\sigma^a)^{\alpha\beta}(\tilde\sigma_a)^{\gamma\delta}=-2\epsilon^{\alpha\beta\gamma\delta}\ .  
\end{align}
Products of pairs of sigma matrices yield one new structure, as 
\begin{align}
(\sigma_{ab})_\alpha{}^\beta:= (\sigma_{[a})_{\alpha\gamma}(\tilde\sigma_{b]})^{\gamma\beta}\ ,\qquad (\tilde\sigma_{ab})^\alpha{}_\beta:=  (\tilde\sigma_{[a})^{\alpha\gamma}(\sigma_{b]})_{\gamma\beta}\ ,
\end{align}
are related, viz.,
\begin{align}
(\sigma_{ab})_\alpha{}^\beta=-(\tilde\sigma_{ab})^\beta{}_\alpha\ .
\end{align}
Products of triplets of sigma matrices yield two independent structures, viz., 
\begin{align}
(\sigma_{abc})_{\alpha\beta}=(\sigma_{[a})_{\alpha\rho_1}(\tilde\sigma_b)^{\rho_1\rho_2}(\sigma_{c]})_{\rho_2\beta}\ ,\qquad 
(\tilde\sigma_{abc})^{\alpha\beta}= (\tilde\sigma_{[a})^{\alpha\rho_1}(\sigma_b)_{\rho_1\rho_2}(\tilde\sigma_{c]})^{\rho_2\beta}\ ,
\end{align}
which are separately symmetric, i.e.,
\begin{align}
(\sigma_{abc})_{\alpha\beta}=(\sigma_{abc})_{\beta\alpha}\ ,\qquad (\tilde\sigma_{abc})^{\alpha\beta}=(\tilde\sigma_{abc})^{\beta\alpha}\ .
\end{align}
The relations obtained so far imply the Fierz identities 
\begin{align}
(\sigma^{ab})_\alpha{}^\beta (\sigma_a)_{\gamma\delta}&=-\epsilon_{\alpha\gamma\delta\rho}(\tilde\sigma^b)^{\rho\beta}+2  (\sigma^b)^{\phantom{\beta}}_{\alpha[\gamma}\delta^\beta_{\delta]}\\
&= 4 (\sigma^b)^{\phantom{\beta}}_{\alpha[\gamma}\delta^\beta_{\delta]}+\delta_\alpha^\beta (\sigma^b)_{\gamma\delta}\ ,\\ 
(\sigma^{ab})_\alpha{}^\beta (\tilde\sigma_a)^{\gamma\delta}&=\epsilon^{\beta\gamma\delta\rho}(\sigma_b)_{\rho\alpha}-2(\tilde\sigma^b)^{\beta[\gamma}\delta_\alpha^{\delta]}\\
&=-4(\tilde\sigma^b)^{\beta[\gamma}\delta_\alpha^{\delta]}-\delta_\alpha^\beta (\tilde\sigma^b)^{\gamma\delta}\ ,\\ 
(\sigma^{ab})_\alpha{}^\beta(\sigma_{ab})_\gamma{}^\delta&= 10 \delta_{[\alpha\gamma]}^{\beta\delta}-6\delta_{(\alpha\gamma)}^{\beta\delta}=2\delta_\alpha^\beta\delta_\gamma^\delta-8\delta_\gamma^\beta\delta_\alpha^\delta\ ,\\
(\tilde\sigma^{abc})^{\alpha\beta}(\sigma_a)_{\gamma\delta}&=
-4\delta^{(\alpha}_{[\gamma}(\sigma^{bc})^{\phantom{(}}_{\delta]}{}^{\beta)}\ ,\\
(\sigma^{abc})_{\alpha\beta}(\sigma_{ab})_\gamma{}^\delta&=-8\delta^\delta_{(\alpha}(\sigma^c)_{\beta)\gamma}\ ,\\
(\tilde\sigma^{abc})^{\alpha\beta}(\sigma_{ab})_\gamma{}^\delta&=-8\delta^{(\alpha}_\gamma(\tilde\sigma^c)^{\beta)\delta}\ ,\\\label{A.82}
(\sigma^{abc})_{\alpha\beta}(\tilde\sigma_{abc})^{\gamma\delta}&=-48\delta^{\gamma\delta}_{(\alpha\beta)}\ ,\\
(\sigma^{abc})_{\alpha\beta}(\sigma_{abc})_{\gamma\delta}&=0\ .
\end{align}
Combining \eqref{A.82} with the chirality condition \eqref{A.62} yields the chirality properties
\begin{align}
\epsilon_{abcdef}(\sigma^{def})_{\alpha\beta}=-6 (\sigma_{abc})_{\alpha\beta}\ ,\qquad 
\epsilon_{abcdef}(\tilde\sigma^{def})^{\alpha\beta}=6 (\tilde\sigma_{abc})^{\alpha\beta}\ .
\end{align}
Thus $(\sigma_{abc})_{\alpha\beta}$ is anti-self-dual and $(\tilde\sigma_{abc})^{\alpha\beta}$ is self-dual. This agrees with the convention in the main text that symmetric anti-chiral and chiral bi-spinors represent anti-self-dual and self-dual three-forms, respectively; in particular, the three-form entering \eqref{5.8} is self-dual.
One also has
\begin{align}
(\sigma^a)_{\alpha\rho}(\tilde\sigma_a)^{\rho\beta}={}&6\delta_\alpha^\beta\ \ ,\qquad  (\sigma^{ab})_\alpha{}^\rho(\sigma_a)_{\rho\beta}=-5(\sigma^b)_{\alpha\beta}\ \ ,\qquad  (\sigma^{abc})_{\alpha\rho}(\tilde\sigma_a)^{\rho\beta}=4(\sigma^{bc})_\alpha{}^\beta\ ,\\
(\sigma^{ab})_\alpha{}^\rho (\sigma_{ab})_\rho{}^\beta={}&-30\delta_\alpha^\beta\ \ ,\qquad  (\sigma^{abc})_{\alpha\rho} (\tilde\sigma_{bc})^\rho{}_\beta=-20 (\sigma^c)_{\alpha\beta}\ \ ,\qquad  
(\sigma^{abc})_{\alpha\rho} (\tilde \sigma_{abc})^{\rho\beta}=-120\delta_\alpha^\beta\ ,
\end{align}
and
\begin{align}
(\sigma^{ab})_{\alpha}{}^\beta(\sigma_{cd})_{\beta}{}^{\alpha}=-8\delta^{[ab]}_{[cd]}\ \ ,\qquad  
(\sigma^{abc})_{\alpha\beta}(\tilde\sigma_{def})^{\alpha\beta}=-24\delta^{[abc]}_{[def]}\ .    
\end{align}

Other useful identities are 
\begin{align}
    &(\sigma_{[a})_{\alpha\beta}(\tilde\sigma_{b]})^{\gamma\delta}=-2
    \delta^{[\gamma}_{[\alpha}(\sigma_{ab})_{\beta]}{}^{\delta]} \ ,\qquad  
    (\sigma_{[a})_{\alpha\beta}(\sigma_{bc]})_{\gamma}{}^{\delta}= \frac{1}{3}\delta^\delta_{[\alpha}(\sigma_{abc})_{\beta]\gamma}-\frac13\epsilon_{\alpha\beta\gamma\rho}(\tilde\sigma_{abc})^{\rho\delta} ,\\ 
    &(\tilde\sigma_{[a})^{\alpha\beta}(\tilde\sigma_{bcd]})^{\gamma\delta}=-\frac{1}{4} \epsilon^{\alpha\beta(\gamma|\rho}(\sigma_{abcd})_{\rho}{}^{|\delta)}\ .
\end{align}
Similar ones can also be found using \eqref{raiseandlowersigma}\ .

\section{Superalgebra decompositions}\label{Sec:conventions}

\subsection{\texorpdfstring{$Spin(2,6)$}{Spin(2,6)} covariant basis}

The superconformal algebra $\mathfrak{osp}(8^\ast|8)$ consists of conformal generators 
\begin{align}\label{1.4} M_{\hat{a}\hat{b}}={}&\frac14 (\widehat C\widehat \Gamma_{\hat a\hat b})^{\hat\alpha\hat\beta}  M^{(+)}_{\hat\alpha\hat\beta}\ ,\qquad  M^{(+)}_{\hat\alpha\hat\beta}=\frac14 (\widehat \Pi^{(+)}\widehat \Gamma^{\hat a\hat b}\widehat C^{-1})_{\hat\alpha\hat\beta}  M_{\hat a\hat b}\ ,
\end{align} 
symplectic Majorana--Weyl supercharges 
\begin{align}
Q^{(+)}_{\hat{\alpha}I}=\widehat \Pi^{(+)}_{\hat\alpha}{}^{\hat\beta}Q^{(+)}_{\hat{\alpha}I}\ ,
\end{align}
and R-symmetry generators $N_{IJ}$, obeying 
\begin{align}
&[ Q^{(+)}_{\hat{\alpha}I},{ Q}^{(+)}_{\hat{\beta} J}] = \eta_{IJ} M^{(+)}_{\hat \alpha\hat \beta}+ ( \widehat C^{-1})_{\hat\alpha\hat\beta} N_{IJ}\ ,\\
&[ M_{\hat \alpha\hat \beta}, Q^{(+)}_{\hat{\gamma}}I]_\star=2i ( \widehat C^{-1})_{[\hat \beta|\hat \gamma}  Q^{(+)}_{|\hat{\alpha}]I}\ ,\qquad
[N_{IJ}, Q^{(+)}_{\hat{\alpha}K}]_\star=2i \eta^{\phantom{(}}_{(J|K} Q^{(+)}_{\hat{\alpha}|I)}\ ,\\
\label{1.7}
&[ M_{\hat \alpha\hat \beta}, M_{\hat \gamma\hat \delta}]_\star=4i( \widehat C^{-1})_{[\hat \beta|[\hat \gamma|} M_{\hat \alpha]|\hat \delta]}\ ,\qquad 
[N_{IJ},N_{KL}]_\star=4i\eta_{(J|(K|}N_{I)|L)}\ .
\end{align}
In the vectorial basis, the conformal algebra reads
\begin{align}\label{1.9}
[ M_{\hat a\hat b}, M_{\hat c\hat d}]_\star=4i\widehat{\eta}_{[\hat b|[\hat c|} M_{\hat a]|\hat d]}\ ,
\end{align}
which is equivalent to \ref{1.7} via \ref{1.4} and the Fierz identity, which would not hold without the projectors,
\begin{align}
(\widehat{\Pi}^{(\pm)}\widehat\Gamma^{[\hat{a}|\hat{b}}\widehat C^{-1})_{\hat{\alpha}\hat{\beta}}(\widehat{\Pi}^{(\pm)}\widehat\Gamma_{\hat{b}}{}^{|\hat{d}]}\widehat C^{-1})_{\hat{\gamma}\hat{\delta}}=4(\widehat{\Pi}^{(\pm)}\widehat C^{-1})_{[\hat{\gamma}|[\hat{\beta}}(\widehat{\Pi}^{(\pm)}\widehat \Gamma^{\hat{a}\hat{d}}\widehat C^{-1})_{\hat{ \alpha}]|\hat{\delta}]}\ .
\end{align}
The reality conditions are given by
\begin{align}
( M_{\hat{a}\hat{b}})^\dagger={}& M_{\hat{a}\hat{b}}\ ,\qquad  ( M^{(+)}_{\hat\alpha\hat\beta})^\dagger=(\widehat C\widehat A)^{\hat\alpha\hat\gamma} (\widehat C\widehat A)^{\hat\beta\hat\delta}  M^{(+)}_{\hat\gamma\hat\delta}\ ,\\
(N_{IJ})^\dagger={}&\Omega^{IK}\Omega^{JL} N_{KL}\ ,\qquad ( Q^{(+)}_{\hat{\alpha}I})^\dagger =\Omega^{IJ}(\widehat C\widehat A)^{\hat\alpha\hat\beta} Q^{(+)}_{\hat{\beta} J}\ .
\end{align}

\subsection{\texorpdfstring{$Spin(1,6)$}{Spin(1,6)} covariant basis}

Under $O(1,6)$, the conformal charges decompose into seven-dimensional Lorentz generators ${M}_{\underline{ab}}$ and transvections, viz., 
\begin{align}
{P}_{\underline{a}}:={M}_{\underline{a}0'}\ ,\qquad [{P}_{\underline{a}},{P}_{\underline{b}}]_\star=i{M}_{\underline{ab}}\ ,
\end{align}
and, going to the Weyl basis \eqref{3.48}, the supercharges become  symplectic Majorana spinors of $Spin(1,6)$, viz.,
\begin{align}
Q^{(+)}_{\hat{\alpha}I}=:\left[\begin{array}{c} {Q}_{I\underline{\alpha}}\\ 0\end{array}\right]\ ,\end{align}
with closure relation
\begin{align}
[Q_{I,\underline{\alpha}},Q_{J,\underline{\beta}}]_\star=\eta_{IJ}(\underline{M}{}_{\underline{\alpha\beta}}+\underline{P}{}_{\underline{\alpha\beta}})+(\underline{C}^{-1})_{\underline{\alpha\beta}}N_{IJ}\ ,\end{align}
where the bi-spinorial 
\begin{align}
\underline{M}{}_{\underline{\alpha\beta}}:=\frac14 (\underline{\Gamma}^{\underline{ab}})_{\underline{\alpha\beta}} M_{\underline{ab}}\ ,\qquad \underline{P}{}_{\underline{\alpha\beta}}:=-\frac12 (\underline{\Gamma}^{\underline{a}})_{\underline{\alpha\beta}} P_{\underline{a}}\ ,\end{align}
whose closure relations involve projectors, viz.,
\begin{align}
[\underline{M}_{\underline{\alpha\beta}},\underline{M}_{\underline{\gamma\delta}}]_\star&=\frac{i}{16} (\underline{\Gamma}^{\underline{ab}})_{\underline{\alpha\beta}} (\underline{\Gamma}_{\underline{b}}{}^{\underline{c}})_{\underline{\gamma\delta}} (\underline{\Gamma}_{\underline{ac}})^{\underline{\epsilon\varphi}} M_{\underline{\epsilon\varphi}} \ ,\\
[\underline{M}_{\underline{\alpha\beta}},\underline{P}_{\underline{\gamma\delta}}]_\star&=-\frac{i}{16} (\underline{\Gamma}^{\underline{ab}})_{\underline{\alpha\beta}} (\underline{\Gamma}_{\underline{b}})_{\underline{\gamma\delta}} (\underline{\Gamma}_{\underline{a}})^{\underline{\epsilon\varphi}} P_{\underline{\epsilon\varphi}} \ ,\\
[\underline{P}_{\underline{\alpha\beta}},\underline{P}_{\underline{\gamma\delta}}]_\star&=\frac{i}{16} (\underline{\Gamma}^{\underline{a}})_{\underline{\alpha\beta}} (\underline{\Gamma}^{\underline{b}})_{\underline{\gamma\delta}} (\underline{\Gamma}_{\underline{ab}})^{\underline{\epsilon\varphi}} M_{\underline{\epsilon\varphi}} \ .
\end{align}
These gamma-matrix projectors retain the distinct exchange symmetries of the $MM$, $MP$, and $PP$ commutators. In particular, the mixed commutator cannot be replaced by an expression symmetric under exchange of the two spinor-index pairs.

\subsection{\texorpdfstring{$Spin(1,5)\times O(1,1)$}{Spin(1,5)xO(1,1)}-covariant basis}

Under $O(1,5)\times O(1,1)$, we decompose the conformal $\mathfrak{so}(2,6)$ into $M_{ab}$ and 
\begin{align}\label{1.15}
D:=P_6\ ,\qquad T_a:=M_{6a}+P_a  \ ,\qquad K_a:=M_{6a}-P_a  \ , 
\end{align}
i.e., we use the ``yellow-book'' \cite{DiFrancesco} conventions, in which
\begin{align}\label{1.16}
[D,T_a]_\star=iT_a\ \ ,\qquad  [D,K_a]_\star=-iK_a\ \ ,\qquad 
[K_a,T_b]_\star=2i(\eta_{ab}D-M_{ab})\ ;
\end{align}
in the bi-spinoral basis,
\begin{align}
M_{\underline{\alpha\beta}}=\left[\begin{array}{cc} T_{\alpha\beta}& M_\alpha{}^\beta+\frac12\delta_\alpha^\beta D\\
M^\alpha{}_\beta-\frac12\delta^\alpha_\beta D& K^{\alpha\beta}\end{array}\right]\ ,\qquad M^\alpha{}_\beta=-M_\beta{}^\alpha\ ,
\end{align}
where 
\begin{align}
T_{\alpha\beta}:= -\frac12(\sigma^a)_{\alpha\beta} T_a\ ,\qquad M_\alpha{}^\beta=\frac14 (\sigma^{ab})_\alpha{}^\beta M_{ab}\ ,\qquad K^{\alpha\beta}:= \frac12(\tilde{\sigma}^a)^{\alpha\beta} K_a\ ,  
\end{align}
with commutation rules  
\begin{align}
[M_{\alpha}{}^{\beta},M_{\gamma}{}^{\delta}]_\star={}&i(\delta_\gamma^\beta M_{\alpha}{}^{\delta}-\delta_\alpha^\delta M_{\gamma}{}^{\beta})\ ,\qquad [D,M_\alpha{}^\beta]_\star=0\ ,\\
[M_{\alpha}{}^{\beta},T_{\gamma\delta}]_\star={}&2i\left(\delta_{[\gamma|}^\beta T^{\phantom{\beta}}_{\alpha|\delta]}-\frac14 \delta_\alpha^\beta T_{\gamma\delta}\right)\ ,\qquad [D,T_{\alpha \beta}]_\star=iT_{\alpha\beta}\ ,\\
[M_{\alpha}{}^{\beta},K^{\gamma\delta}]_\star={}&-2i\left(\delta_{\alpha|}^{[\gamma|} K^{\alpha|\delta]}_{\phantom{|}}-\frac14 \delta_\alpha^\beta K^{\gamma\delta}\right)\ ,\qquad [D,K^{\alpha \beta}]_\star=-iK^{\alpha\beta}\ ,\\
[T_{\alpha\beta},K^{\gamma\delta}]_\star={}&4i\delta_{[\beta}^{[\gamma}\left(M_{\alpha]}{}^{\delta]}+\frac12\delta_{\alpha]}^{\delta]} D\right)\ ,
\end{align}
and inverse relations
\begin{align}
T_a= \frac12 (\tilde\sigma_a)^{\alpha\beta}T_{\alpha\beta}\ ,\qquad M_{ab}=-\frac12 (\sigma_{ab})_\alpha{}^\beta M_\beta{}^\alpha\ ,\qquad K_a=-\frac12 (\sigma_a)_{\alpha\beta}K^{\alpha\beta}\ .
\end{align}
Under $Spin(1,5)\times O(1,1)$, the supercharges split into a pair of symplectic Majorana-Weyl spinors of $Spin(1,5)$ of opposite chiralities and conformal weights, viz., 
\begin{align}
Q^{(\pm)}_{\underline\alpha I}:=\Pi^{(\pm)}_{\underline\alpha}{}^{\underline\beta}Q_{\underline\beta I}\ ,\qquad \Pi^{(\pm)}:=\frac12 (1\pm \Gamma)\ .
\end{align}
In the Weyl basis \eqref{A.60}, we have 
\begin{align}
\qquad Q^{(+)}_{\underline\alpha I}=\left[\begin{array}{c}Q_{\alpha I}\\0\end{array}\right]\ ,\qquad Q^{(-)}_{\underline\alpha I}=\left[\begin{array}{c}0\\ S^{\alpha}{}_I\end{array}\right]\ ,    
\end{align}
identified as the supercharges and special  supercharges, respectively, in Poincar\'e backgrounds, with non-trivial closure relations
\begin{align}
[Q_{\alpha I},Q_{\beta J}]_\star={}&\eta_{IJ} T_{\alpha\beta}\ ,\qquad [S^{\alpha}{}_I,S^{\beta}{}_{J}]_\star=\eta_{IJ} K^{\alpha\beta} \ ,\\
[Q_{\alpha I},S^{\beta}{}_J]_\star={}&\eta_{IJ} \left(M_{\alpha}{}^{\beta}+\frac12\delta_\alpha^\beta D\right)+\delta_\alpha^\beta N_{IJ}\ ,\\
[T_{\alpha\beta},S^\gamma{}_I]_\star={}&2i\delta_{[\alpha}^\gamma Q^{\phantom{\gamma}}_{\beta] I}\ ,\qquad [K^{\alpha\beta},Q_{\gamma
I}]_\star=-2i\delta_{\gamma\phantom{I}}^{[\alpha} S^{\beta]}{}_I\ ,\\
[M_\alpha{}^\beta,Q_{\gamma I}]_\star={}&i\left(\delta_\gamma^\beta Q_{\alpha I}-\frac14\delta_\alpha^\beta Q_{\gamma I}\right)\ ,\qquad [D,Q_{\alpha I}]_\star=\frac{i}2 Q_{\alpha I}\ ,\\
[M_\alpha{}^\beta,S^{\gamma}{}_I]_\star={}&-i\left(\delta^\gamma_\alpha S^{\beta}{}_I-\frac14\delta_\alpha^\beta S^{\gamma}{}_I\right)\ ,\qquad  [D,S^{\alpha}{}_I]_\star=-\frac{i}2 S^{\alpha}{}_{I}\ .
\end{align}

\end{appendix}

{
\bibliographystyle{abe}
\bibliography{mybib}{}
}

\end{document}